\documentclass[twocolumn,tighten]{aastex63}
\usepackage{natbib}
\usepackage{amsmath}
\usepackage{listings}
\usepackage[normalem]{ulem}
\usepackage[section]{placeins}
\usepackage{xcolor}

\shorttitle{The erratic path to coalescence of LISA massive black hole binaries in CNDs}
\shortauthors{Souza Lima et al.}

\begin{document}

\title{{\Large \bf The erratic path to coalescence of LISA massive black hole binaries\\ in sub-pc resolution simulations of smooth circumnuclear gas disks}}


\correspondingauthor{Rafael Souza Lima}
\email{rafael@physik.uzh.ch}

\author[0000-0002-9282-2938]{Rafael Souza Lima}
\affiliation{Center for Theoretical Astrophysics and Cosmology, Institute for Computational Science, University of Zurich, Winterthurerstrasse 190, CH-8057 Z\"urich, Switzerland}

\author[0000-0002-7078-2074]{Lucio Mayer}
\affiliation{Center for Theoretical Astrophysics and Cosmology, Institute for Computational Science, University of Zurich, Winterthurerstrasse 190, CH-8057 Z\"urich, Switzerland}

\author[0000-0002-1786-963X]{Pedro R. Capelo}
\affiliation{Center for Theoretical Astrophysics and Cosmology, Institute for Computational Science, University of Zurich, Winterthurerstrasse 190, CH-8057 Z\"urich, Switzerland}

\author[0000-0001-9458-821X]{Elisa Bortolas}
\affiliation{Center for Theoretical Astrophysics and Cosmology, Institute for Computational Science, University of Zurich, Winterthurerstrasse 190, CH-8057 Z\"urich, Switzerland}

\author{Thomas R. Quinn}
\affiliation{Department of Astronomy, University of Washington, Seattle, WA 98195-1580, USA}


\begin{abstract}
We report on high-resolution simulations that explore the orbital decay of massive black hole (MBH) pairs 
with masses between $10^5$ and $10^7 M_{\odot}$ embedded in a circumnuclear gas disk (CND). An adiabatic
equation of state is adopted, with a range of adiabatic indices, which maintains a smooth flow.
Mergers between MBHs in this mass range would be detectable by the upcoming Laser Inteferometer Space Antenna (LISA). 
The orbital evolution is followed from the CND scale ($100$~pc) down to separations of $0.1$--$0.01$~pc at which 
a circumbinary disk (CBD) could form. The decay is erratic and strongly dependent on the gas flow within the disk, 
that ultimately determines the net torques experienced by the sinking MBH. Overall, we can identify 
three different evolutionary stages:  (i) an initially slow decay that leads to no significant change 
in the orbital angular momentum, resulting in some circularization; (ii) 
a fast migration phase in which the orbital 
angular momentum decreases rapidly; and (iii) a final, very slow decay phase, 
in which orbital angular momentum can even increase, and a CBD can form. 
The fast migration phase owes to disk-driven torques originating primarily 
from the co-orbital region of the secondary MBH, at a distance of 1--3 Hill radii. 
We find strong analogies with fast Type III migration for massive planets in protoplanetary disks. 
The CBD forms only when the decay rate becomes small enough to allow it enough time to carve a cavity around the primary MBH, 
at scales $\lesssim 1$~pc; when this happens, the MBH separation nearly stalls in our higher-resolution run. 
We suggest an empirically modified gap-opening criterion that takes into account such timescale 
effects as well as other deviations from standard assumptions made in the literature. 
Interestingly, a CBD does not form in the lower-resolution runs, resulting in a faster decay rate at sub-pc separations. 
Our findings indicate that the orbital decay at sub-pc separations in gaseous disks is an inefficient and fragile process. 
We quantify how additional mechanisms, such as hardening by three-body encounters with stars, 
might become crucial, even in these gas-rich environments, in order for the binary to reach the gravitational wave emission stage.
\end{abstract}



\section{Introduction}\label{sec:Introduction}

Binaries embedded in gaseous disks are a common configuration in the Universe. Some typical examples are, in the lower range of binary mass ratios, planet--star systems in protoplanetary disks \citep{ward97, armitage07}, and, in the upper range, young stellar binaries \citep{shu87, orosz12}. They are also expected in the form of massive black hole (MBH) binaries in the nuclei of gas-rich galaxy mergers \citep[e.g.][]{mayer07, mayer10,Pfister_et_al_2017}. Most, if not all galaxies host a MBH, and the larger galaxies are assembled via mergers \citep{mundy17}; thus, a multitude of MBHs in the resulting merger remnants could provide observational evidence for the effects of their interaction with the environment, for instance via simultaneous activity of active galactic nuclei (AGN; e.g. \citealt{Capelo_et_al_2017,DeRosa_et_al_2020}) and ultimately with the strong emission of gravitational waves (GWs) when two of them inspiral and coalesce into one. Therefore, a detailed description of such interactions supplies substantial insight about the physics happening at scales now still only marginally resolved for most astronomical sources. The most detailed theoretical description of the corresponding dynamical systems to date is obtained with the aid of computer simulations, which have been proved as a suitable tool due to their flexibility, adaptability, and generality.

The interaction between a background gaseous disk and a perturber has been an active field of research for decades, receiving special attention in the context of planet migration \citep[e.g.][]{goldreich80, lin86, crida06, papaloizou07, malik2015, FlemingQuinn17} and, more recently, also in the context of MBH pairs (e.g. \citealt{dotti06, mayer10, dotti12, mayer13, goicovic16, souzalima17}, hereafter Paper~I; \citealt{tang17}). Early work have established the types I and II of migration regimes for planets in protoplanetary disks, followed by the type III in the mid 2000's, that correspond, respectively, to low-mass planets that induce perturbations in the disk, large-mass planets that can cause a gap to form in the host disk, and partial gap-forming planets in more massive disks \citep{papaloizou07}. For MBHs, the main migration mechanism originally considered was dynamical friction against a stellar background \citep{chandrasekhar43, begelman80}, which could then be further assisted by its variant for a gaseous background \citep{ostriker99, escala05}. It was also noted then that, at binary hardening scales, the operational conditions for migration would fail, but further efficient migration could be re-triggered by the emission of GWs once the tightly bound binary reached small enough separations. Since then, the field has developed further, as a range of scenarios were studied with the advent of better computer simulations, pointing towards a more diverse picture than originally surmised \citep{mayer13}.

The planned launch into space of the Laser Interferometer Space Antenna (LISA; \citealt{Amaro-Seoane_et_al_2017,Barack_et_al_2019}), a GW observatory, calls for a better understanding of the dynamics of MBH pairs in its detection window ($10^5$--$10^7~M_\odot$). However, there has been debate on the typical outcome in the transition scale bridging migration at larger separations and the GW regime which, in this mass range, happens around and below 1~pc, an issue often referred to as the `last (or final) parsec problem' \citep[e.g.][]{begelman80,Milosavljevic_Merritt_2001}. It is possible that gas can further assist MBH pairs to coalesce at sub-pc scales \citep{escala05}, but the details of their joint interaction may vary with the system configuration \citep{roedig12}. One can gather insight regarding the possible configurations as outcomes of the MBH--disk interaction from larger scales, by referring to planet formation literature and drawing analogs by re-scaling their systems -- for example, when trying to determine whether this interaction can cause a gap to form.

In \citet{crida06}, they developed a criterion to determine whether a perturber could cause a gap in a protoplanetary disk by comparing opening and closing (viscous) torques. Their work was expanded by \citet{malik2015}, who accounted also for the perturber's migration timescale, which should be large enough in comparison to the timescale to carve a gap, otherwise the perturber would be dislocated to another orbit before the gas response could build up. More recently, \citet{muller18} have tested those criteria in a set of semi-analytical simulations, and found that they can hold as long as a fudge factor is introduced. New criteria were introduced by \citet{kanagawa18}, who studied the migration of gap-opening planets in more detail, also with the aid of two-dimensional simulations, motivated by divergences between the results of contemporary simulations and the picture established by the standard type II migration regime.

In the context of MBH binaries in self-gravitating disks, there are additional challenges, such as a larger and less constrained parameter space. For example, the mass ratios for the binary can range from values closer to one, which fall in the ranges commonly investigated on binary star formation and dynamics, to some that could be scaled to those typically found in the planet migration literature (in the order of a hundredth), to other combinations that are atypical in either of those fields, since astrophysical MBHs have known masses ranging from a few times to tens of billions times the mass of the Sun. Despite of this conundrum, some work has been done to disentangle key aspects of the gas--MBH interactions and MBH pair migration regimes: \citet{delvalle12} established a link between the cavity opening conditions and the mass geometry induced in the wake of the perturber, supplemented by follow-up studies \citep{delvalle14, delvalle15}.

In the past decade, the effects of gravitational torques in the gas-assisted evolution of MBH binaries have been studied in more detail (see, e.g. \citealt{roedig12, tang17}, and references therein). Justifiably, these works have emphasized the crucial phase of sub-pc binary separations, where, as we pointed before, some mechanisms that can drive faster orbital decays at larger scales are no longer effective means for an eventual MBH binary coalescence \citep{begelman80, chapon13}. However, fully three-dimensional (3D) smoothed particle hydrodynamical (SPH) simulations of such systems fall short in numbers, and those few studies typically only considered the evolution in the last phase of the decay, where the computational domain and initial conditions (ICs) are set for stable or marginally stable configurations, for which the binary lives in a cavity of a circumbinary disk (CBD) that extends up to a few times the separation between the two bodies.

Our work differs from those by considering 3D SPH simulations of a MBH pair in a self-gravitating circumnuclear disk (CND), where their mutual evolution is followed across hundreds of pc down to the limits of our force resolution, which is, in the best cases, a fraction of a pc, in an attempt to bridge the larger-scale picture with the one from the works mentioned above. Along this evolution, we confront some of the gap-opening prescriptions with the results from our simulations, to test their predictability for the various conditions experienced by our systems, for example the configurations of orbital potential and disk height met along the orbit of the migrating perturber. We also follow the general aspects of the fundamental gravitational torques caused by the gas on the MBH pair from different ranges and at different separation scales, in an attempt to spot particular characteristics of the MBH--disk interactions in different regimes.

We applied some of the current gap-opening criteria on a set of simulations of a CND, to test their applicability in a wider range of scenarios than originally envisioned by their underlying assumptions. The same criteria will be applied to study the opening of the CBD cavity at small separation because the mass ratio between the secondary and the primary is large enough (1:20) that the motion of the primary can be neglected in first approximation (gap-opening criteria assume implicitly that the gap is opened by a perturbed moving in a static central potential). The description of the ICs and simulations, and of the formalism utilized in the gap-opening analysis are described in Section~\ref{sec:methods}. The results of the simulations, sub-divided into the analysis of the orbital decay and torques responsible for it, and the gap-opening analysis, are presented in Section~\ref{sec:Results}. In Section~\ref{sec:Discussion}, we discuss the implication of the results on the predictability range of the gap-opening criteria, on some aspects of the gas-assisted orbital decay of a MBH pair, and on the detection rates of dual AGN and future GW missions, besides the conclusions and limitations of this work.


\section{Methods}\label{sec:methods}

The numerical simulations were performed using the Tree-SPH $N$-body code ChaNGa \citep{jetley08, menon15}, which shares the same SPH algorithm and sub-grid physics modules of Gasoline \citep[][]{stadel01,wadsley04,wadsley17}, used in our previous works on CNDs with embedded MBH pairs (e.g. \citealt{fiacconi13}; \citealt{roskar15}; Paper~I). In the remainder of the section, we describe the setup of the ICs and the analysis strategy.

\subsection{Initial conditions and simulations}\label{sec:ic}

\tabletypesize{\footnotesize}
\begin{deluxetable}{lllll}
	\tablecolumns{4}
	\tablecaption{Technical specifications of the simulations described in this paper. Column (1): name. Column (2): gravitational softening. Column (3): type of SPH kernel (cubic spline versus Wendland~C$^2$) and number of SPH neighbors. Column (4): total number of SPH particles. Column (5): adiabatic index.
	\label{tab:simulations}
	}
	\tablehead{
		\colhead{Label\tablenotemark{a}} & \colhead{$\epsilon$ [pc]} & \colhead{kernel/neighbors} & \colhead{$N_{\textrm gas}$} & \colhead{$\gamma_\mathrm{ad}$}
	}
	\startdata
	{\it LReps50g14}			    	& 0.5	& spline/32		& $2 \times 10^{5}$		& 1.4\\
	{\it LReps50g12}		        	& 0.5	& spline/32		& $2 \times 10^{5}$		& 1.2\\
	{\it LReps50g12A}	    & 0.5	& spline/32		& $2 \times 10^{5}$		& 1.2\\
	{\it MReps50g14}			    	& 0.5	& spline/32		& $2 \times 10^{6}$		& 1.4\\
	{\it MReps23g14R}	    	& 0.23	& Wendland/64	& $2 \times 10^{6}$		& 1.4\\
	{\it MReps23g11R}			        	& 0.23	& Wendland/64	& $2 \times 10^{6}$		& 1.1\\
	{\it HReps05g14}		    	& 0.05	& Wendland/64	& $2 \times 10^{7}$		& 1.4
	\enddata
	\tablenotetext{a}{Runs {\it LReps50g14}, {\it LReps50g12}, and {\it LReps50g12A} were named {\it naad g14}, {\it naad g12}, and {\it ad g12}, respectively, in Paper~I. Run {\it LReps50g12A} includes gas accretion onto MBHs. See Paper~I for a description of the implementation of BH physics.}
	\vspace{-20pt}
\end{deluxetable}

We used an initial CND model similar to that in Paper~I. For this work, we ran a large suite of simulations at varying mass and force resolution. Only the low-resolution runs were previously presented in Paper~I. The CND models adopt a \citet{mestel1963} disk for the underlying density profile, and are similar to the models used in \citet{escala05}, \citet{dotti07}, and \citet{fiacconi13}. The disk is represented by $2 \times 10^5$ (low resolution; LR), $2 \times 10^6$ (medium resolution; MR), or $2 \times 10^7$ (high resolution; HR) SPH (gas) particles, depending on the run (see Table~\ref{tab:simulations}), with total mass $M_{\textrm{d}} = 10^8 M_{\odot}$, initialized with a surface density following a \citet{mestel1963} profile with scale radius $R_{\textrm{d}} = 100$~pc and maximum radius $\sim$150~pc, and a Gaussian vertical structure of scale height $z_{\textrm{d}} = h R$, where $h=0.05$ is the aspect ratio. The initial temperature of the gas particles is set to approximately $T_0 = 4.9 \times 10^3$~K in all but the low-resolution runs, where $T_0 = 2 \times 10^4$~K. The disk is embedded in a concentric \citet{plummer1911} spheroid (bulge) represented by $10^6$ dark matter particles, with total mass $M_{\rm b} = 5 \times 10^8 M_{\odot}$, scale radius $r_{\rm b} = 50$~pc, and maximum radius $\sim$500~pc. The spheroid helps to stabilize the disk and represents the inner part of a galactic bulge. In the center of the system, the primary MBH (MBH1) is placed at rest, with a mass $M_1 = 10^7 M_\odot$.

In the low-resolution runs ({\it LReps50g14, LReps50g12, LReps50g12A}), the system (CND plus primary MBH) is left to relax for a few orbits, i.e. around 20~Myr. Then, a secondary MBH (MBH2) with mass $M_2 = 5 \times 10^5 M_{\odot}$ is placed close ($\sim$2~pc) to the midplane at 80~pc separation from the center, which initially coincides with the location of the primary MBH. Its initial velocity is determined by the modulus of the circular velocity at the given position and by imposing the radial-to-tangential velocity ratio equal to 1, thus enforcing an initial eccentricity $e = 0.707$. The medium-resolution \emph{MReps50g14} run is initialized similarly. However, the relaxation time is 11~Myr and MBH2's initial separation from the center is 100~pc (and less than 1~pc from the midplane). The high-resolution \emph{HReps05g14} run also follows the same procedure, but the relaxation time is 2--3~Myr and the initial MBH2's separation from the center is 5~pc. The other (medium-resolution) runs -- \emph{MReps23g14R} and \emph{MReps23g11R} -- start from a snapshot of the \emph{MReps50g14} run, occurring around 27~Myr past MBH2's original placement. The differences from their parent run, as well as the force softening parameter for the MBHs and gas particles, are described in Table~\ref{tab:simulations}.

\subsection{Gap-opening analysis}\label{sec:gap-criteria}

According to the semi-analytical torque balance criterion by \citet{crida06} and \citet{malik2015}, a perturber of mass $M_2$ (typically a planet) will open a gap in a disk dominated by the gravitational influence of the central primary object (of mass $M_\star$; typically a star) if it satisfies

\begin{equation}
\label{eq:crida}
\frac{3}{4}\frac{H}{R_{\rm H}} + \frac{50 \nu_{\rm p}}{q \Omega_{\rm p} r^2} \lesssim 1,
\end{equation}

\noindent where $H$ is the disk's full width at half maximum (FWHM), $R_{\rm H} \simeq r (q/3)^{1/3}$ is the Hill radius of the perturber, $r$ is its distance to the disk's center, $q = M_2/M_{\star}$ is the mass ratio between the perturber and the central primary object, $\nu_{\rm p} = \alpha c_{\rm s} H$ is the physical disk kinematic viscosity according to the $\alpha$-disk model in \citet{shakura73}, with $\alpha$ being a dimensionless stress parameter and $c_{\rm s}$ the speed of sound, and $\Omega_{\rm p}$ is the Keplerian angular frequency for a circular orbit at $r$.

The assumptions that lead to the inequality above are not met in our self-gravitating disk with a migrating perturber \citep[see][]{muller18}, as standard gap-opening criteria have been derived for perturbers on a fixed orbit in a cold, thin, laminar disk. A self-gravitating disk develops gravitoturbulence, which thickens the disk and effectively increases the pressure support against gravity (turbulence introduces {\it de facto} a non-thermal pressure). Nonetheless, we can investigate phenomenologically how predictive  the inequality \eqref{eq:crida} is for a range of parameters that are relevant to our system. Adapting and rearranging its terms, the criterion reads

\begin{equation}
\label{eq:c}
c \leq \left( \frac{g}{H} - b \beta \right) \frac{1}{\gamma},
\end{equation}

\noindent where we substituted the right-hand side of inequality \eqref{eq:crida} with $g$, which is a fudge factor to account, at least phenomenologically, for the deviations from the assumptions behind the criterion by \citet{crida06}. In addition, we substituted $M_{\star}$ by $M_{\rm{enc}}(r)$, the enclosed mass at radius $r$, as a proxy for the main source of gravitational acceleration at that distance, and defined $c = 50\,\alpha\,c_s\,M_2^{-1}\,G^{-1/2}$, $b = 3(3/M_2)^{1/3}/4$, $\beta = M_{\rm{enc}}^{1/3} / r$, and $\gamma = (M_{\mathrm{enc}}/r)^{1/2}$, $G$ being the gravitational constant. Since all the additional physics mentioned above has been shown, by previous work, to go in the direction of stifling gap opening, we expect that our numerical results will be more easily matched by using $g < 1$ (a small value of $g$ indeed corresponds to restricting the parameter space for which the conditions expressed in inequality~\ref{eq:c} can be satisfied).

\subsection{Analysis of disk-driven torques}

The orbit of the secondary MBH can shrink, leading to the formation of a binary and its subsequent hardening, depending on the magnitude and sign of the gravitational torques exerted by the surrounding matter. As the MBHs are initially placed in the disk plane, and remain within the disk for the whole orbital evolution (their maximum separation from the disk plane is 2 pc, never exceeding the disk's scale height), and the mass density is dominated by the gas in the disk plane, it is expected that torques from the gas in the disk will be dominant relative to torques from the stars and dark matter, which have a lower mass density and  are more extended by construction. Due to our choice of masses for the MBHs, we focus our torque analysis on the secondary MBH  since it is the source of essentially all the orbital angular momentum in the MBH pair. We recall that, in order to decay, the secondary MBH has to lose orbital energy, and as such it will lose also orbital angular momentum. However, the secondary MBH might lose or gain orbital angular momentum even if its orbit does not decay, thus maintaining constant orbital energy. This would result in a change of orbital eccentricity. The character of the torques will determine the character of the decay, whether or not the eccentricity of the secondary MBH is modified, and whether or not the orbital distance from the primary MBH decreases or not. 

Throughout the paper, the gravitational torque from the gas onto the secondary MBH is computed via

\begin{equation}
    \mathbf{\tau_{\mathrm{gas}}} = \sum_{\rm particles}{\mathbf{r_\mathrm{CoM}} \times \mathbf{F_\mathrm{g}}},
    \label{eq:torque}
\end{equation}

\noindent where $\mathbf{r_\mathrm{CoM}} = \mathbf{r_\mathrm{MBH2}} - \mathbf{r_\mathrm{CoM,0}}$ is the position of MBH2 relative to the center of mass of the MBH pair, $\mathbf{r_\mathrm{CoM,0}}$ is the center of mass of the MBH pair, and $\mathbf{F_\mathrm{g}}$ is the gravitational force from each gas particle over which the summation occurs. Hereafter, we refer to torques along the orbital angular momentum $\mathbf{L}$ of the secondary MBH  -- a ``negative'' (``positive'') torque means that its projection along the $\mathbf{L}$ direction is anti-parallel (parallel) to $\mathbf{L}$ -- at a given instant.


\section{Results}\label{sec:Results}

Our focus is on understanding the nature of the orbital decay of a massive MBH from the edge of the CND, at $\sim$100~pc scales, down to below sub-pc scales, eventually approaching the stage at which a cavity might be carved around a tight MBH binary, thus leading to a CBD as that assumed in the ICs of the last stage of MBH binary evolution just preceding GW emission \citep[e.g.][]{tang17}. For the sake of clarity, we mainly use five of the runs described in Table~\ref{tab:simulations}, to interpret the results of our analysis, unless otherwise stated. This is because these runs have the largest differences within the set, representing cases with different mass and/or force resolution, different equation of state (EoS), and even different initialization procedure. These reference runs are {\it LReps50g14, MReps50g14, MReps23g14R, MReps23g11R}, and the {\it HReps05g14} run, which is our highest-resolution run.

\begin{figure*}
    \centering
	\includegraphics[width=.95\linewidth]{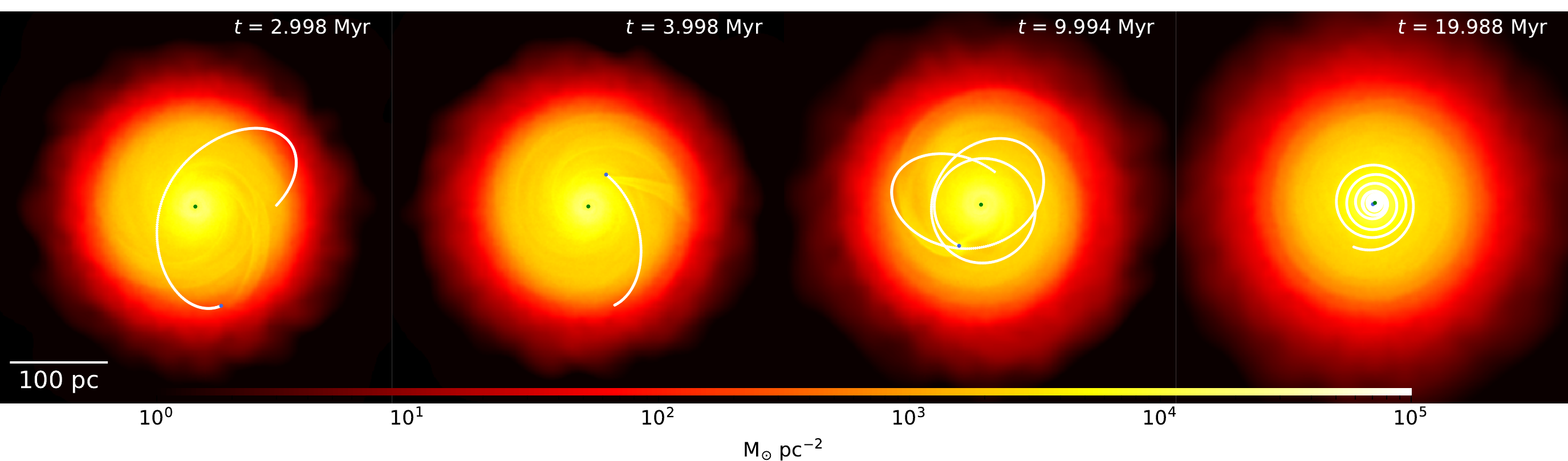}\\
	\includegraphics[width=.95\linewidth]{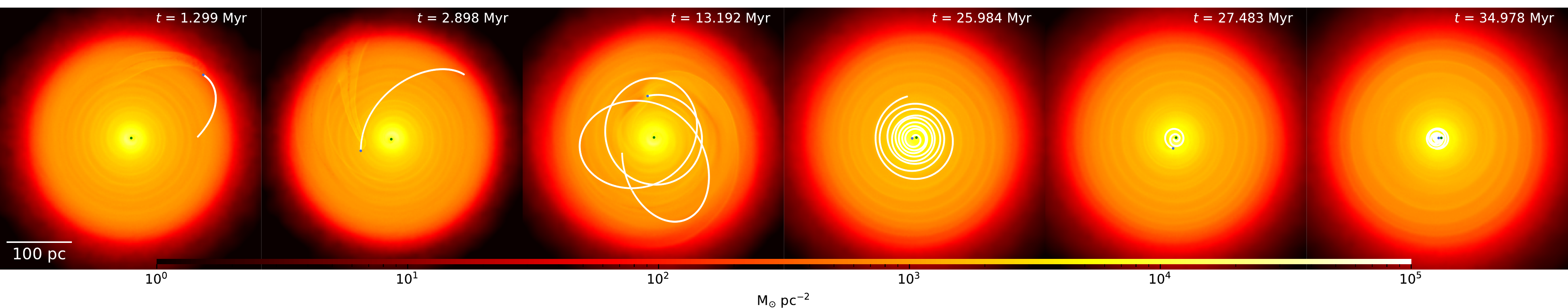}\\
	\includegraphics[width=.95\linewidth]{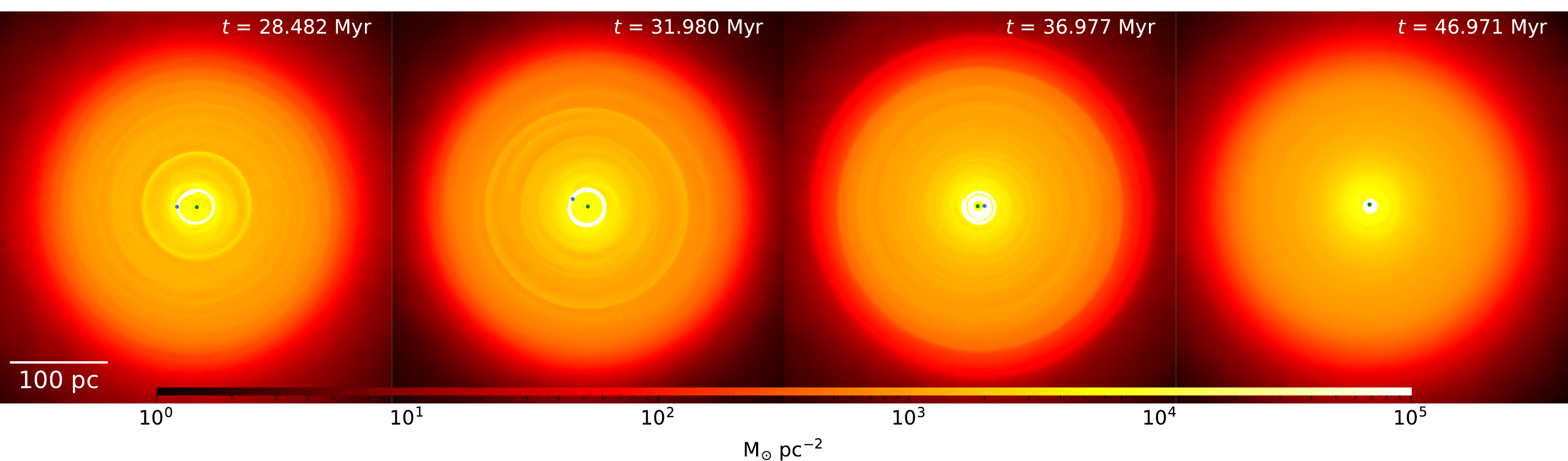}\\
    \includegraphics[width=.95\linewidth]{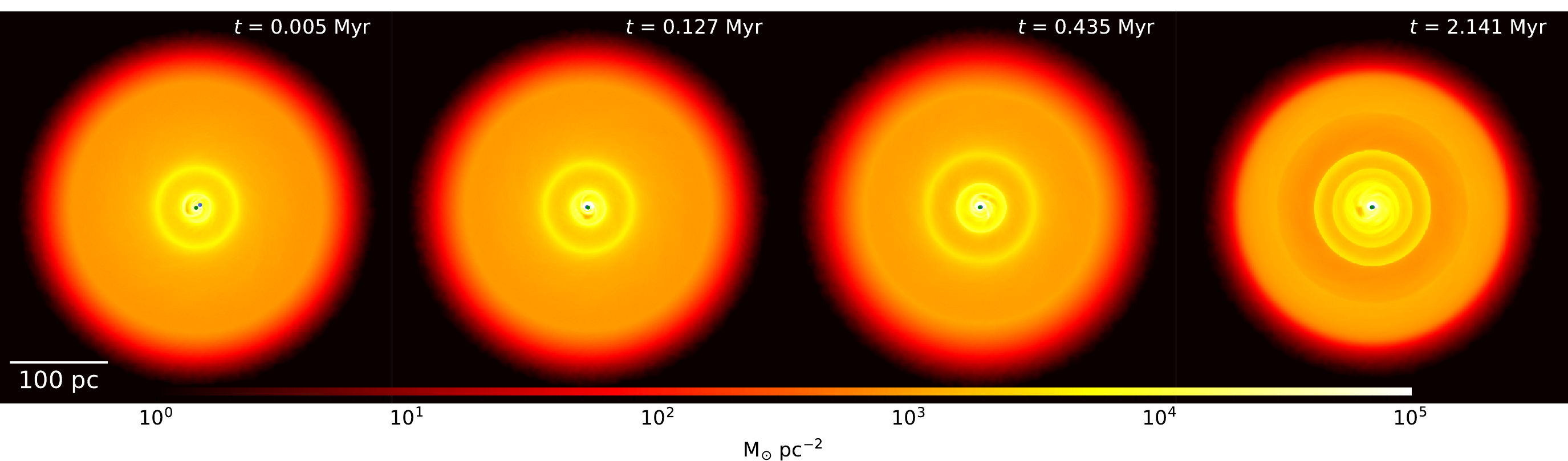}
    \caption{Large-scale, mass-weighted integrated gas density maps for a subset of the simulations described in Table~\ref{tab:simulations}. From top to bottom, each row corresponds to \emph{LReps50g14}, \emph{MReps50g14}, \emph{MReps23g14R}, and \emph{HReps05g14}. The MBHs are indicated with filled circles. Note that run \emph{MReps23g14R} (as well as run \emph{MReps23g11R}, not shown here) started from the snapshot shown in the fifth panel of the second row.
    }
    \label{fig:simcomparison-rho}
\end{figure*}

\begin{figure*}
    \centering
    \includegraphics[width=.95\linewidth]{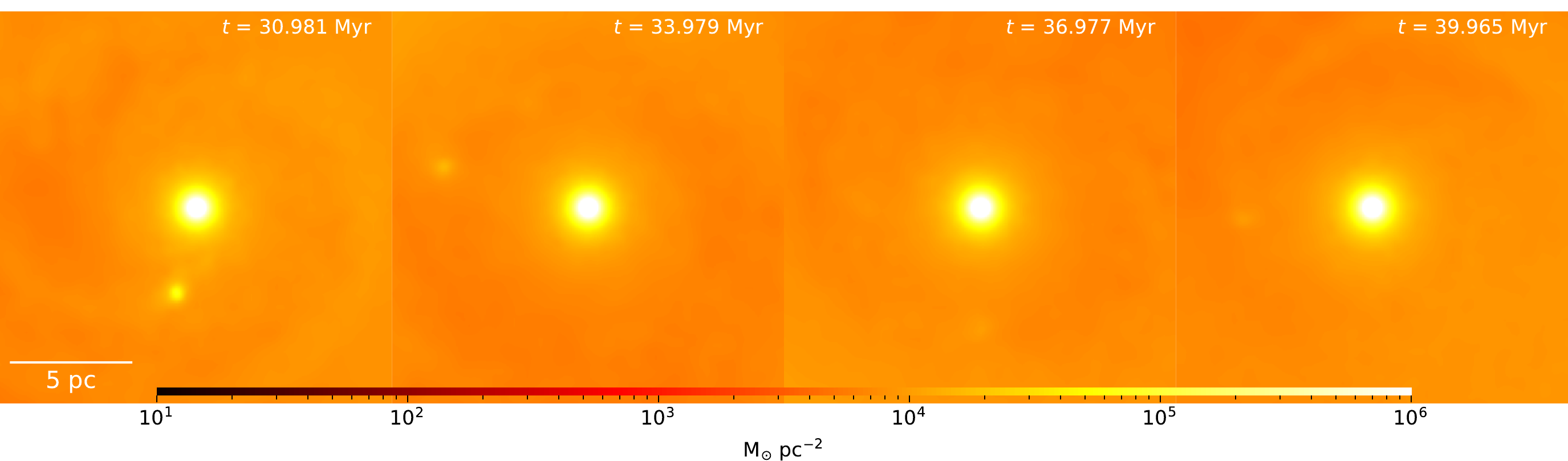}\\
    \includegraphics[width=.95\linewidth]{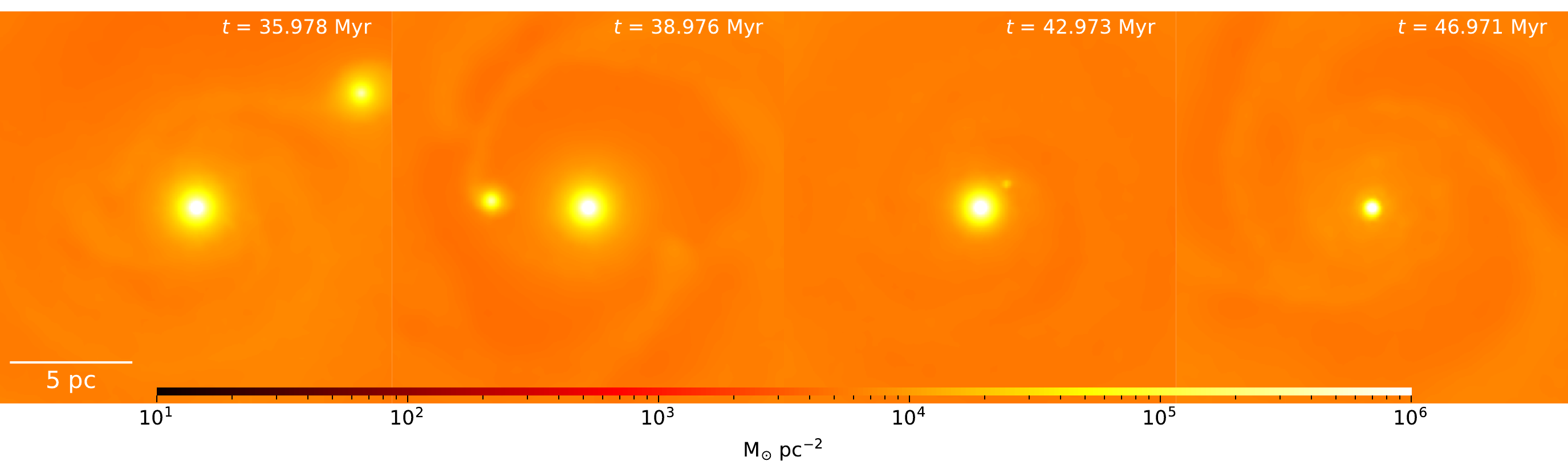}\\
    \includegraphics[width=.95\linewidth]{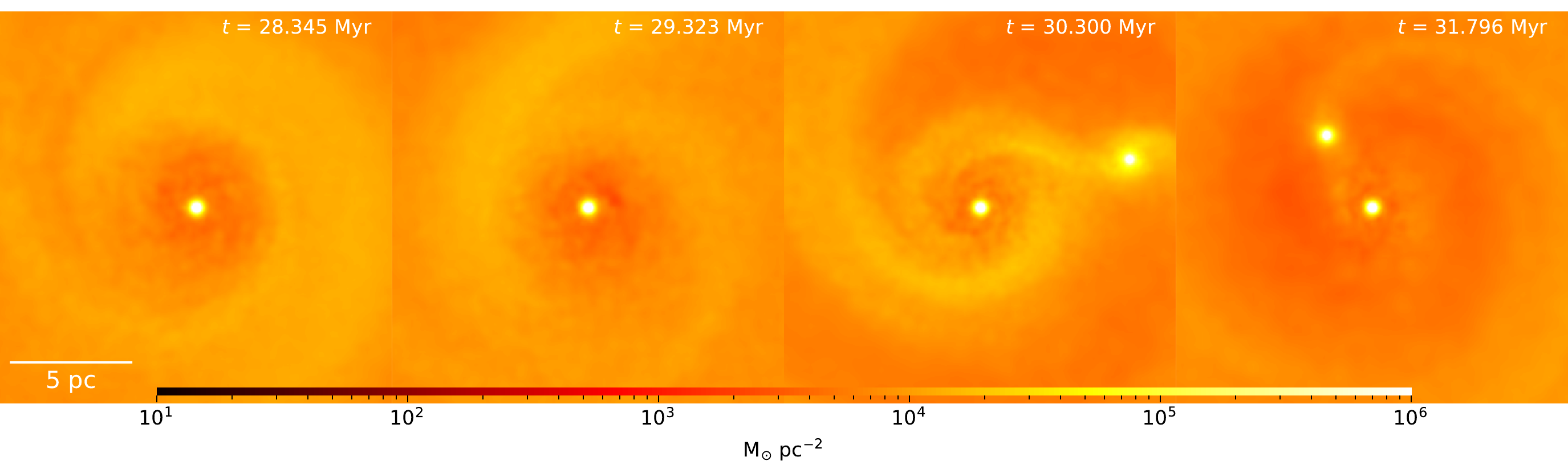}\\
    \includegraphics[width=.95\linewidth]{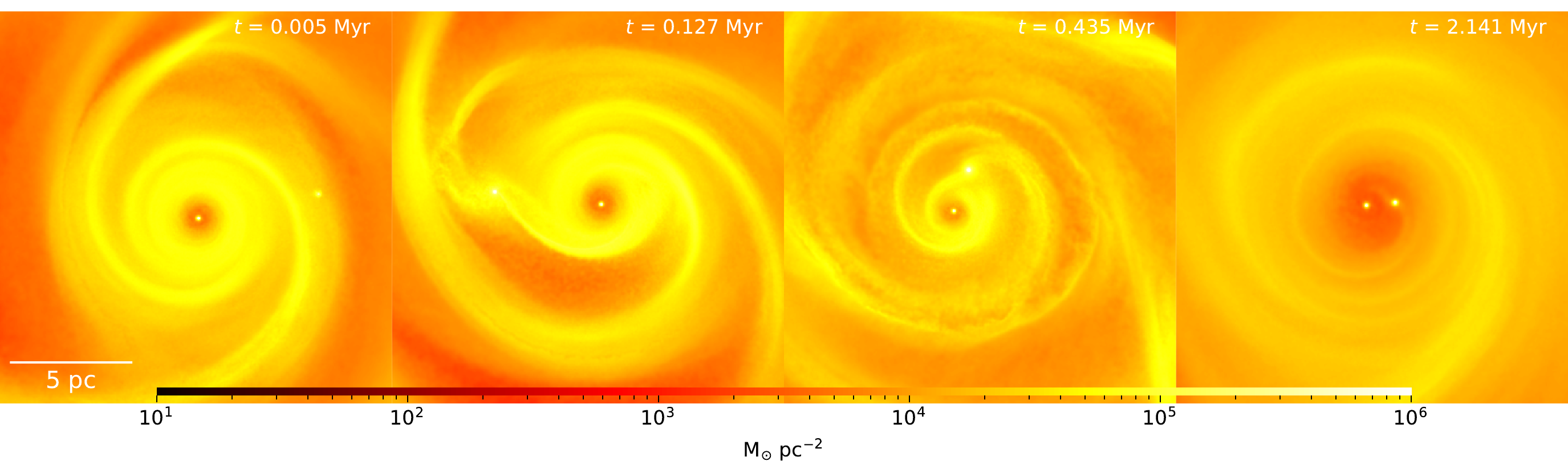}
    \caption{Small-scale, mass-weighted integrated gas density maps for a subset of the simulations described in Table~\ref{tab:simulations}. The ranges of distances and densities are chosen to highlight the environment on scales of a few pc from the primary MBH, where the final phases of the secondary MBH's orbital decay occur in our set of simulations, and a circumbinary disk forms in some cases. From top to bottom, each row corresponds to the runs \emph{MReps50g14}, \emph{MReps23g14R}, \emph{MReps23g11R}, and \emph{HReps05g14}. The secondary MBH lies out of the scale in the first two panels of the third row (run {\it MReps23g11R}), but up to the last snapshot it displays signs of density wave excitation characteristic of a relatively efficient migration phase (see also Figure~\ref{fig:separation}). A considerable drop of density for the gas within the MBH's orbit is observed at the last snapshot of the {\it HReps05g14} run, and partially also for the \emph{MReps23g11R} run, signaling the formation of a central cavity (see also Figure~\ref{fig:sigma-profiles}).
    }
    \label{fig:density-maps}  
\end{figure*}

\begin{figure*}
    \centering
    \includegraphics[width=.95\linewidth]{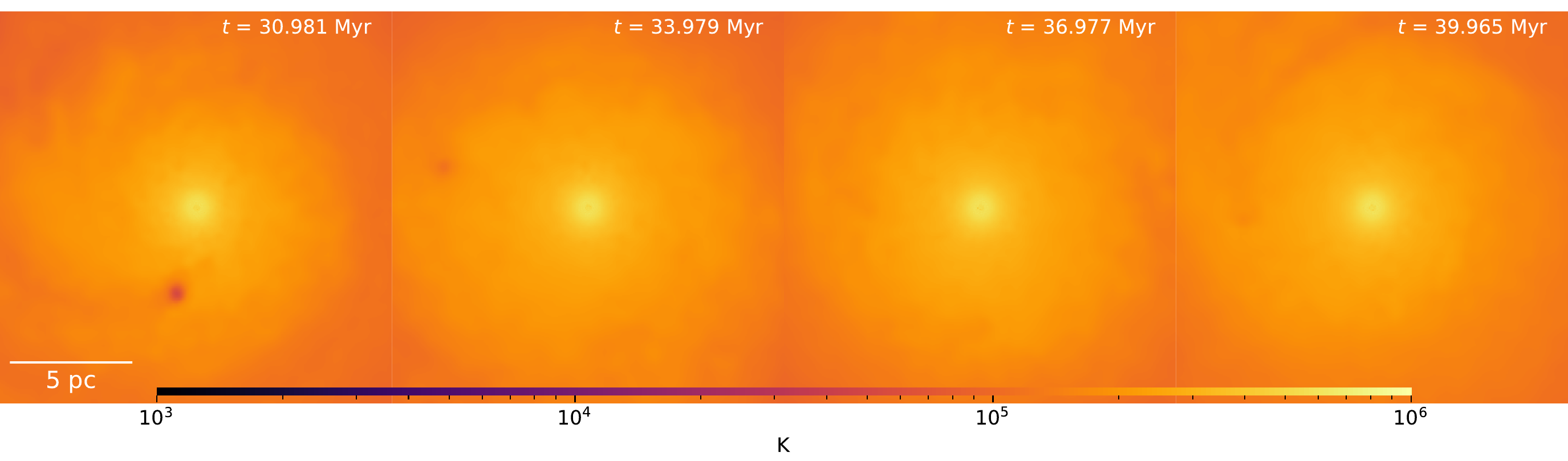}\\
    \includegraphics[width=.95\linewidth]{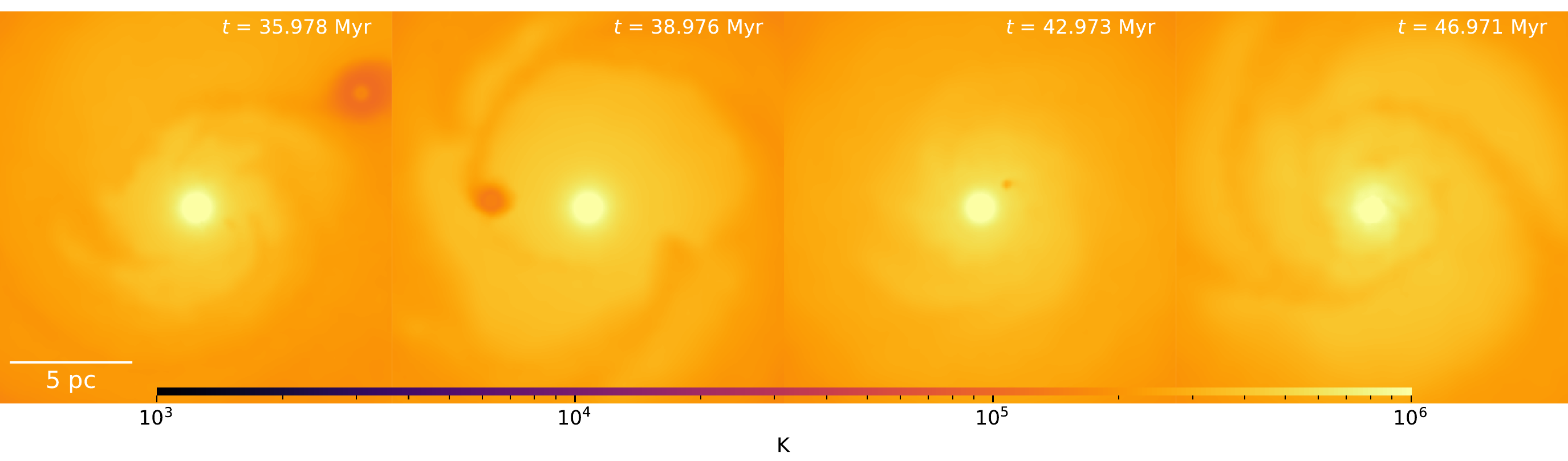}\\
    \includegraphics[width=.95\linewidth]{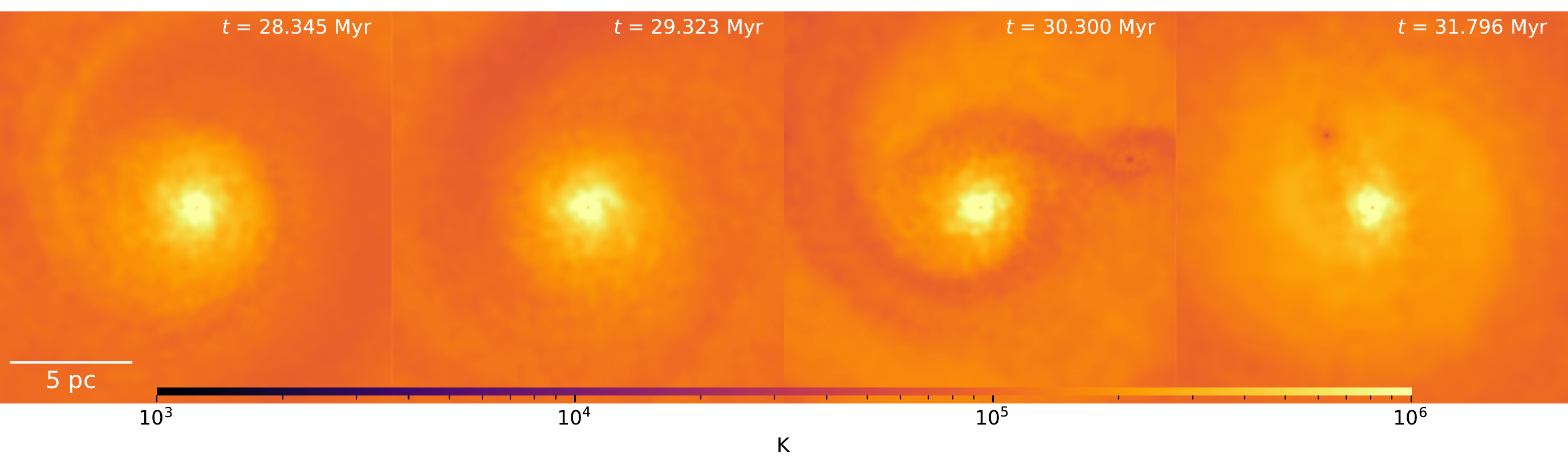}\\
    \includegraphics[width=.95\linewidth]{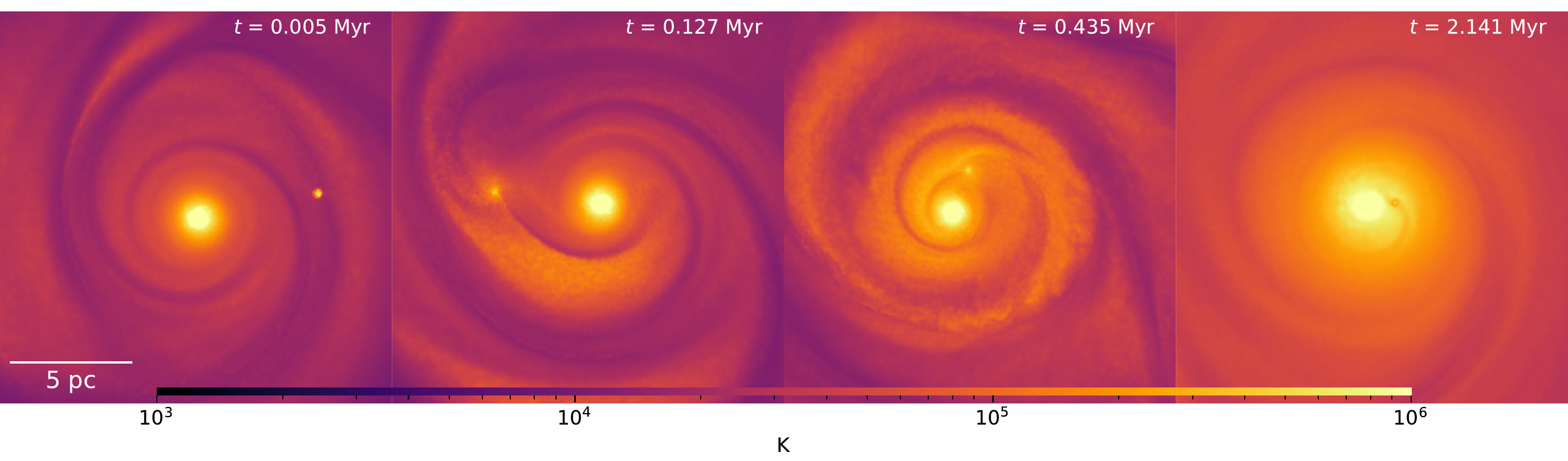}
    \caption{The gas temperature maps (density-weighted, integrated along the line of sight) are shown for the same runs as in  Figure~\ref{fig:density-maps}. In the panels of run \emph{HReps05g14} (bottom row), the temperature increases as the gap develops, probably as a result of tidal shocking by the binary. We refer the reader to Figure~\ref{fig:flagship-at-2255} for more details on the central region of the bottom-right panel.
    }
    \label{fig:temperature-maps}
\end{figure*}

We begin by inspecting the gas density maps at different scales and times, shown in Figures~\ref{fig:simcomparison-rho} and \ref{fig:density-maps}. From the inspection of these figures, it is evident that, whereas the gas flow appears similar in all CND runs at large scales (Figure~\ref{fig:simcomparison-rho}), which is not surprising, given that all simulations adopt similar ICs, there are marked differences at small scales (Figure~\ref{fig:density-maps}). Also, in some runs, the MBH pair shrinks its separation to as small as the spatial resolution imposed by the gravitational softening, whereas in others the decay appears to slow down or be suppressed (Figure~\ref{fig:density-maps}). In the highest-resolution run (bottom panels of Figure~\ref{fig:density-maps}), the gas flow shows a much more complex non-axisymmetric structure, with more marked spiral density waves. The same is true for the run with the stiffest EoS ({\it MReps23g11R}), which maintains lower temperatures due to the lower adiabatic index (Figure~\ref{fig:temperature-maps}), thus enhancing the spiral disturbance triggered by the sinking secondary MBH. Another remarkable difference is the accumulation of gas around the primary MBH, which is much more prominent in the lower-resolution runs (top panels of Figure~\ref{fig:density-maps}), probably reflecting the higher numerical diffusivity (driven by artificial viscosity). All these differences should of course have an impact on the torques experienced by the secondary MBH.

In the next subsection, we discuss the orbital evolution of the secondary MBH in different runs, and then report on the torque analysis in the different simulations. Armed with the knowledge of how the torque behaves, we will then study the eventual opening of a gap or central cavity as the secondary MBH sinks towards the center. We will use the same conditions that apply to gap opening also to study the carving of a circumbinary cavity at small separations because of the large mass ratio between the primary and the secondary MBH, which allows us to assume that the primary MBH moves negligibly with respect to the center of mass. In standard planet migration theory, the gap-opening stage signals  a transition of torque regimes, as after a gap is opened the orbital decay of the perturber should proceed on the disk viscous timescale, which can be 2--3 orders of magnitude longer than the linear-torque (Type~I) dynamical-friction timescales \citep[these are similar for typical background disks, see][]{mayer13}, and even smaller compared to non-linear torque regimes as those observed in studies of migration in massive self-gravitating disks \citep[e.g.][]{malik2015}. Fast orbital decay without gap opening, however, has been observed in simulations of secondary MBHs migrating in adiabatic CNDs with resolution comparable to the lowest-resolution runs considered in this paper \citep[][]{mayer13}. It will be important to clarify if, under which conditions, and at what separations, the orbital decay enters the latter, slower stage, as it is ultimately relevant to determine the coalescence timescale of MBHs. Note that, in addition, current studies of orbital evolution for MBH pairs already embedded in a circumbinary cavity are highly inconclusive on both the timescale and the direction of migration \citep[e.g.][]{tang17,Munoz_et_al_2019}.

\newcommand{\wrr}{0.28898954703832752\textwidth}
\newcommand\w{.22\textwidth}
\newcommand{\hrr}{4.3207331042382586cm}
\newcommand\h{4cm}
\newcommand{\addpictopmid}[1]{\includegraphics[trim={0 .7cm 2.7cm 0}, clip, height=\h, width=\w] {#1}}
\newcommand{\addpicbotmid}[1]{\includegraphics[trim={0 .0 2.7cm 0}, clip, height=\hrr, width=\w] {#1}}
\newcommand{\addpicright}[1]{\includegraphics[trim={0 .7cm 0 0}, clip, height=\h, width=\wrr] {#1}}
\newcommand{\addpicbotright}[1]{\includegraphics[trim={0 0 0 0}, clip, height=\hrr, width=\wrr] {#1}}

\begin{figure*}
    \centering
	{\addpictopmid{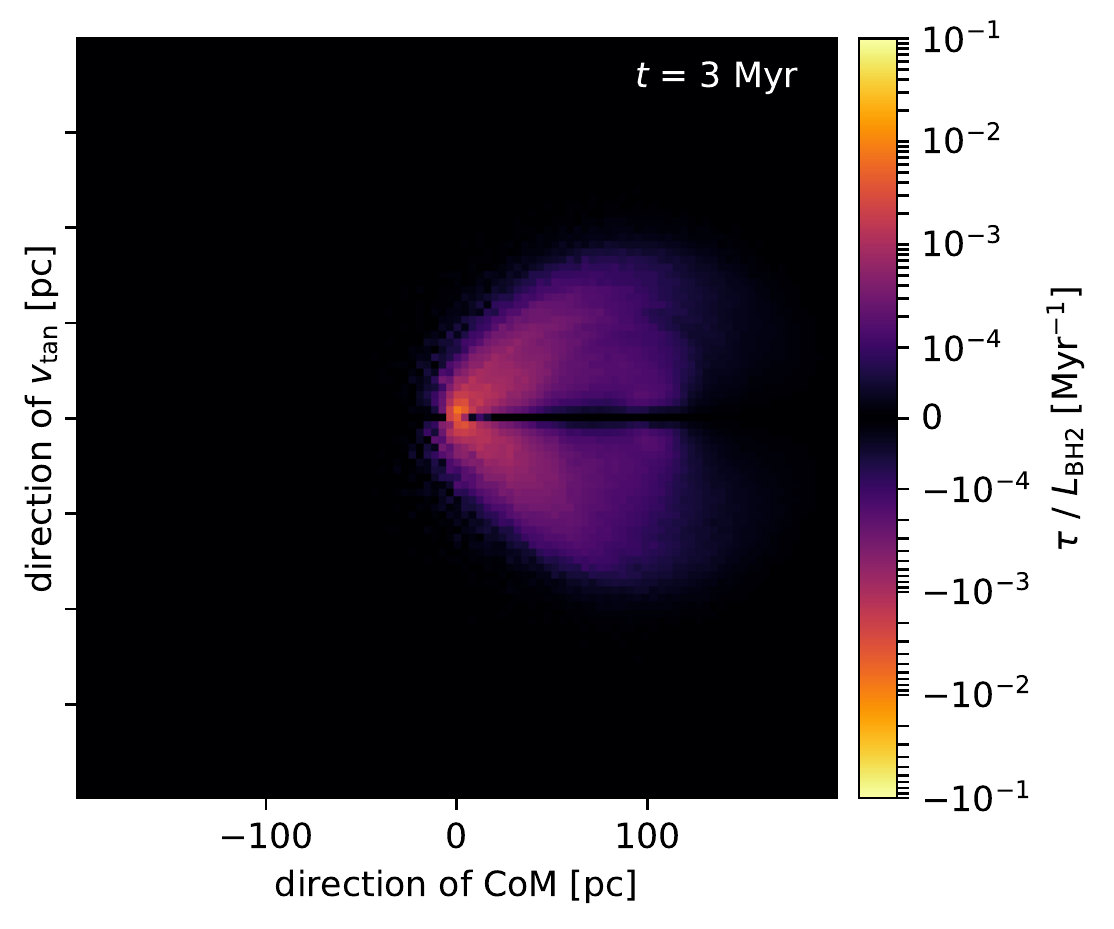}}
	{\addpictopmid{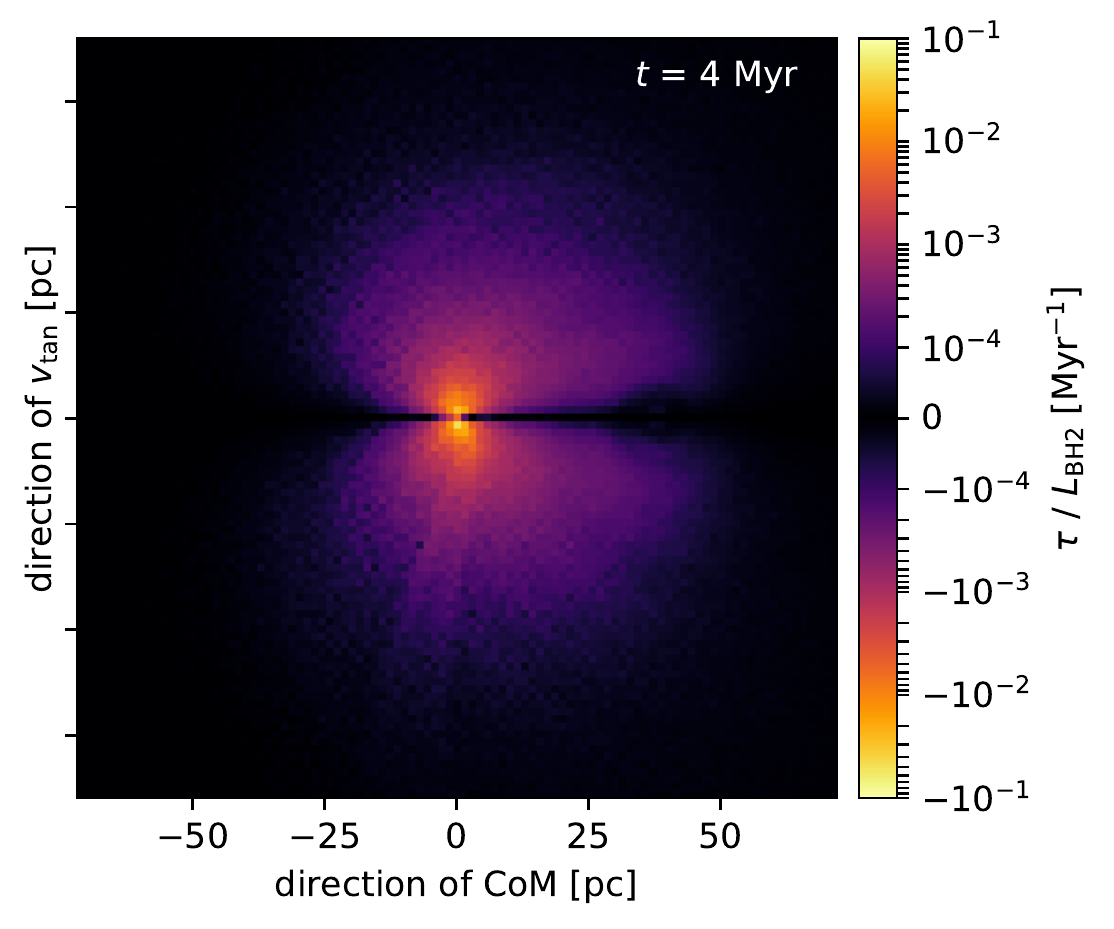}}
	{\addpictopmid{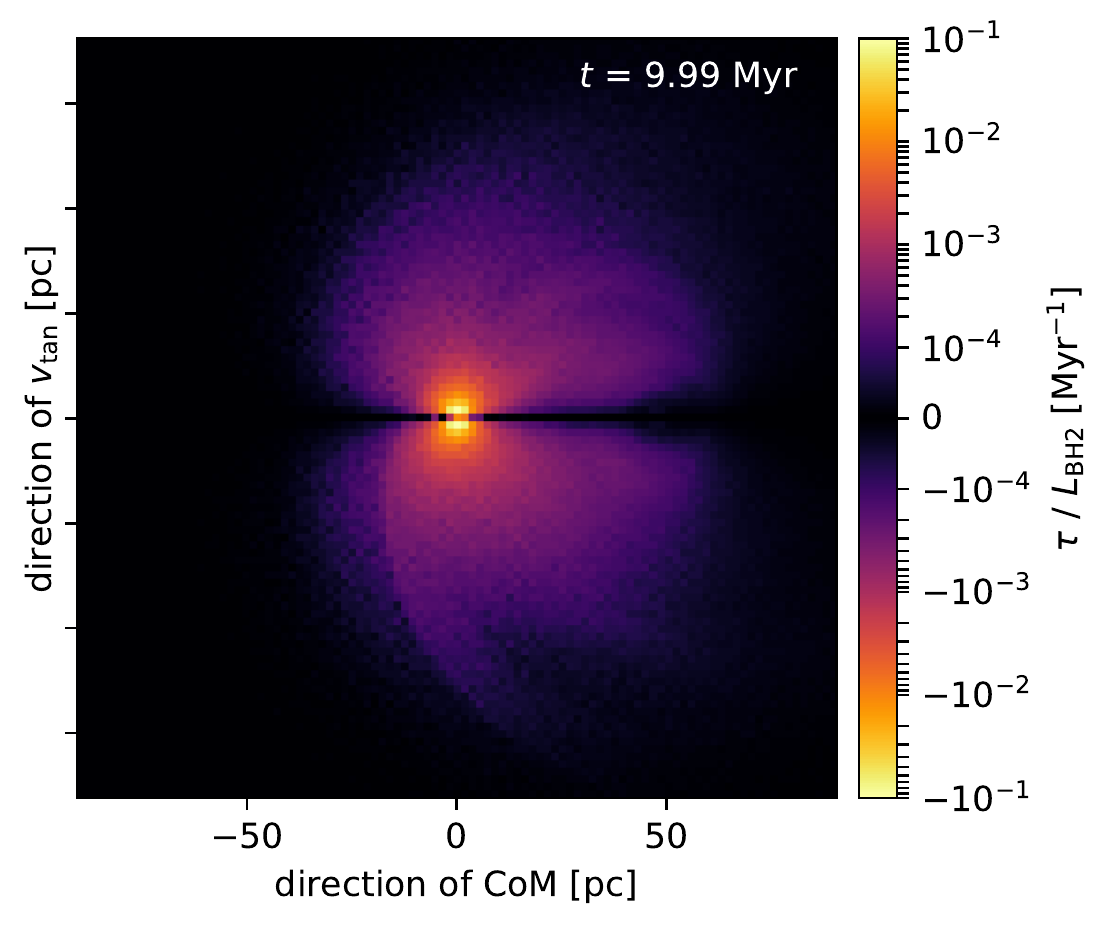}}
	{\addpicright{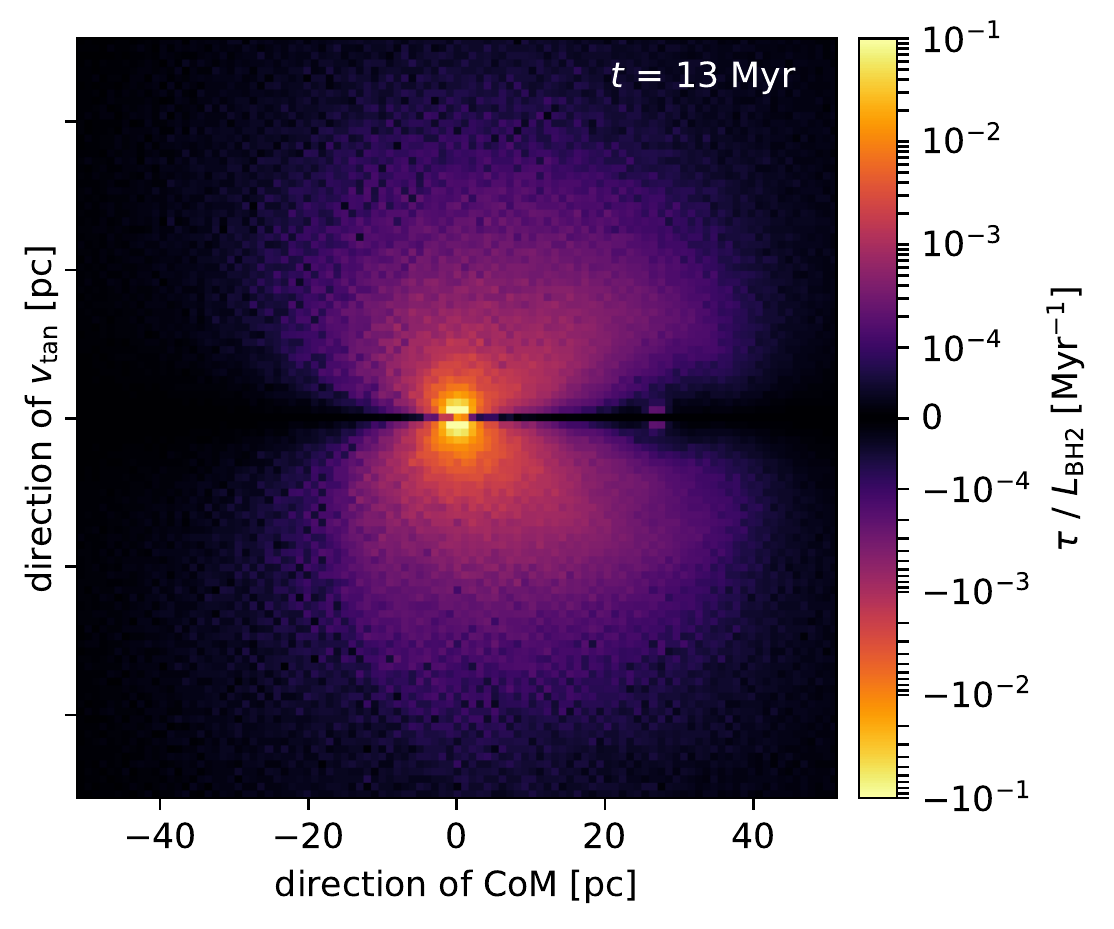}}
	\\
	{\addpictopmid{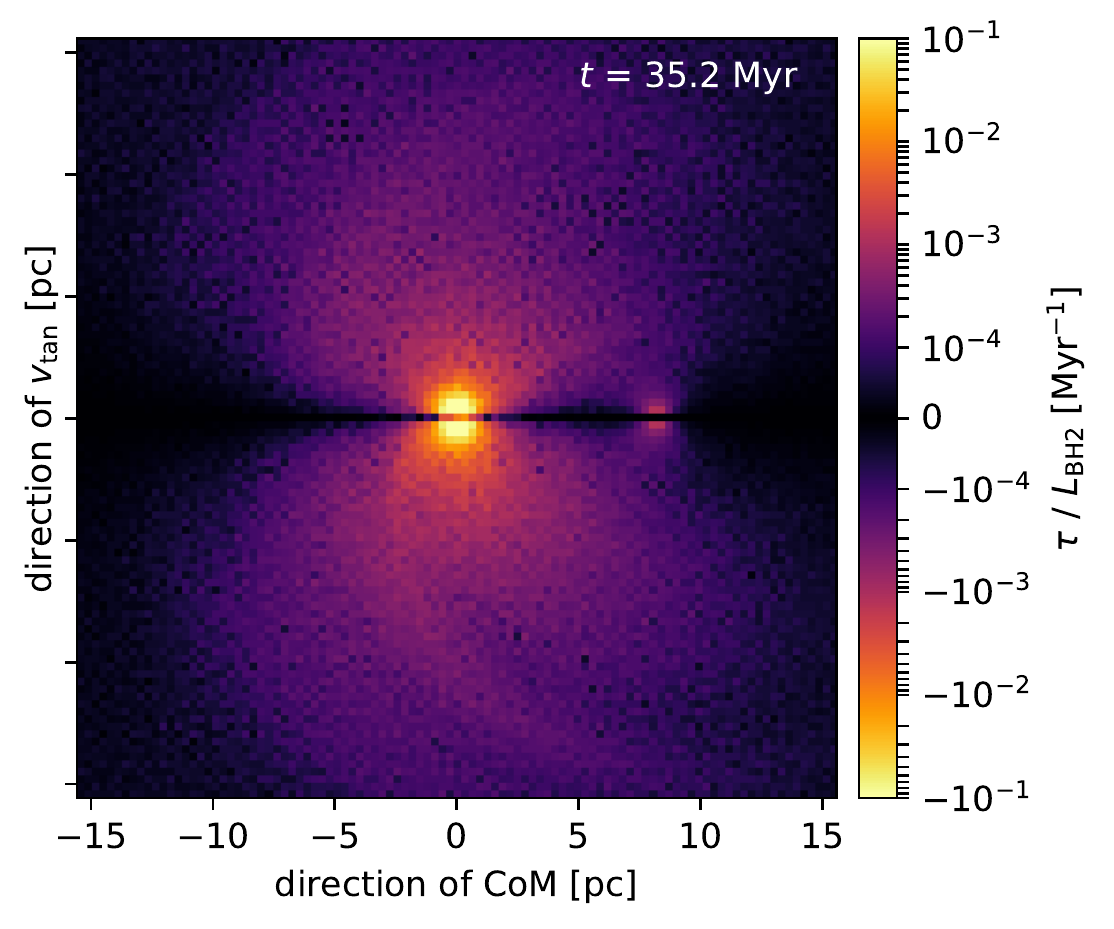}}
	{\addpictopmid{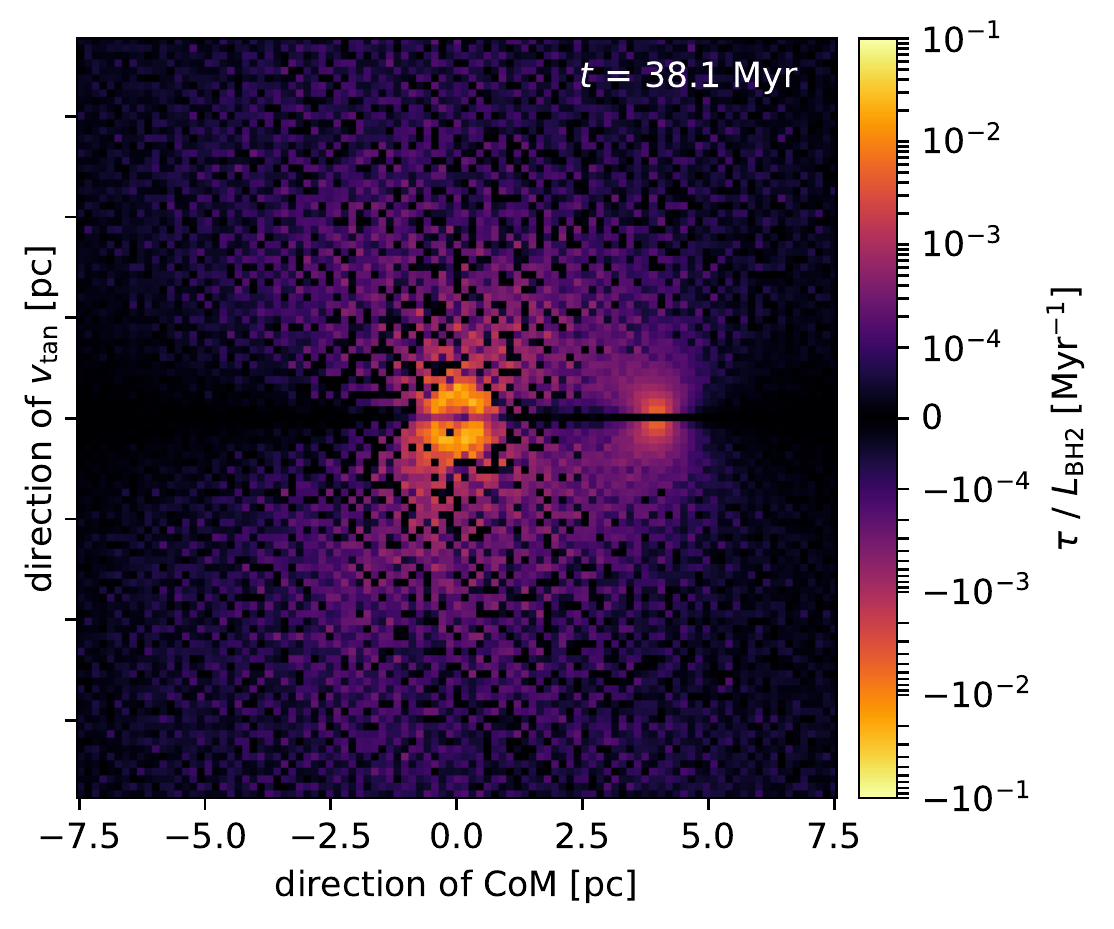}}
	{\addpictopmid{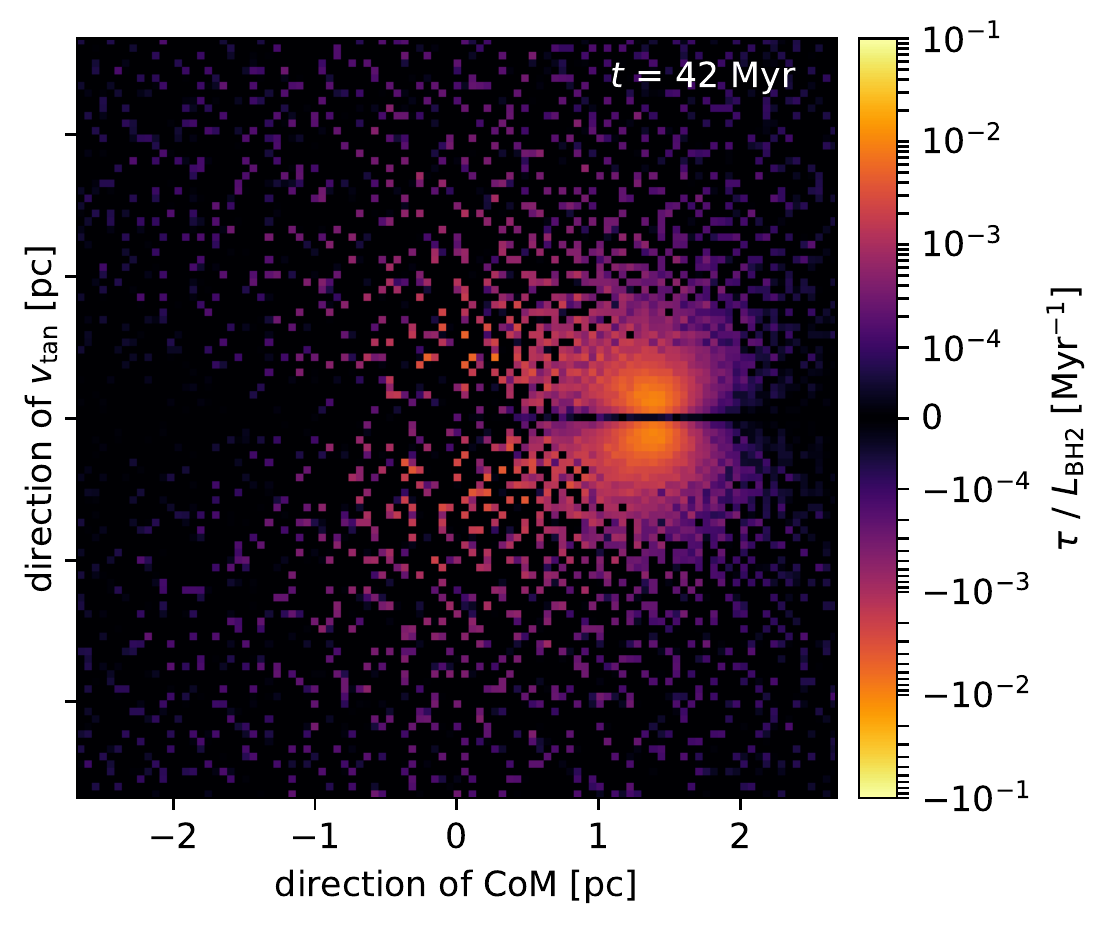}}
	{\addpicright{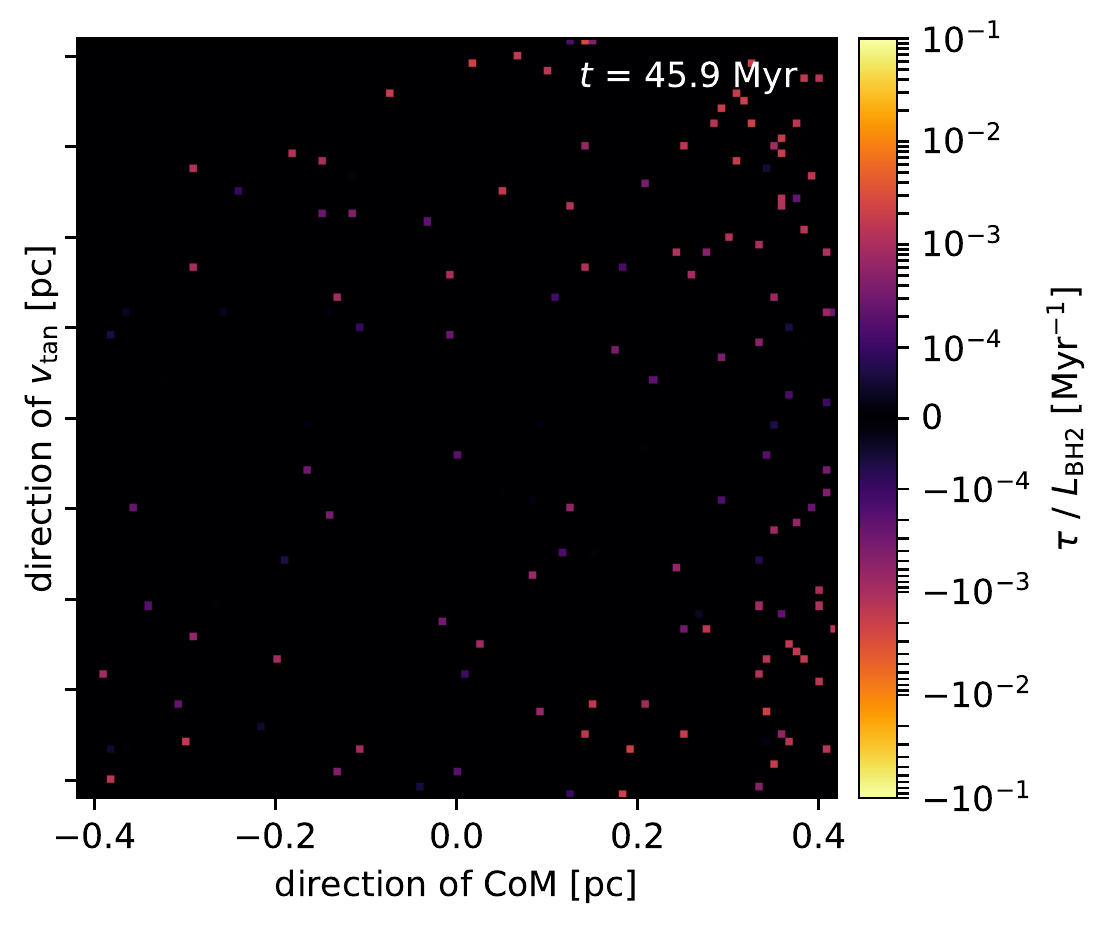}}
	\\
	{\addpictopmid{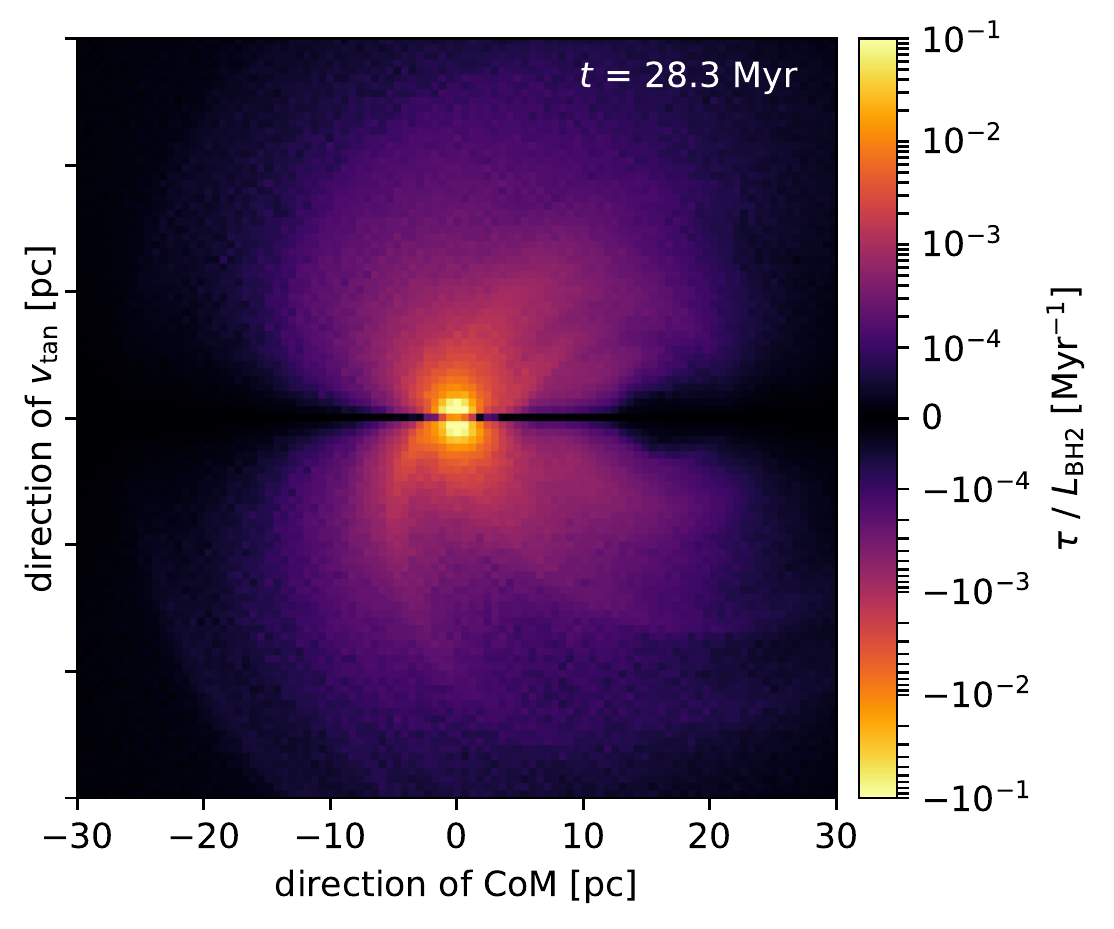}}
	{\addpictopmid{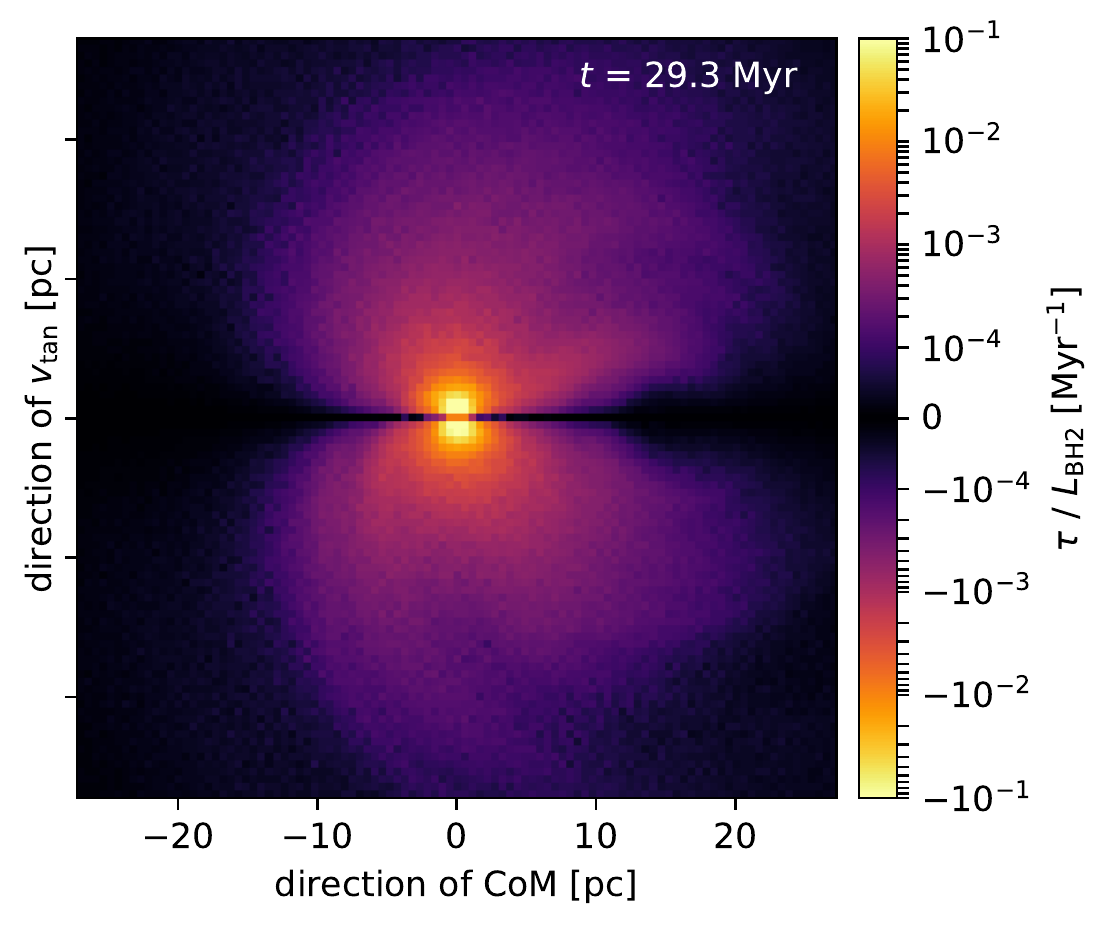}}
	{\addpictopmid{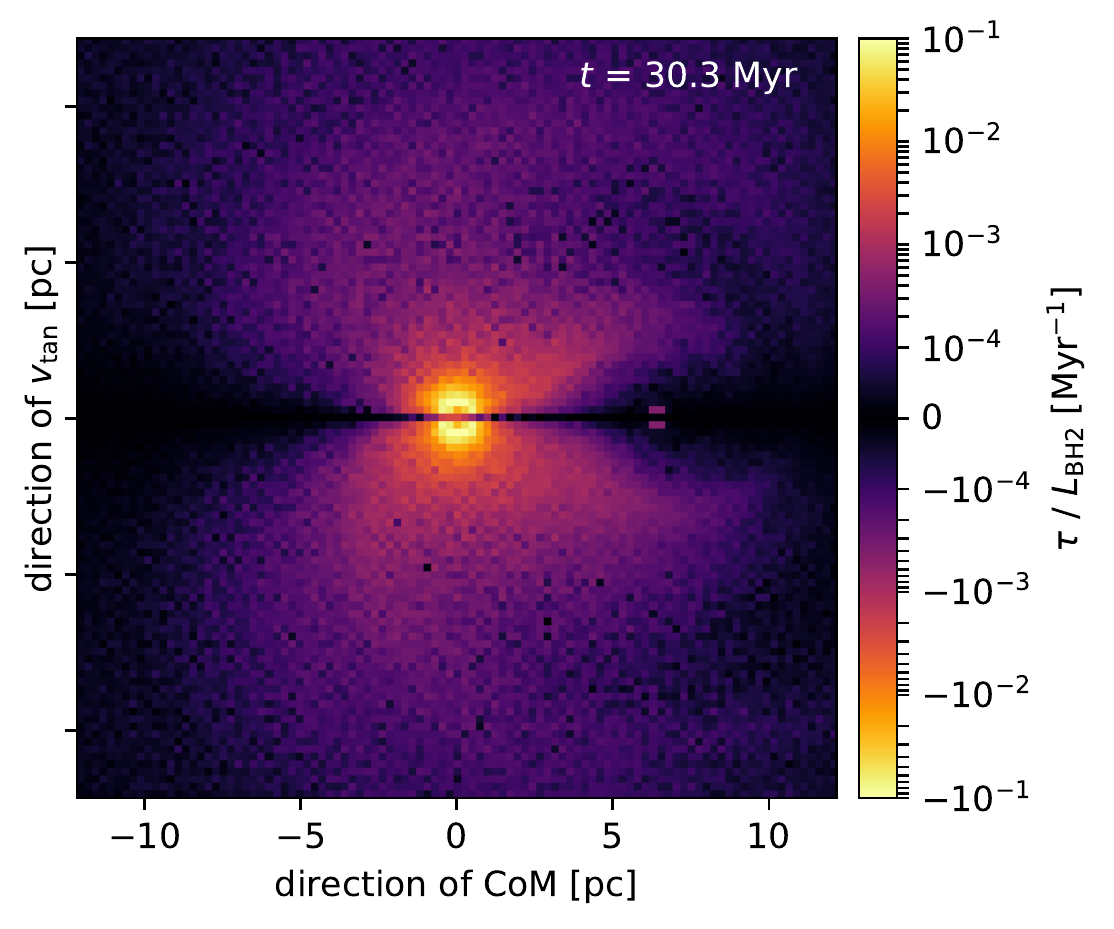}}
	{\addpicright{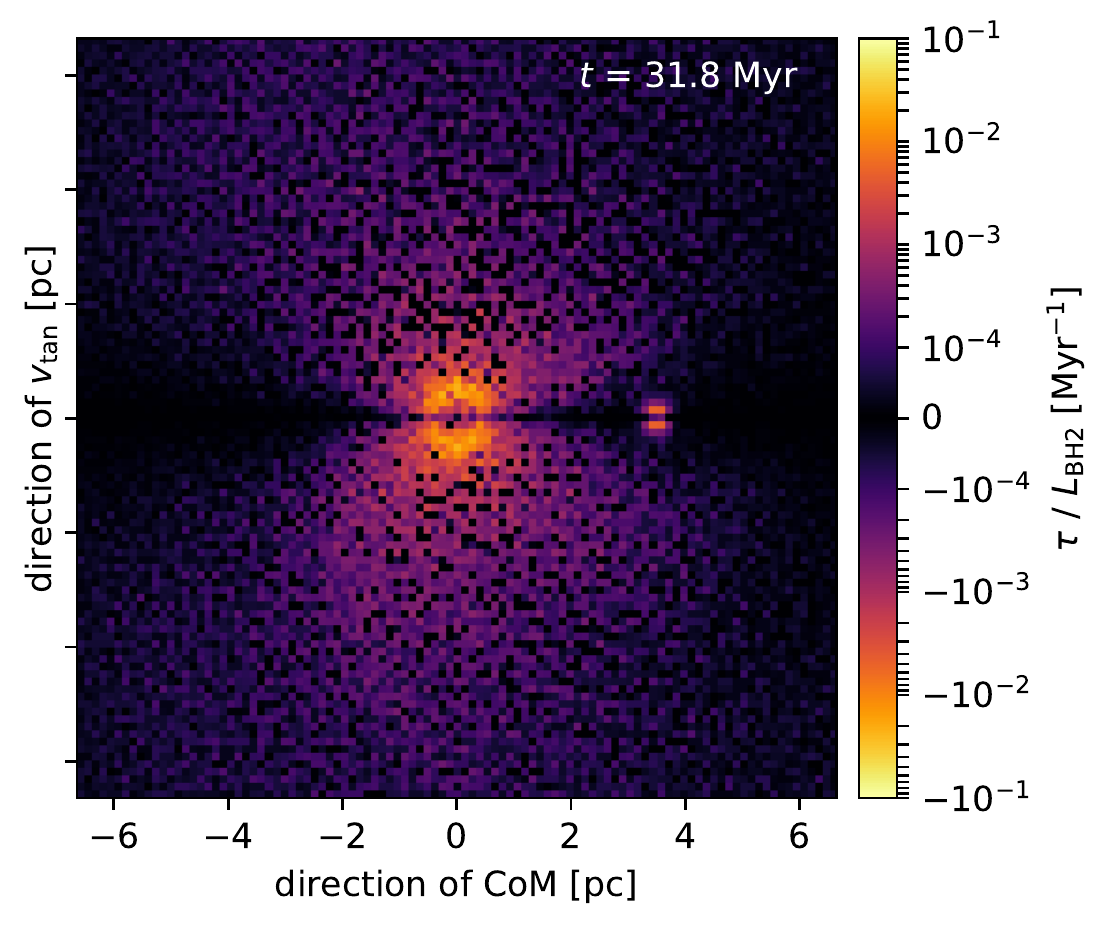}}
	\\
	{\addpicbotmid{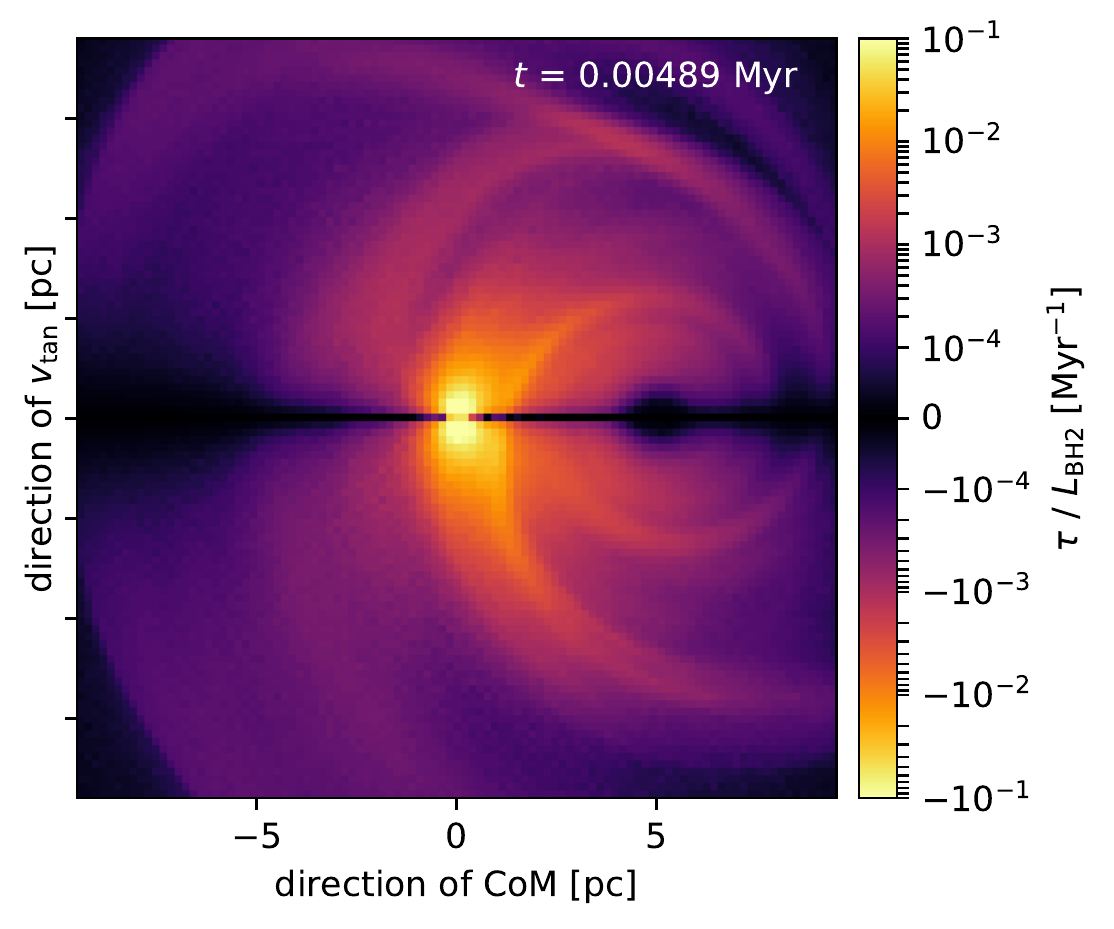}}
	{\addpicbotmid{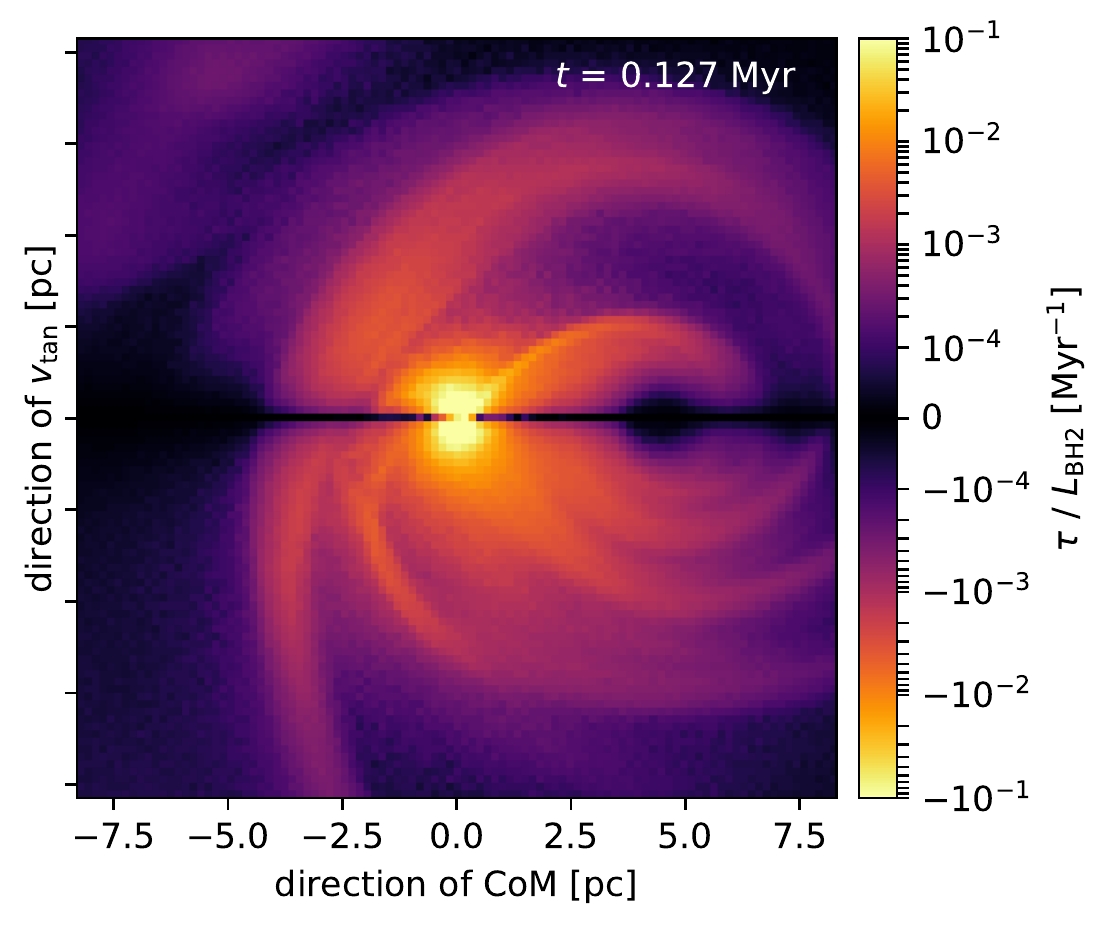}}
	{\addpicbotmid{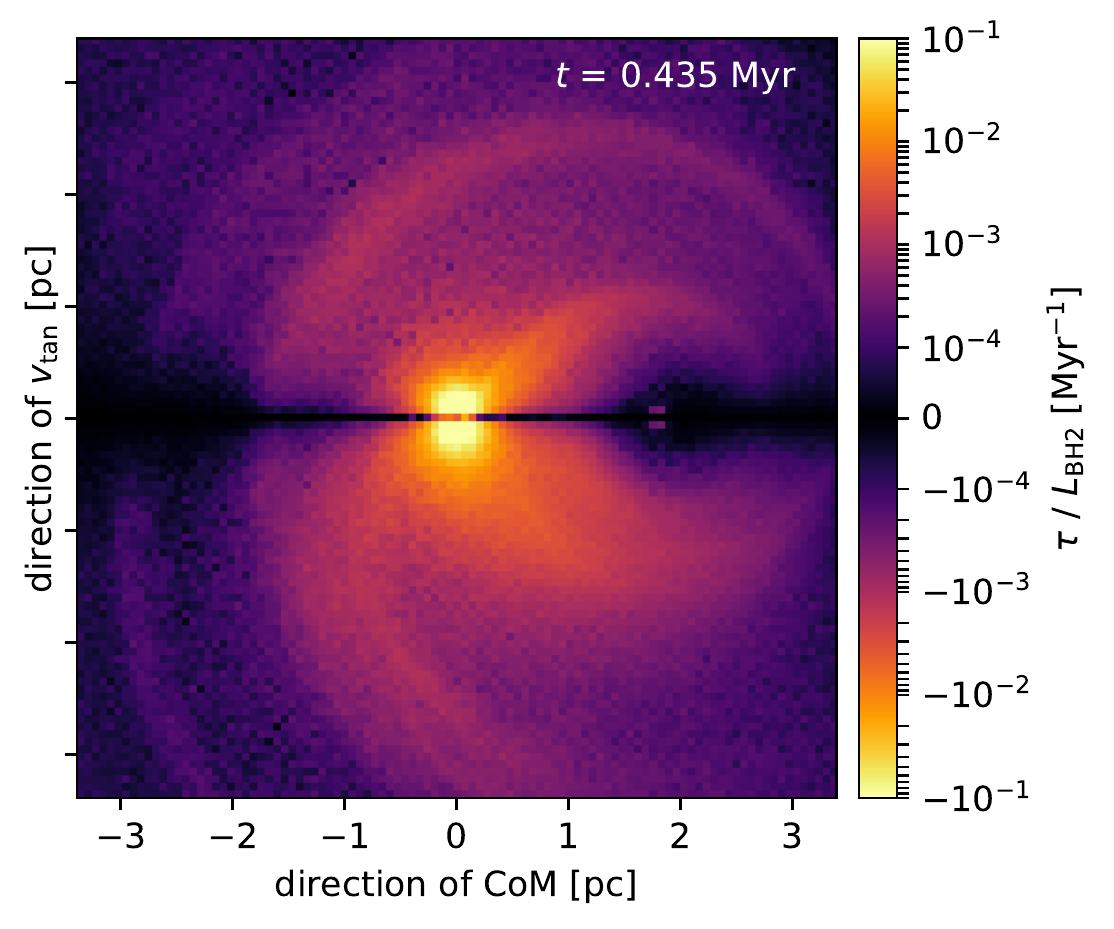}}
	{\addpicbotright{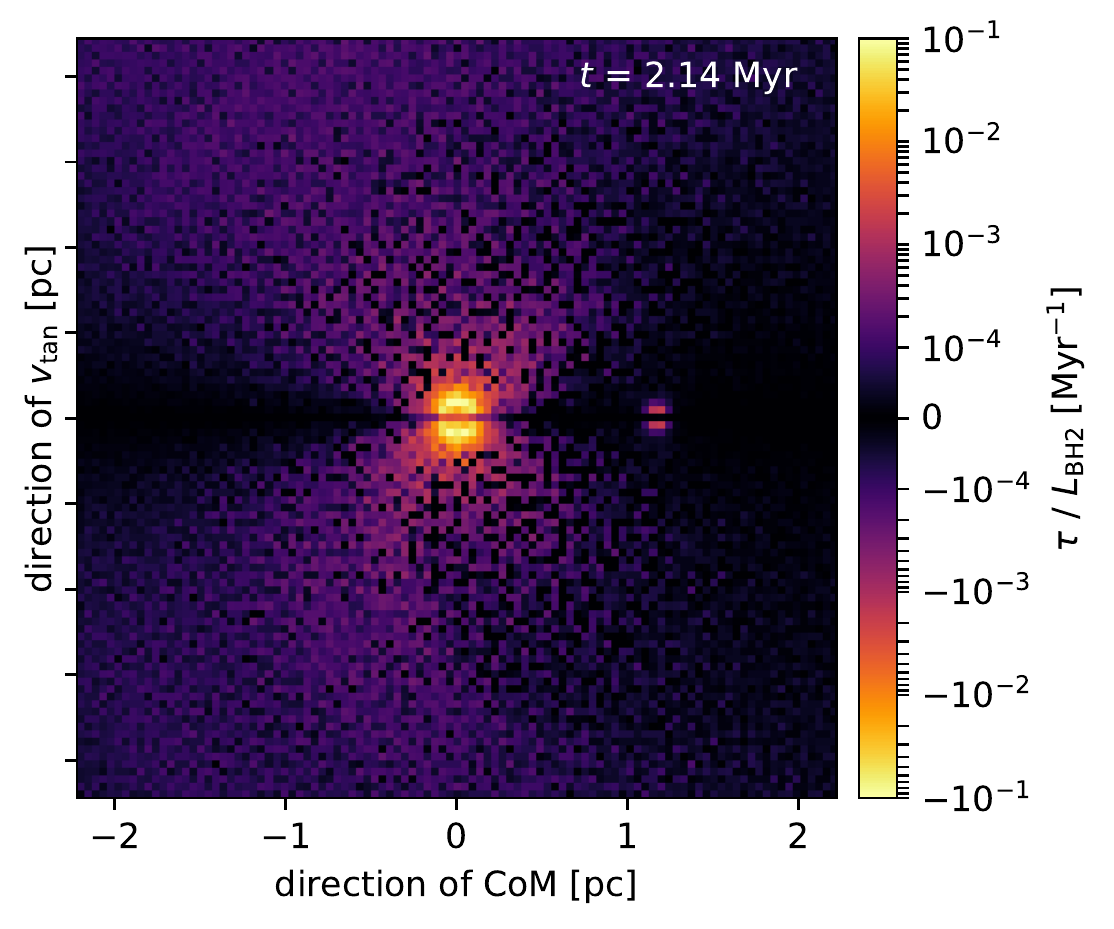}}
    \caption{Map of the torque relative to MBH2's angular momentum (recall the notes regarding Equation~\ref{eq:torque}) for runs \emph{LReps50g14}, \emph{MReps23g14R}, \emph{MReps23g11R}, and \emph{HReps05g14}. In this representation, particles positioned with $y > 0$ ($y < 0$) cause positive (negative) torques. The distance represented on the horizontal axis is along the direction to the CoM, $\left|\,\mathbf{r_\mathrm{CoM}}\,\right|$, and the range is thus chosen accordingly. The choice of the snapshots in the first row has been made to highlight the three  decay phases evident in Figure~\ref{fig:simcomparison-rho}, as described in the text. Note that, in the last panel of the second row, the size scale is comparable to the softened region. Since the unresolved torques are not displayed, the particles near the orbital plane are omitted in this view, causing only a few particles from outside the sphere of 0.46~pc radius to be displayed. In order to assess the angular distribution of the intensity of the torque, thus inferring the directionality of the fluid elements that are the source of the torque itself, we refer the reader to Figure \ref{fig:torque-directional-map-row}.
    }
\label{fig:torque-maps}
\end{figure*}

\subsection{Orbital decay of the secondary MBH}\label{sec:orbital-decay}

\begin{figure}
    \centering
    \includegraphics[trim={0cm .6cm 1cm 1cm}, clip, width=.98\linewidth] {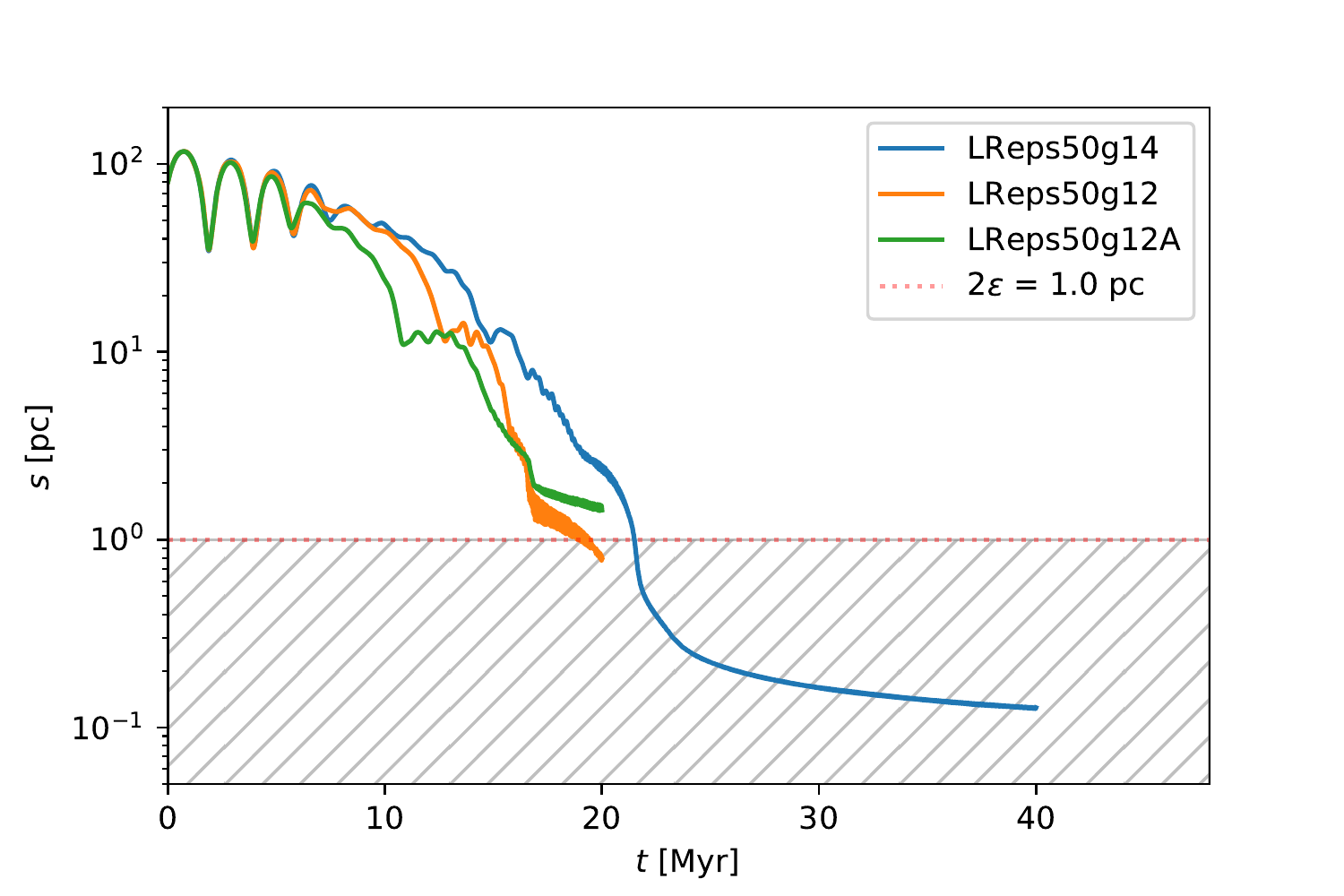}
    \includegraphics[trim={0cm .6cm 1cm 1cm}, clip, width=.98\linewidth] {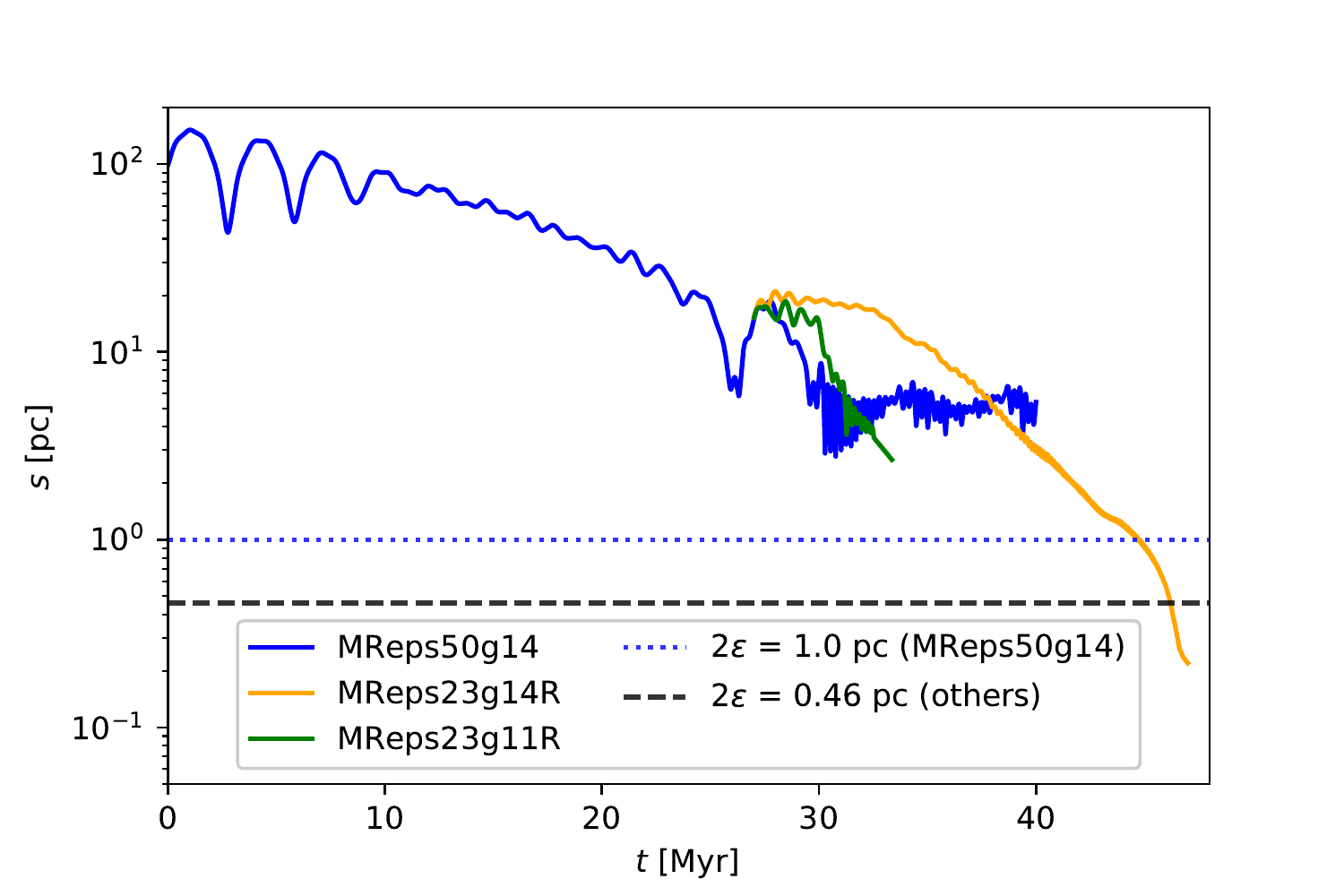}
    \includegraphics[trim={0cm 0cm 1cm 1cm}, clip, width=.98\linewidth] {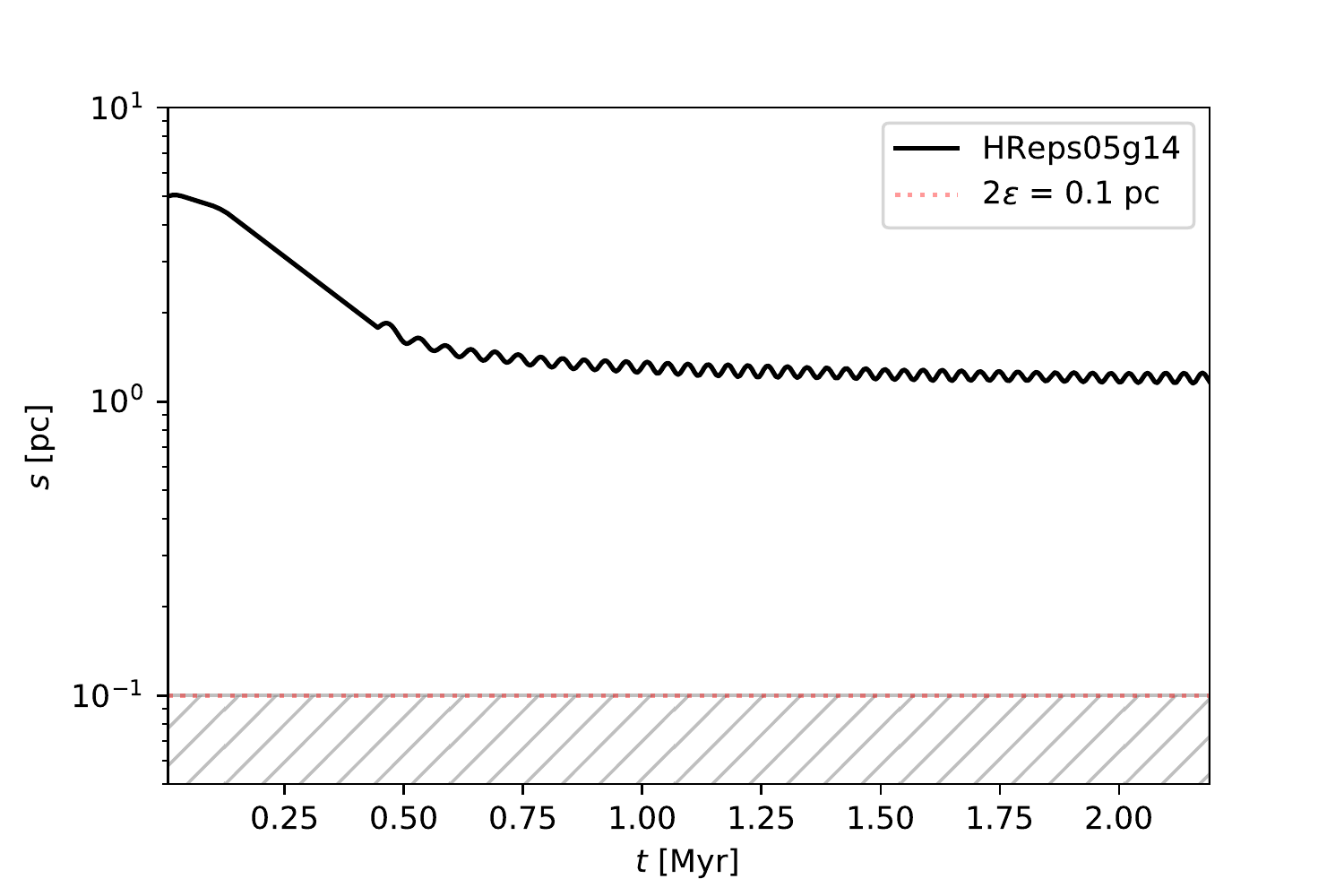}
    \caption{
    The orbital separation of the secondary MBH versus time is shown. The top panel displays the low-resolution runs \emph{LReps50g14}, \emph{LReps50g12}, and \emph{LReps50g12A}. The orbit of the secondary MBH continues to decay below the resolution limit, corresponding to $2 \epsilon$, where $\epsilon$ is given in Table~\ref{tab:simulations}. The central  panel shows {\it MReps50g14}, {\it MReps23g14R}, and {\it MReps23g11R}. The latter two started from a snapshot of run {\it MReps50g14} at $\sim$27~Myr. The bottom panel displays the {\it HReps05g14} run. The orbit stalls around a separation of $\sim$1.5~pc (see Section~\ref{sec:torque-results})  A comparison of the orbital-decay curves for scales below 2~pc is shown in Figure~\ref{fig:separation-comparison-sub2pc}.
    }
    \label{fig:separation}
\end{figure}

We begin by providing a general qualitative overview of the orbital decay in the various runs. We recall that, in our sample, we have runs reaching as low as 0.23 and 0.05~pc in spatial resolution (gravitational softening) and employing up to $2 \times 10^7$ gas particles. These runs intentionally approach the resolution of simulations starting with the MBHs embedded in an already formed CBD \citep[e.g.][]{Roedig_et_al_2014,tang17}. Figure~\ref{fig:separation} shows the orbital decay curves for our references runs. Generally speaking, the decay can be divided in three phases. There is first an initial slow decay-phase down to a separation of about 50--60~pc, followed by a rapid decay-phase down to just above pc separations, and then by a third slowest decay-phase at separations $\lesssim$1~pc. The first phase typically lasts for $\sim$10--15~Myr, during which the secondary MBH decays less than about a third of the way from the center, whereas in the second phase the secondary MBH covers nearly two thirds of the remaining path in less than 10~Myr. In the cases in which the secondary MBH is able to reach a (resolved) separation well below pc, as small as $\sim$0.2--0.3~pc, as in the {\it MReps23g14R} run, it does so in about 50~Myr.

While the first two phases were already identified in previous work \citep[e.g.][]{escala05, dotti07, fiacconi13, mayer13}, we remark that the third phase is addressed possibly for the first time in this work, as previously published CND simulations did not have enough mass and spatial resolution. The first phase of the decay corresponds to the stage in which the dominant drag comes from the local trailing wake produced by the secondary MBH, consistent with the description provided by the theory of dynamical friction in a gaseous medium, as illustrated extensively in \citet{mayer13} and \citet{chapon13}. The trailing wake is visible in Figure~\ref{fig:simcomparison-rho} (see, e.g. the top panel). In this phase, the orbit circularizes in all runs, an effect that is well understood and that has been quantitatively described for the first time via a recent semi-analytical model \citep{Bonetti2020b}; circularization is well quantified by the evolution of the orbital angular momentum in the simulations (see the analysis in the next subsection). The second phase will be referred to as the {\it migration} phase, as it is driven by torques generated by the gas flow perturbation induced by the secondary MBH, such as a triggered global spiral density wave or local asymmetry in the gas near the Hill sphere, analogously to the case of planet migration. These torques are nearly by-symmetric as tidal torques, as opposed to the dynamical friction wake, which is an inherently asymmetric disturbance (only trailing). By inspection of the curves, it is evident that there is no obvious dependence of the  decay rate  with resolution. The slower decay-phase, or third stage, as we will refer to from now on, is what differs the most between the runs. In some cases, the secondary MBH appears to stall. This is clearly not an effect of poor resolution, as the stalling is more evident precisely in the {\it HReps05g14} run, which has the highest resolution. Finally, Figure~\ref{fig:simcomparison-rho} shows that the decay is faster with a softer EoS, which approaches the isothermal case, in the first and second phase of the decay, consistent with previous work \citep[e.g.][]{chapon13}, but not necessarily in the last phase.

\subsubsection{Gravitational torque analysis}\label{sec:torque-results}

In order to understand the nature of the three orbital-decay phases suggested by Figure~\ref{fig:separation}, we turn to analyze the gravitational torques exerted by the fluid elements onto the secondary MBH. The gravitational torques from the gas particles acting on the secondary MBH were computed following Equation~\eqref{eq:torque}. The contribution from the particles within the softened region (a sphere centered on MBH2 with radius equal to two softening lengths) is not accounted for when we refer to a curve or map as ``resolved'', but we additionally show the  result including particles in the non-resolved regions by referring to it as ``unresolved''. We note that, while in principle one should only consider resolved components, the unresolved components can still  exert a torque. Even more, the literature on planet migration, especially for massive planets, has often shown cases in which the most important contribution to the torque was indeed coming from fluid elements within the co-orbital region of the perturber, which were close to the resolution limit. A typical case where this is relevant is the fastest known migration regime -- Type III (\citealt{Masset_Papaloizou_2003}; \citealt{Lin_Papaloizou_2010}; \citealt{baruteau14}) -- which has been shown to be the dominant mode for massive perturbers in self-gravitating disks (with a perturber-to-disk mass ratio of $\sim$10$^{-2}$, see \citealt{malik2015}), the case that applies to this paper. We visualize the results of the torques analysis in various ways, which  are useful to a different degree for enabling the quantitative understanding of the orbital decay process.

\begin{figure*}
	\centering
	\includegraphics[width=.95\linewidth]{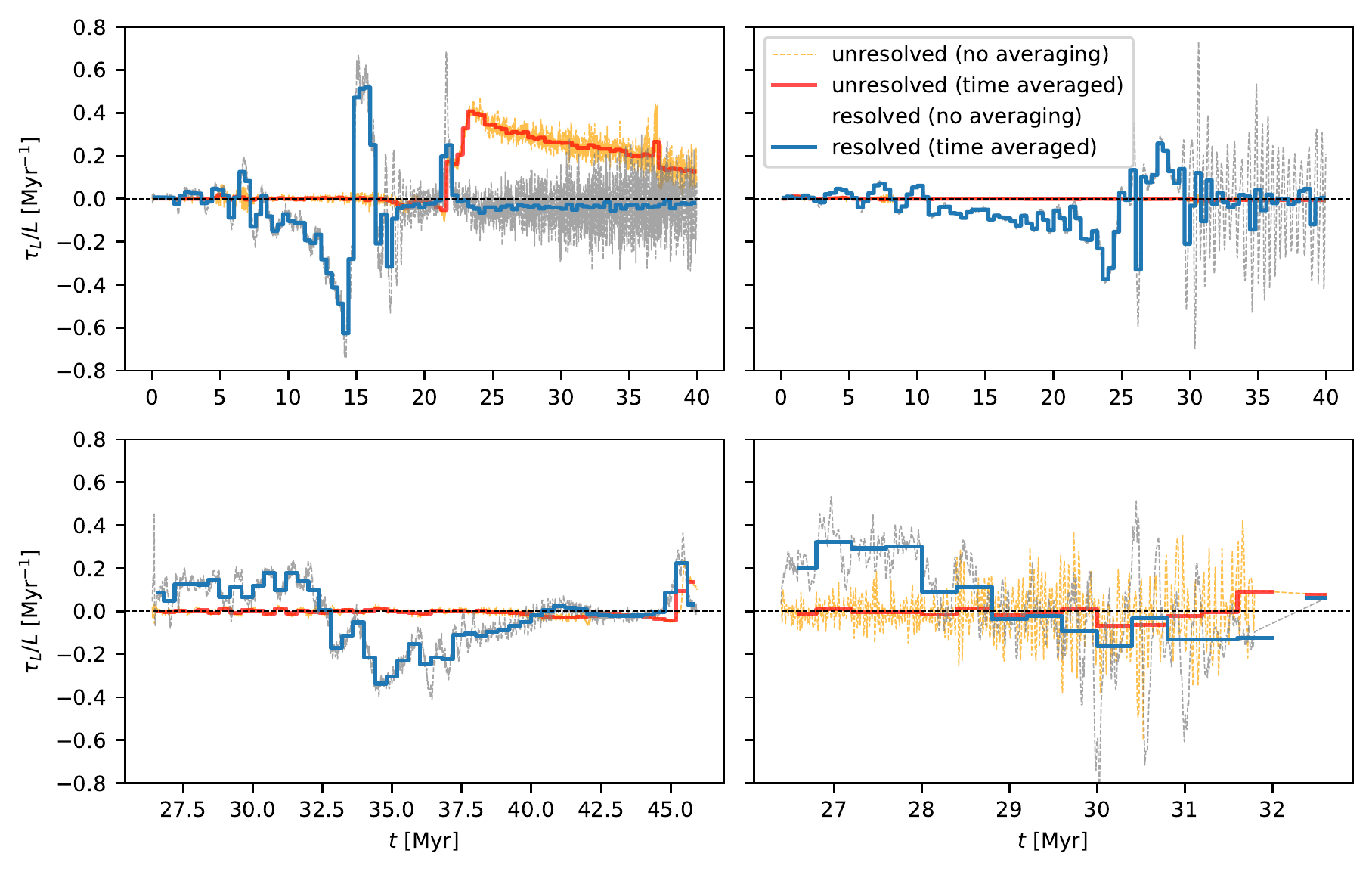}
	\includegraphics[width=0.48\linewidth]{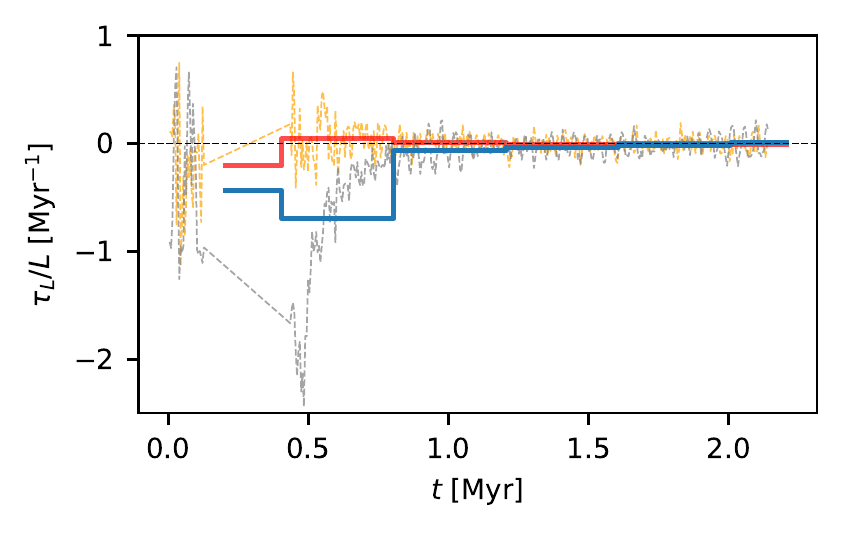}
	\caption{
	    The time evolution of the gravitational torques acting on the secondary MBH normalized to the secondary MBH's orbital angular momentum is shown for, respectively, runs {\it LReps50g14} (top-left panel), {\it MReps50g14} (top-right), {\it MReps23g14R} (middle-left), {\it MReps23g11R} (middle-right), and {\it HReps05g14} (bottom). The torque is evaluated in two separate regions, one for unsoftened (resolved) and one for softened (unresolved) interactions between the fluid elements and the secondary MBH, and further time-averaged (blue and red lines only). The timescale chosen for the averaging is 0.4~Myr. In terms of a reference orbital time for fluid elements, this corresponds to the period of a Keplerian orbit relative to the secondary MBH at about 2~pc. This distance is chosen for being a multiple of the softening length even in the low-resolution cases. This is done to average out torque fluctuations due to gas in the close vicinity of the secondary MBH, which, by construction, occur on timescales shorter than that chosen for the time-averaging. The grey, dashed line indicates instantaneous resolved torques, namely above the scale of the gravitational softening; the blue, solid line shows the latter torque averaged over 0.4 Myr as just 
outlined. The dashed yellow (not time-averaged) and red (time-averaged) lines indicate estimates for the torques when considering a Plummer potential with $2 \epsilon$ as the size of the spherical region chosen to soften the interaction between fluid elements and the secondary MBH.
    }
	\label{fig:relative-torque-time-averaged}
\end{figure*}

We begin by studying the total torque as a function of time, from the beginning to the end of a given simulation. The torque evolution, both measured instantaneously
at any snapshot and after applying a time-averaging over a reference timescale of $0.4$ Myr,
 is shown in Figure~\ref{fig:relative-torque-time-averaged} for five representative runs ({\it LReps50g14, MReps50g14, MReps23g14R, MReps23g11R}, and {\it HReps05g14}), which clearly display a three-stage behaviour corresponding to that noted in the orbital separation plots. In the first stage, dominated by the dynamical friction wake, the net torque is small and fluctuating, being only slightly negative at pericenter but positive otherwise, a behaviour that leads to a slow decay but significant orbit circularization, as seen in the orbital-decay curves. In the second stage, the torque becomes increasingly more negative. This is the fast {\it migration} stage, in which the orbit shrinks rapidly at nearly constant eccentricity (the orbit is already nearly circular at this point, see Figure~\ref{fig:separation}). 

\begin{figure}
    \centering
    \includegraphics[width=\linewidth]{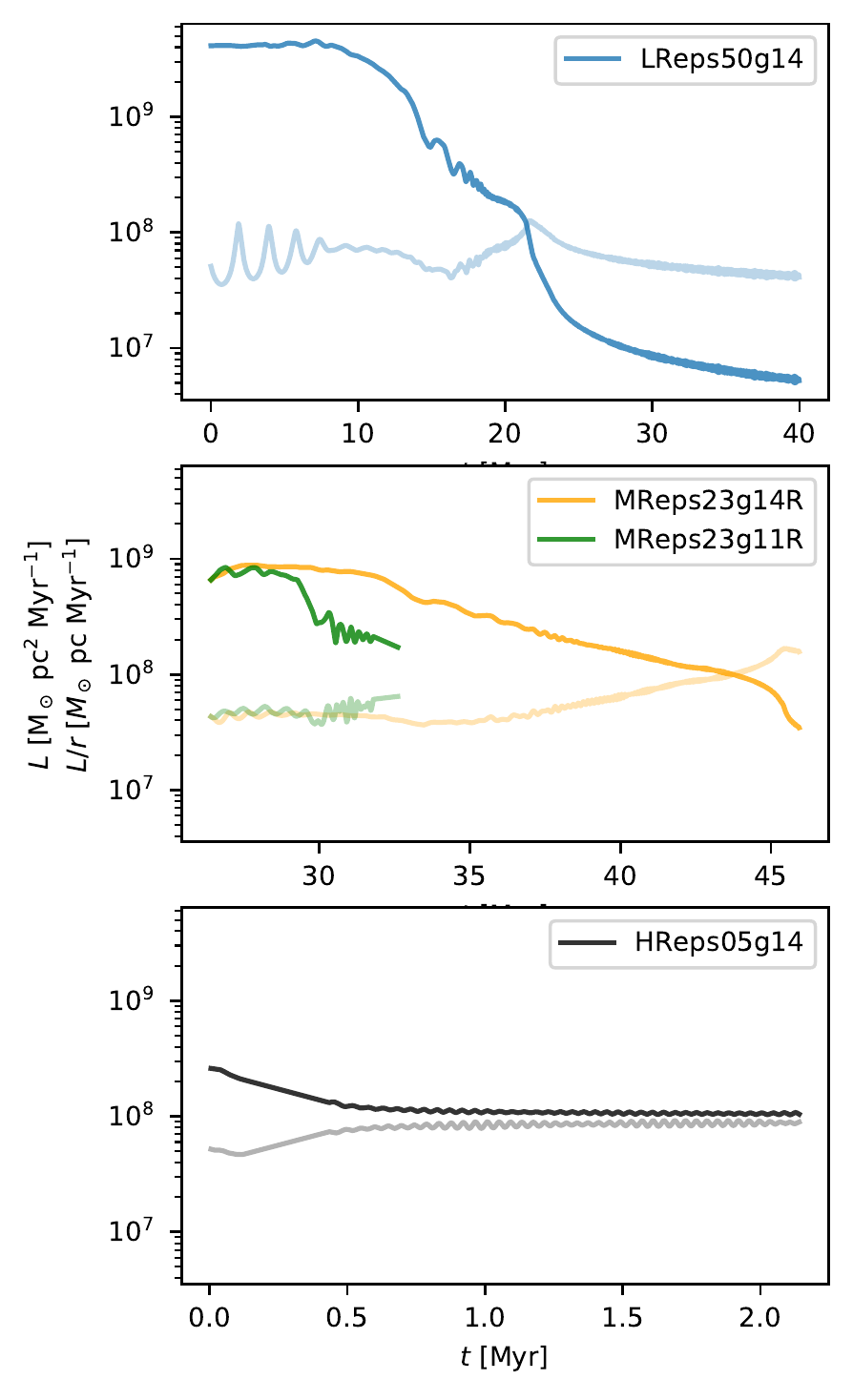}
    \caption{
        Evolution of the orbital angular momentum of the secondary MBH (thick solid curves) and tracer for tangential velocity (estimated from $L/r$, thin solid curves) for runs \emph{LReps50g14} (top panel), \emph{MReps23g14R} (middle panel), and \emph{HReps05g14} (bottom panel). The rapid decay phases shown in Figure~\ref{fig:separation} correspond naturally to drops in angular momentum. The stalling phase seen in the bottom panel is also associated to a plateau on the torque evolution (Figure~\ref{fig:relative-torque-time-averaged}).
    }
    \label{fig:angmom}
\end{figure}

Circularization phases are very clear if we inspect the orbital angular momentum and azimuthal velocity plots in Figure~\ref{fig:angmom}. These correspond to phases in which the orbital angular momentum remains nearly constant over time, despite the fact that the secondary MBH is decaying. This occurs in the initial decay phase, as we recalled several times, but also towards the end of the simulations (in the {\it HReps05g14} run, it fluctuates slightly over time at the end, probably reflecting a periodic excitation and de-excitation of eccentricity due to the interaction of the binary with the surrounding CBD).

When reaching about 1~pc separation, the behaviour of the different runs diverges. In some cases (e.g. in the low-resolution runs and the medium-resolution run {\it MReps23g14R}), the net torque becomes zero or even positive (even more if we consider the torque including the contribution below the softening), consistent with a slower decay and further circularization of the orbit. The softened potential causes an artificially extended gas distribution around the primary MBH to arise, and it seems to become the dominant region for the torques experienced by the secondary MBH. In the {\it HReps05g14} run, the net orbit-averaged torque becomes essentially zero when the circumbinary cavity is finally carved (Figure~\ref{fig:relative-torque-time-averaged}). In this case, thus, the binary appears to stall, at least temporarily.

\begin{figure*}
	\centering
	\includegraphics[width=.95\linewidth]{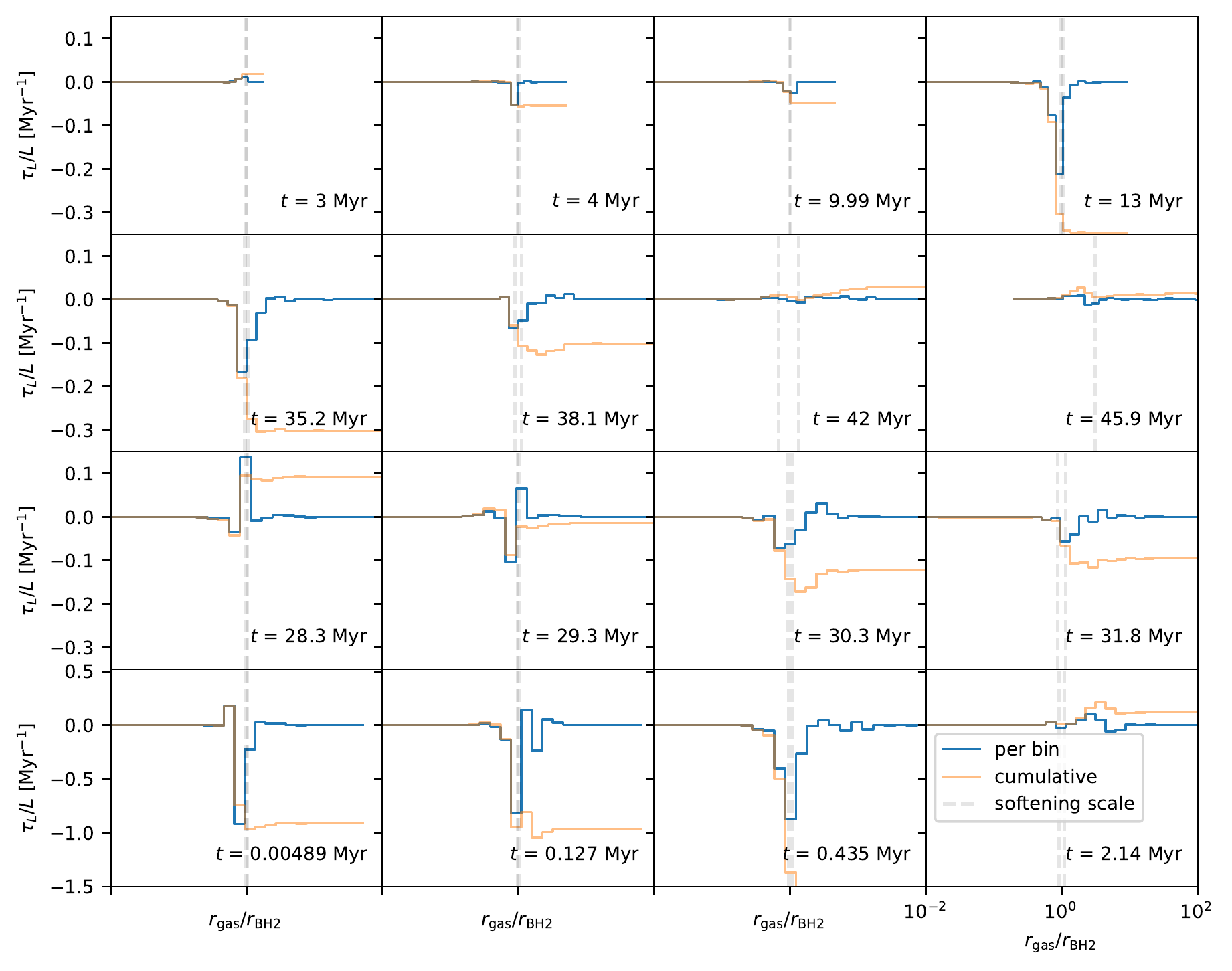}
	\caption{
    	Histogram of the differential (blue) and cumulative (orange) radial torque profile, normalized to the orbital angular momentum $L$, obtained by logarithmically  binning the disk in radius  from an origin coinciding with the primary MBH's location up to the outskirts of the disk ($\sim$150--200~pc). In particular, the horizontal axis represents the distance of fluid elements sourcing the torque relative to the distance of the secondary, both measured in the reference frame of the primary  and projected on the disk plane. Thus, plots that are truncated to the right refer to instances in which the secondary is still close to the disk edge, and those truncated to the left are such because of a binning effect, namely arising when the range in separations is smaller than the size of the first bin. From top to bottom, the simulations are \emph{LReps50g14}, \emph{MReps23g14R}, \emph{MReps23g11R}, and \emph{HReps05g14} (at the same times as in Figure~\ref{fig:torque-maps}). Note that the total cumulative torque shown here at a given time corresponds to a single data point in the torque evolution plots of Figure~\ref{fig:relative-torque-time-averaged}. We have chosen time snapshots according to the different decay phases described in the text and in Figure~\ref{fig:relative-torque-time-averaged}, although fluctuations in the radial profiles occur in between. For example, at many time instances, the negative torque, which causes decay, appears to be strongest in the co-orbital region, even if we have neglected the gas within two softening lengths. Figure~\ref{fig:torque-profiles-by-orbital-frequency} in the appendix shows analogous plots in which the binning is taken over orbital frequencies instead of the radial coordinate.}	
\label{fig:torque-profiles}
\end{figure*}

Further information about the nature of the torques is provided by their spatial distribution, which shows where the most important contribution of the torque originates from. To this aim, we show, for the usual set of representative runs, the relative torque maps in the reference frame of the secondary MBH in the plane perpendicular to the orbital angular momentum vector (Figure~\ref{fig:torque-maps}), and the azimuthally summed torque radial profile (differential and cumulative, Figure~\ref{fig:torque-profiles}); Figures~\ref{fig:torque-directional-map-row}--\ref{fig:torque-directional-map-panel} in the appendix show the directionality of the torque at different times via polar histograms. The relative torque maps in Figure~\ref{fig:torque-maps} clearly show that the torque, irrespective of its sign, is contributed mostly by fluid elements in the vicinity of the secondary MBH (from 1 to 3 Hill radii).
It also shows that the net torque results typically from a very small imbalance between the negative and positive part of the torque, which have very similar magnitude at all times, in line with previous results performing a similar analysis for MBH binaries in CBDs \citep{tang17} and for intermediate mass-ratio inspirals in accretion disks \citep{derdzinski19}. For the {\it HReps05g14} run, the map (bottom panels of Figure~\ref{fig:torque-maps}) highlights considerable more structure in the torque distribution owing to the much higher mass and force resolution, but qualitatively the picture does not change: the torque is still localized around the orbit of the secondary MBH.

The radial torque profiles confirm the local, nearly co-orbital nature of the torque, but also show that the radial torque distribution undergoes significant fluctuations over time. While there is always a stage, corresponding to the fast-decay phase, in which the torque is negative and originates from a region of a couple of Hill radii, at later times, when the secondary MBH is often seen to slow down in its course towards the center, the torque can become positive or simply very small. It is interesting also to note that, when the torque becomes positive, it also originates from  outside the Hill sphere. Finally, Figure~\ref{fig:torque-profiles} also suggests that, in the low-resolution runs, the co-orbital region providing the negative torque is of order of the softening, hence it is not formally resolved. Yet, the character of the torque distribution, in particular its co-orbital localization, does not seem to depend on resolution as it is featured in all the runs, hence across variations of softening scale and mass resolution by 1--2 orders of magnitude.

By combining the information provided by the radial profiles and by the maps, it is clear that, when the torque is strong and negative, it originates from the vicinity of the secondary MBH. This is reminiscent of the behaviour found in planet migration simulations for planets that are massive enough to depart from the description of linear theory yielding Type~I migration but not massive enough to open a deep gap \citep[][]{baruteau14}. In such case, though, various types of co-orbital torque regimes have been identified, some leading to outward and some leading to inward migration, but generally faster than what expected in Type~I migration. The co-orbital fluid material circling around the perturber, in our case the secondary MBH, in absence of self-gravity completes a horseshoe path, which is distorted if migration is already happening or if viscosity is sufficiently high, leading to a torque which is generally positive \citep[][]{papaloizou07}. However, if the mass content of the co-orbital fluid is high enough, asymmetries in its distribution occurring during migration can give rise to a strong negative torque, resulting in the so-called Type~III migration (\citealt{Masset_Papaloizou_2003}; \citealt{baruteau14}). 

\begin{figure}
	\centering
	\includegraphics[width=.95\linewidth]{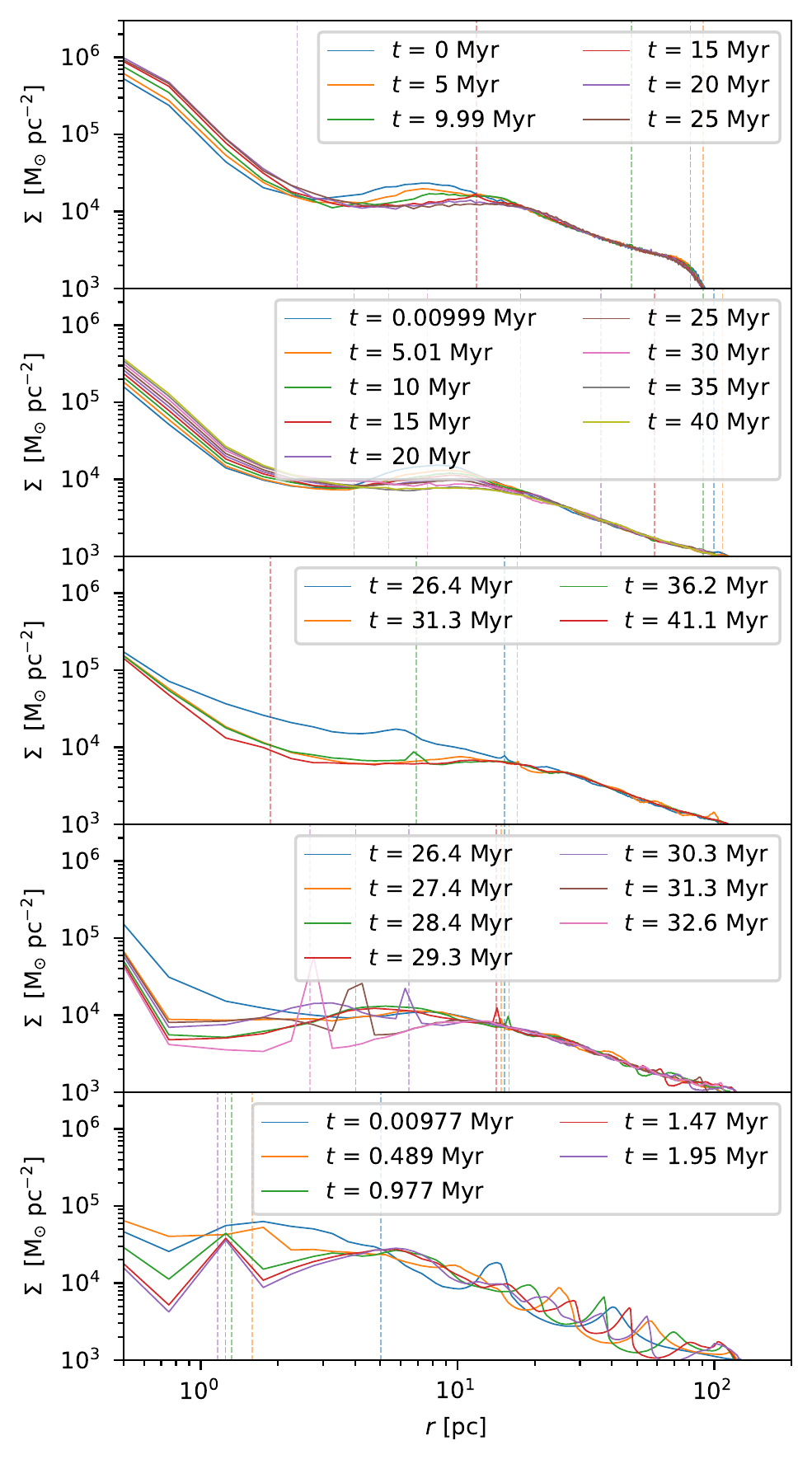}
	\caption{
    	Azimuthally averaged surface density profile at different times for the {\it LReps50g14} (first panel), {\it MReps50g14} (second), {\it MReps23g14R} (third), {\it MReps23g11R} (fourth), and {\it HReps05g14} (fifth) runs. The profiles are centered on MBH1, and equally spaced radial binning was adopted. The MBH pair's projected separations are indicated by vertical lines with the color corresponding to their respective times (note that the last snapshot in the first panel falls out of the $r$ range shown in this figure, hence below the softening length -- see also Figure~\ref{fig:separation}). A gap is signalled by a marked drop in the surface density relative to the initial distribution. This appears to occur only in the two runs corresponding to the two bottom panels, as explained in the text (Section~\ref{sec:gap-results}). Over several orbits, a depression develops, as evidenced by contrasting the curve from later to earlier times.
	}
	\label{fig:sigma-profiles}
\end{figure}

\begin{figure}
    \centering
    \includegraphics[trim={0 1.2cm .8cm 0}, clip, keepaspectratio, height=3.cm] {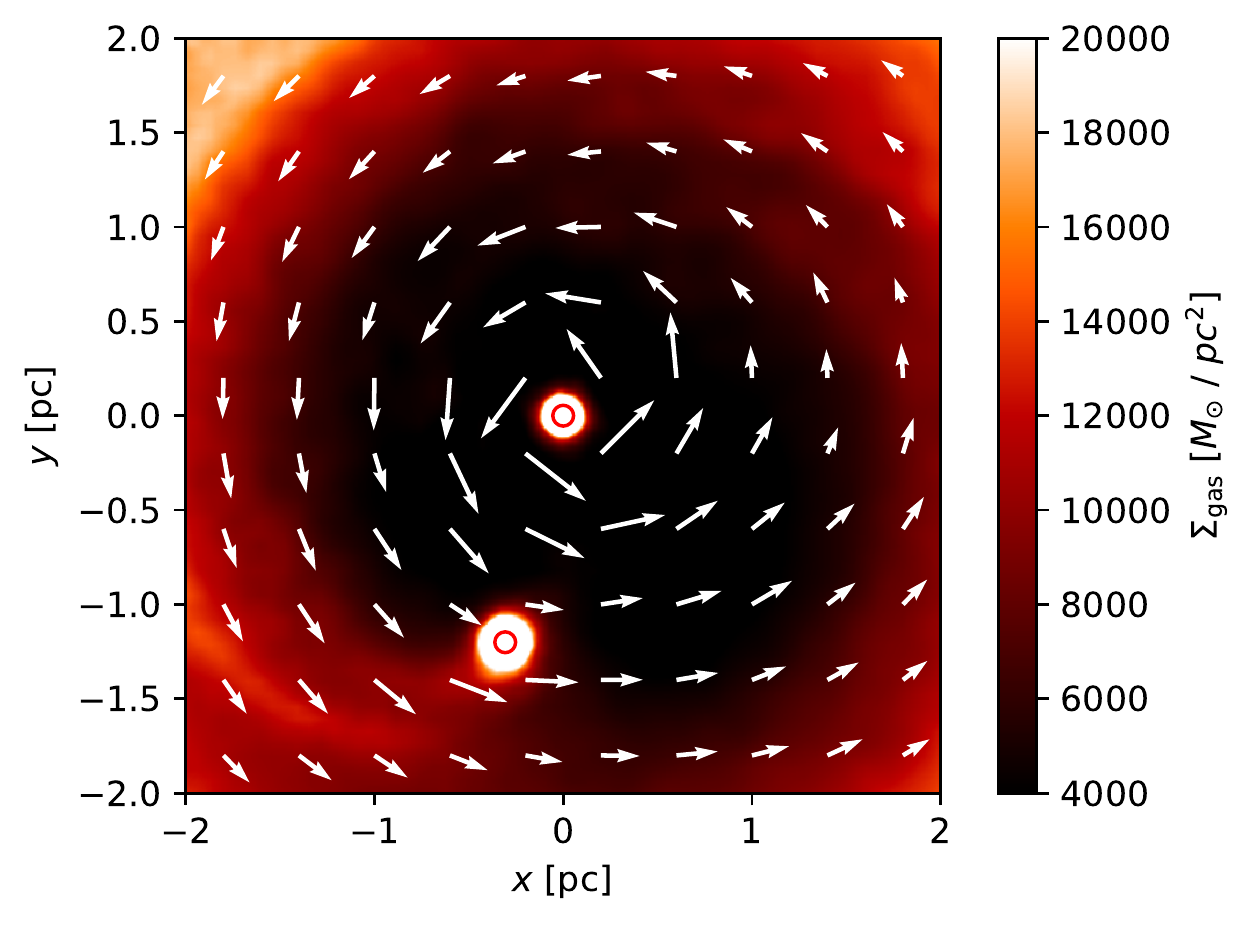}
    \includegraphics[trim={2.0cm 1.2cm 0 0}, clip, keepaspectratio, height=3.cm] {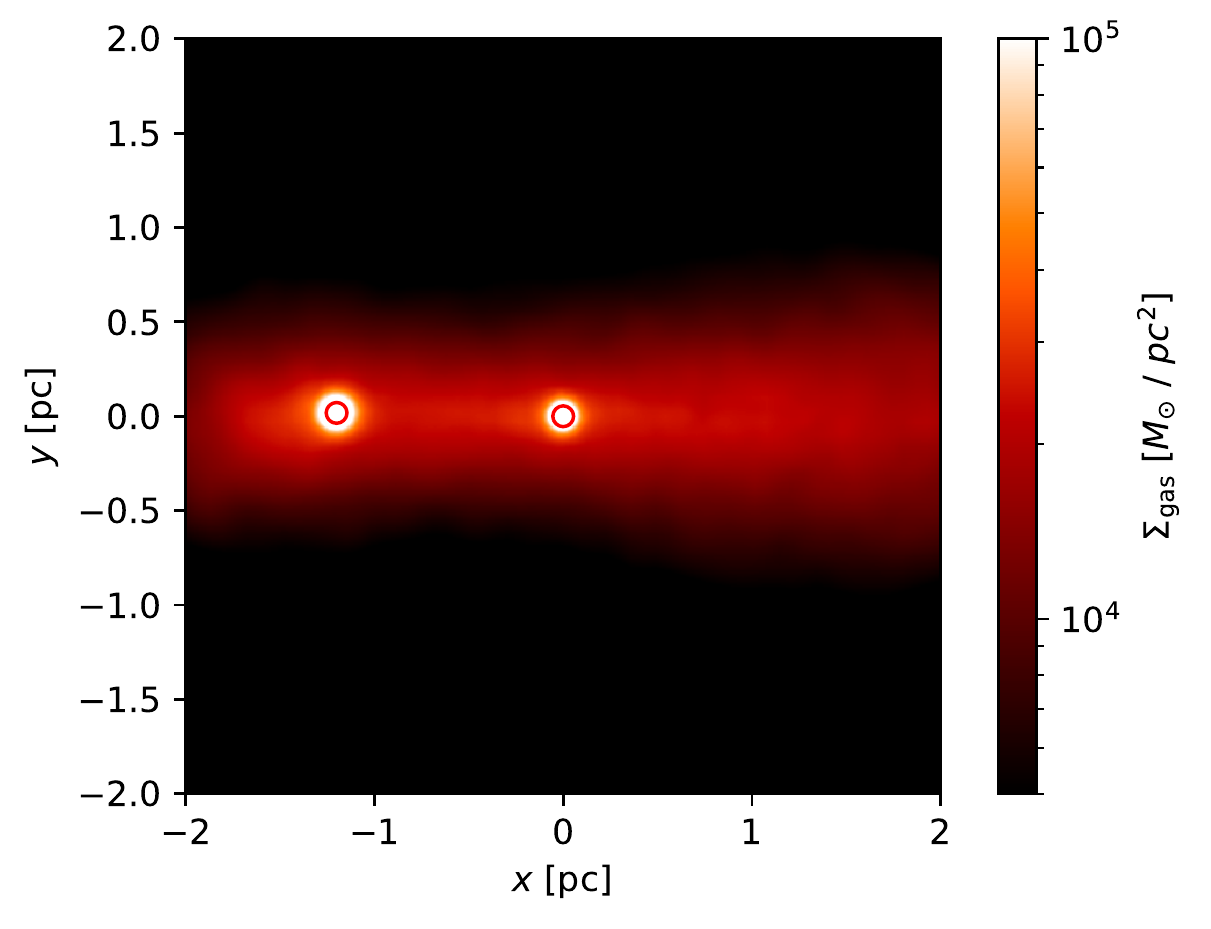}\\
    \includegraphics[trim={0 0 2.4cm 0}, clip, keepaspectratio, height=3.5cm] {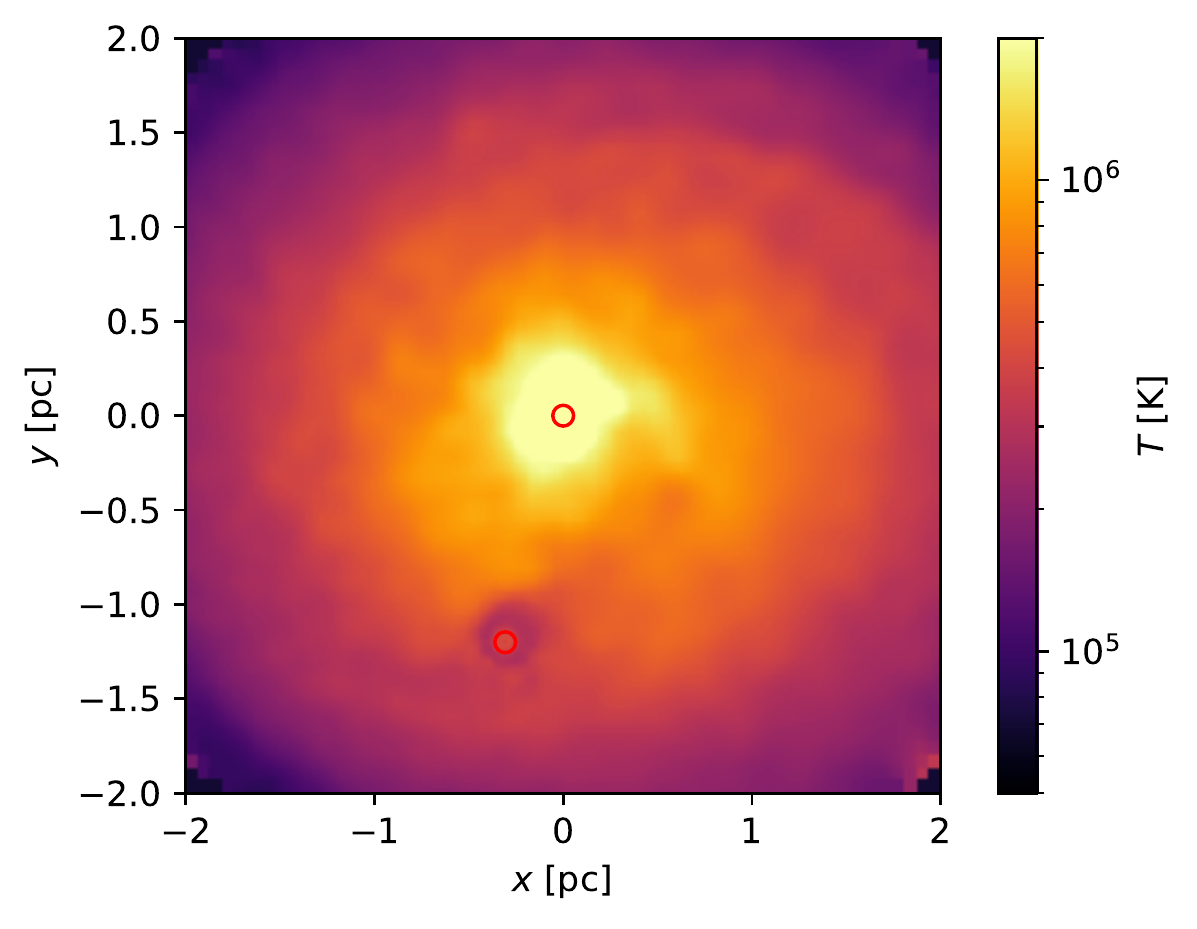}
    \includegraphics[trim={1.8cm 0 0 0}, clip, keepaspectratio, height=3.5cm] {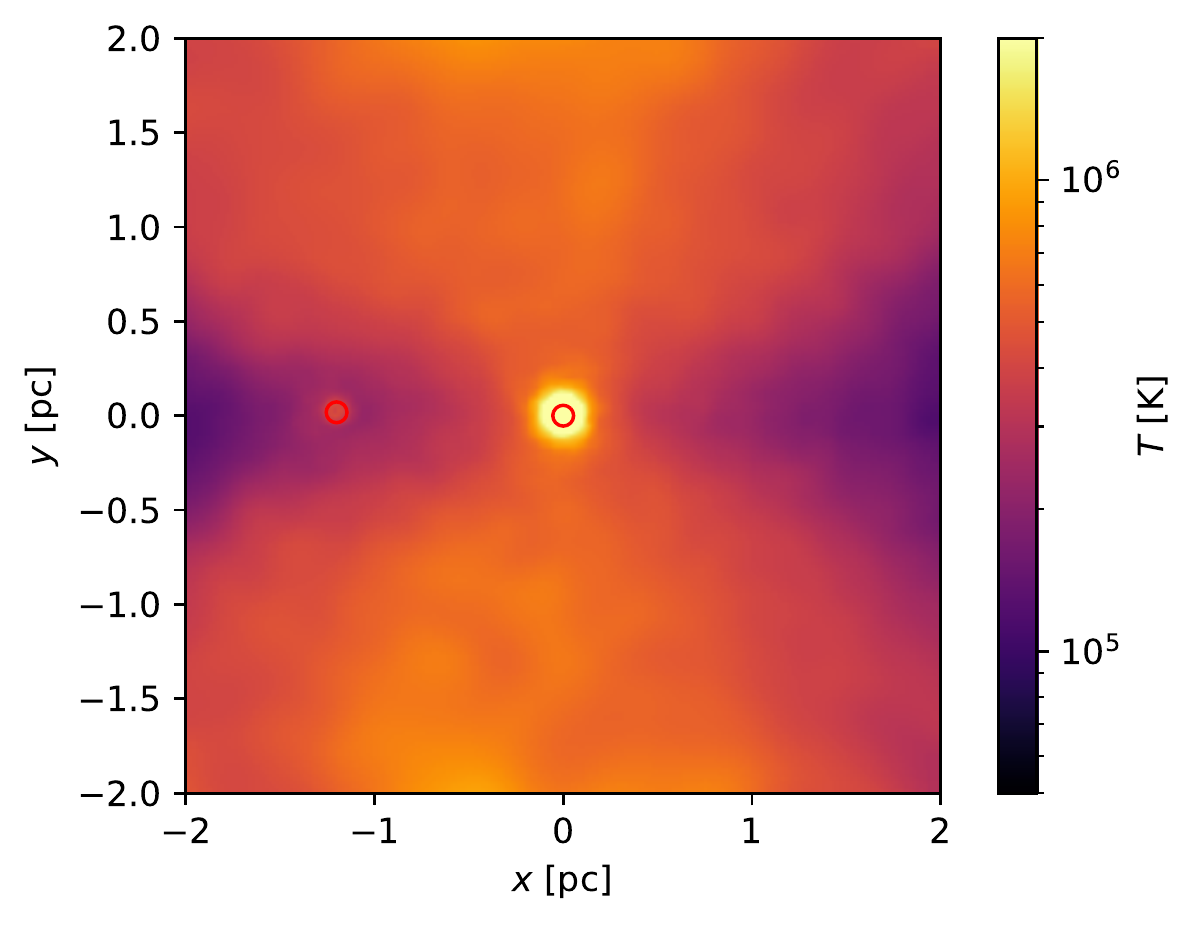}
	\caption{
	Mass-weighted gas density (top panels) and density-weighted temperature (bottom panels) maps for the {\it HReps05g14} run, with face-on (on the left) and edge-on (on the right) views of the disk, at the end of the simulation. In the top-left panel, velocity vectors showing the direction of fluid motion are overlaid. The MBHs are indicated with small red circles.  All panels but the top-left one display the subset of gas particles located within twice the MBH pair's separation, to highlight the values in the central region. A cavity carved by the binary-disk interactions can be seen, which determines the formation of a CBD.
    }
	\label{fig:flagship-at-2255}
\end{figure}

With self-gravity added, fast inward Type~III migration appears to be the most common outcome, as the torque then comes essentially from asymmetries in a disk, the size that assembles around the perturber \citep{Mayer_et_al_2015}. The second, fast decay phase present in all of our runs (see, e.g. Figure~\ref{fig:torque-profiles}) indeed resembles the fast decay in Type~III planetary migration in character. At the opposite extreme, we have the stalling phases (Figures \ref{fig:separation} and \ref{fig:torque-profiles}), such as in the final stage of the {\it HReps05g14} run, when the circumbinary cavity is opened (Figures \ref{fig:sigma-profiles} and \ref{fig:flagship-at-2255}), in which, indeed, the co-orbital torque decreases significantly. Note, however, that a small non-zero torque develops at the end further away from location of the secondary MBH, probably reflecting the modification in gas flow arising as the cavity is opened (Figure~\ref{fig:flagship-at-2255}). Inspection of the torque radial profiles in the latest stage of all runs, which corresponds to a regime of slow decay to near stalling in most of them, also shows, in general, that some small residual torque comes from further away compared to the fast decay phase (the only exception is the {\it LReps50g14} run which, as we know, is also the one that shows less evidence of a slower decay in the late stage).

While a significant reduction of the co-orbital torque seems to be well correlated with the transition from a fast to a slow decay regime, it can also occur temporarily over an orbit, resulting nevertheless in a transitory suppression of the orbital decay. For example, inspection of torque maps over time shows that, sometimes, local asymmetries in the flow induced by the passage of MBH2 outside its Hill sphere, happen to be on the opposite side relative to the location of the secondary MBH half an orbit later, because the orbital frequency of the secondary MBH is smaller than the orbital frequency of the asymmetric pattern, reflecting the relatively deep potential well of the system. This means that an induced  density perturbation that would be typically extracting angular momentum from the secondary MBH has its effect reversed within one orbit, and its overall net effect is therefore nearly cancelled over an orbit.

\begin{figure}
    \centering
    \includegraphics[width=.95\linewidth]{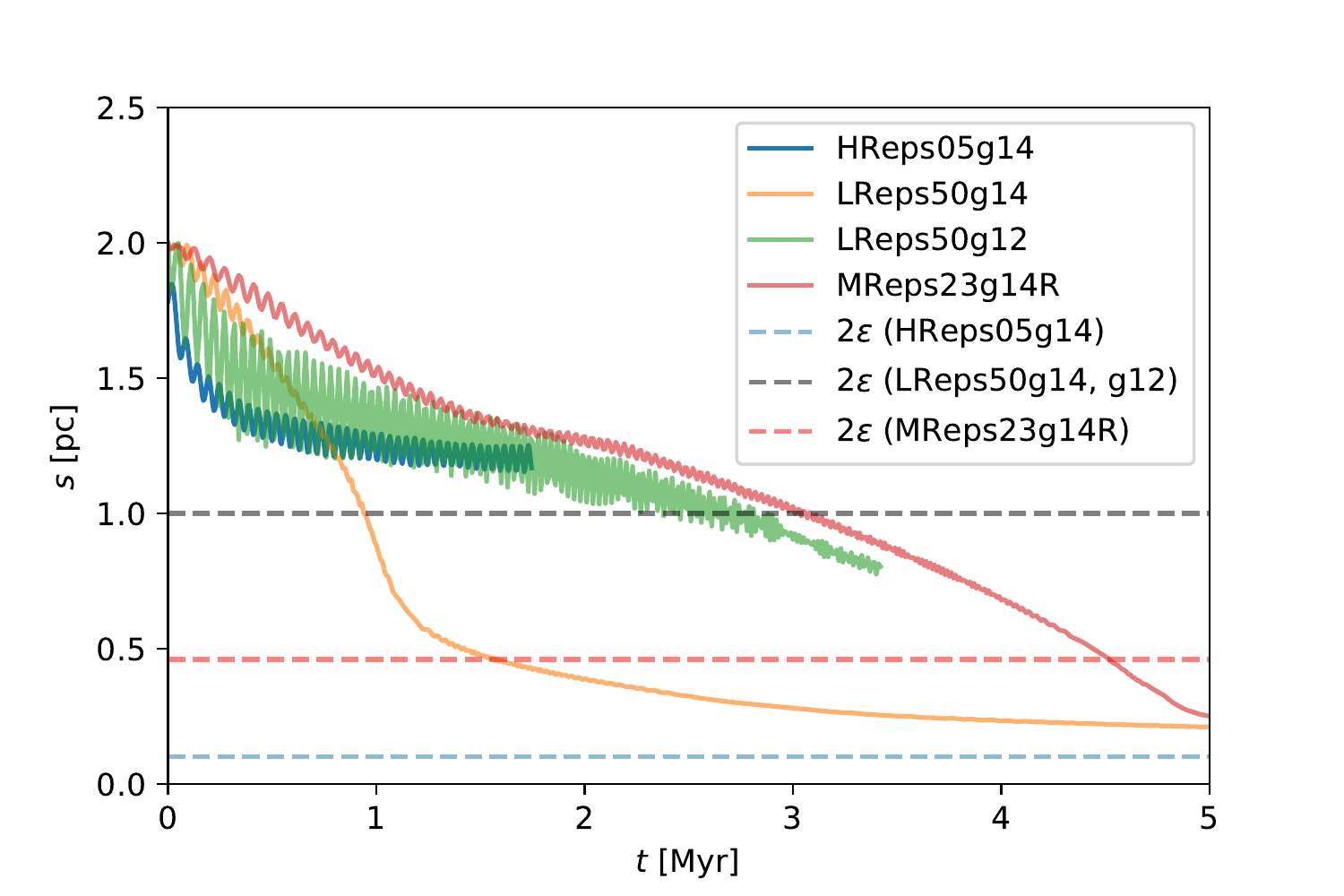}
    \caption{MBH separation versus time for runs of different resolutions. The time coordinate was shifted so that $t = 0$ indicates the time at which the separation $s$ crosses the 2~pc mark. Note that the vertical axis' scale is linear, unlike in Figure~\ref{fig:separation}, to better highlight differences. For the highest-resolution run, the \emph{HReps05g14} run, the slope of the decay curve is the  smallest among those displayed, and the final stalling phase in this run ensues when the separation is about 20 times the scale of the softening, hence being a numerically robust outcome.
    }
    \label{fig:separation-comparison-sub2pc}
\end{figure}

All the evidence we gathered from the analysis of our set of runs shows that, despite some differences, the decay rate is typically reduced when the separation approaches pc scales, and whether or not the decay continues at sub-pc separations or nearly stalls appears to be very sensitive to the detailed conditions of the flow, which is influenced by ICs, resolution, and EoS. This is very reminiscent of previous results for the migration of massive planets and brown dwarfs in protoplanetary disks, for setups with nearly identical mass ratios between the perturber and the (self-gravitating) disk as in our simulations \citep[see, e.g. the runs with 15 Jupiter-mass planets in self-gravitating disks in][]{malik2015}. In the latter study, it was found that, even just changing the initial azimuthal location of the perturber while keeping all the other parameters of the ICs the same, could change significantly the orbital decay rate, and even bring the system from continued migration to stalling regimes.

Finally, another important finding is that, at pc scales, the slow ensuing decay phase can be accompanied by the carving of a circumbinary cavity, producing a CBD. The prototypical case for this transition is the {\it HReps05g14} run, which is our highest-resolution run. Whether the reduction of the torques causes the transition to the cavity opening regime, or whether, on the contrary, it is the ensuing of the CBD that causes a slower decay stage, is not obvious. In the next section, we address this and, more in general, the mechanism of gap/cavity opening in our runs.

\subsubsection{Effects of different resolution}\label{sec:resolution}

The gas density and temperature fields of the inner regions of the single-phase gas disk with an effective adiabatic EoS are shown for simulations of varied number of particles, force resolutions, and adiabatic indices (see Table~\ref{tab:simulations}) in Figures~\ref{fig:simcomparison-rho}--\ref{fig:temperature-maps} and Figure~\ref{fig:flagship-at-2255}. We have already anticipated, at the beginning of this section, the marked differences of the gas flow properties  at 1--10~pc scales highlighted in these figures. In particular, we can infer that a proper description of the transition from CND scales (here, of the order of 100 to 10~pc) into CBD separations (around a pc and below, where the spheres of influence of the MBHs, i.e. the regions where their gravitational forces are dominant, start to overlap) requires a finer sampling of the gas, namely at least $10^6$ particles (and better force resolutions to capture the details of MBH--disk interactions). Indeed, a CBD only forms in the late stages of medium and high mass resolution runs, most notably in the {\it HReps05g14} run, whereas at low resolution the gas flow is smoother and characterized by a much larger accumulation of gas around the primary MBH (see Figure~\ref{fig:simcomparison-rho}). We note that this is the first study of MBH pairs in CNDs in which the CBD formation is resolved. Another important finding, highlighted in Figure~\ref{fig:separation}, is that, contrary to some previous work probing at best pc scales \citep[e.g.][]{escala05,chapon13}, the orbital decay rate at sub-pc scales does not increase with resolution, rather the opposite is observed. This is connected to the fact that, at higher resolution, the gap/cavity opening is resolved, which then changes the torque regime. Here increased mass resolution, rather than the improved force resolution owing to smaller softening, likely plays a major role, as the flow becomes less noisy and less diffusive (artificial viscosity in SPH is reduced with increasing mass resolution, since it depends on the smoothing length).  The reduction of the effect of artificial viscosity at higher mass resolution is strongly suggested by the colder temperature of the flow in the inner disk region as the resolution is increased (top to bottom in Figure~\ref{fig:temperature-maps}). The only exception is the innermost fluid material within a pc from the primary MBH, which remains hot as adiabatic compression is very high in the deepest part of the potential well (note, however, that this would be easily removed by adding radiative cooling and/or accretion onto the primary MBH). The overall resolution-dependent behaviour is reminiscent of  the effect of artificial viscosity and  noisy hydrodynamical forces at lower resolution extensively documented in galaxy-scale simulations \citep[e.g.][]{kaufmann07}.

Finally, in the low-resolution run {\it LReps50g14}, the MBH binary appears to shrink below the distance resolved by the softening, while the decay rate slows down at larger separations in the corresponding run with ten times better mass resolution and equivalent softening. The radial torque profiles indeed show that a significant net negative torque in the co-orbital region persists till the end of the simulation, but it is well below the softening scale, hence it is not trustworthy on a numerical basis. Moreover, we note that the central gas density profile evolves differently in the various runs, leading to a much larger central accumulation of gas around the primary MBH at lower resolution (compare panels from Figure~\ref{fig:sigma-profiles}), due to the well known enhanced loss of angular momentum by artificial viscosity and other spurious numerical effects in low-resolution SPH simulations of shearing flows \citep{kaufmann07, deng17}. The central smooth massive, unresolved envelope that forms this way is likely responsible for the persistent, spurious torque in the {\it LReps50g14} run.

\subsection{Gap-opening analysis and the formation of a circumbinary disk}\label{sec:gap-results}

Inspection of the CNDs' density maps at different stages and scales shows no evidence of gap opening when MBHs are at separations of 10~pc or larger (Figure~\ref{fig:simcomparison-rho}), whereas there are signs of gap opening in some of the runs when the separation of the two MBHs decreases to pc scales. In particular, as we have often recalled, the {\it HReps05g14} run presents a clear case in which a circumbinary cavity is carved, leading to a CBD configuration by the end of the simulation.

We now turn to a more quantitative analysis. In Figure~\ref{fig:sigma-profiles}, we present the density profiles of different representative runs. These gas density profiles were determined from the average value per bin for the particles in a slice around the midplane, excluding the particles too close to the secondary MBH (within three softening lengths), to avoid the overestimation of the disk's background density along its orbit. This, because the sphere of influence of the MBH is resolved in all simulations, causing particles to get captured by it. Those particles are then confined to small volumes, reaching densities orders of magnitude larger than the density of the background flow. The surface density profiles confirm that only in the late stages a depression in the density can form around the secondary MBH, and this is particularly prominent in the {\it HReps05g14} run, as expected. Amongst the runs, the lower-resolution runs do not show evidence of gap opening, likely because, when the conditions would be met at pc scale separations, the resolution is not high enough to correctly model the flow, causing numerical diffusivity by artificial viscosity. At medium resolution, the density depression is more evident in the run with lowest adiabatic index ({\it MReps23g11R}). This is expected, since the gas temperature, and thus its pressure, is lower in this case compared to the standard EoS choice, which goes in the direction of favouring gap formation (see inequality~\ref{eq:crida}).

In order to understand why it is only at small, pc-scale separations that a gap/cavity can be opened, at least in a fraction of our runs, we analyse the gap-opening conditions presented in Section~\ref{sec:gap-criteria}. We should first recall that the standard gap-opening criteria were envisaged for isothermal disks, and neglect both self-gravity and the effect of migration of the perturber. In particular, in self-gravitating protoplanetary disks, gap opening has been shown to be difficult to achieve even for the most massive perturbers, in this case giant planets, because migration is too fast for the gap to be carved in a region where the appropriate conditions would be met \citep[][]{Paardekooper_et_al_2011,malik2015,Paardekooper_Johansen_2018}. Fast orbital decay without gap opening has also been observed in simulations of secondaries migrating in adiabatic CNDs with resolution comparable to the lowest-resolution runs considered in this paper \citep[][]{mayer13}. In the previous subsection, on torque analysis, we have already seen that the second phase of orbital decay, during which the secondary MBH covers most of the distance from the CND boundary to pc-scale separations, is very fast. The prevention of gap opening in fast-migration regimes is caused by the fact that an additional criterion comes into play in such a case, namely the gap-opening timescale has to be shorter than the migration timescale through the region where the gap-opening conditions are satisfied. If this does not occur, the perturber will decay too fast for the gap to open. We will return to this issue below, after the quantitative analysis.

Nevertheless, for now we neglect the aforementioned complications, and simply apply the standard criterion in inequality~\eqref{eq:crida}, re-written in the form of inequality~\eqref{eq:c}, which contains the additional fudge factor {\it g}. We recall that {\it g} can be used, in a phenomenological way, to explore the importance of the additional physics not included in the standard criterion, namely given by the potential effects just outlined. The criterion is a function of essentially three parameters: distance to the center of the potential well, mass enclosed within the secondary MBH's orbit, and disk temperature at its position. The disk surface density, or the mass ratio between the secondary and the primary MBH, do not enter the criterion (but see below). The total enclosed mass was measured in spherical bins, accounting for all particles (bulge, disk, and MBHs), as it is a proxy for the gravitational force in the disk. We note that the gas particles are also subject to hydrodynamic forces, which affect their tangential velocities to some extent. We neglect the latter difference in this analysis (for the {\it HReps05g14} run, we have indeed verified that the effect is small). The temperature or sound speed measurements were taken from partitions with cylindrical radii corresponding to those of the spherical projection. The right-hand side of inequality~\eqref{eq:c} is thus a function of $r$, $M_{\rm{enc}}$, $M_2$, and $H$.  Considering the system described in Section~\ref{sec:ic}, $M_2$ is a fixed parameter for a non-accreting MBH, $r$ takes values within the range 0--200~pc (see Figure~\ref{fig:separation}), and $M_{\rm{enc}}$, which is the mass enclosed within the orbit of the secondary MBH,  has a minimum value equal to $M_1$, the mass of the primary MBH (relevant when the secondary MBH has reached small separations) and a maximum value equal to the total mass of the CND plus the primary MBH, i.e. $6.1 \times 10^8$~M$_{\odot}$ (relevant when the secondary MBH is still orbiting near the edge of the CND). If $H$ can be expressed in terms of $r$ and $M_{\rm{enc}}$, in the limiting case of equality (in inequality~\ref{eq:c}), the points $(M_{\rm{enc}}, r, c(M_{\rm{enc}}, r))$ define the area (along the $c$ axis) below which a companion of mass $M_2$ is expected to open a gap in the gas disk. Hereafter we will refer to the latter
surface as \textit{gap-opening area}. 

\begin{figure*}
	\centering
	\includegraphics[width=.95\linewidth]{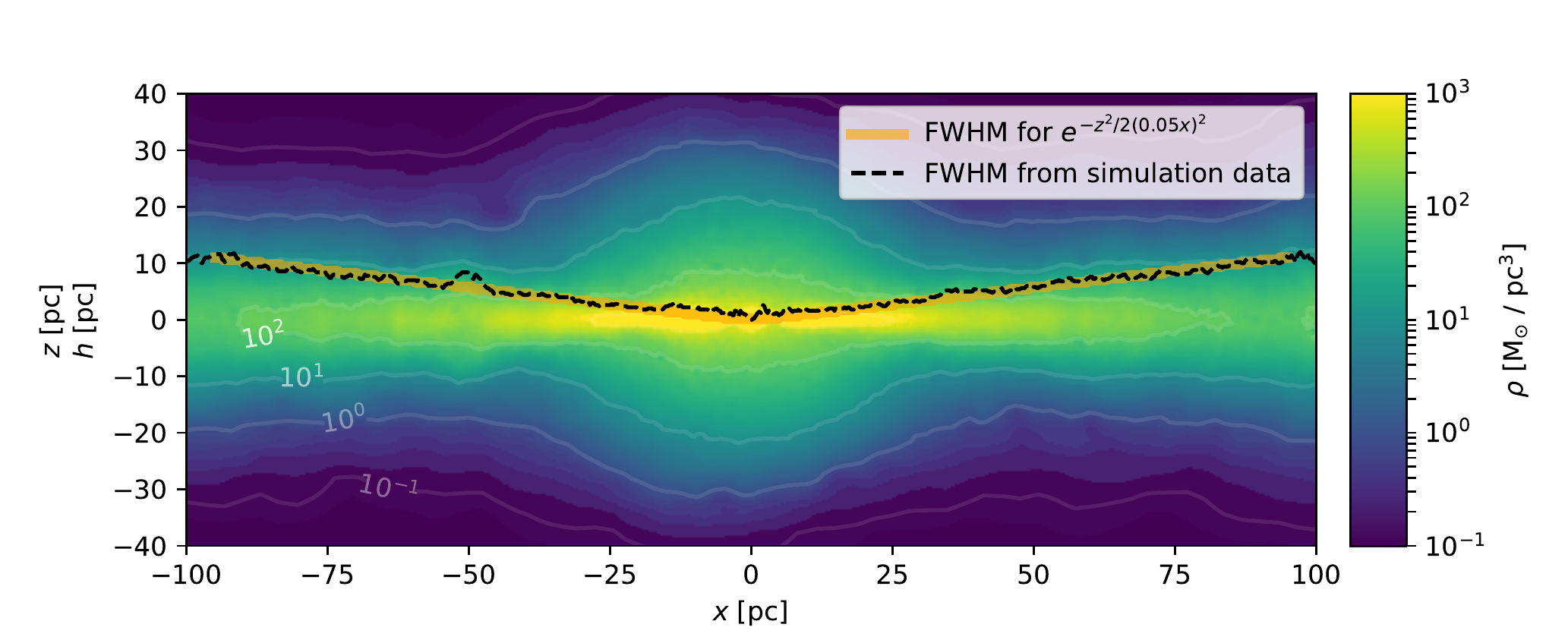}
	\caption{Edge-on view of the disk at an early stage ($\sim$5~Myr) of run \emph{MReps50g14} (see Table~\ref{tab:simulations}) and its height profile. The FWHM is represented by the dashed black line, and the corresponding curve for a Gaussian vertical structure such as  the one used to generate the ICs is plotted in yellow for reference. The secondary MBH is relatively far from this slice, located around $x = 83$~pc and $y = -69$~pc. Both height profiles are still in agreement for radii in the range of 10~pc and above -- here in particular, despite the presence of the perturber -- underlining the applicability of the parameter choice for estimating the surface in Figure~\ref{fig:c-surface} at least on this range.}
	\label{fig:fwhm-relaxed-naad-s32-ml-500}
\end{figure*}

The vertical density distribution profile is shown in Figure~\ref{fig:fwhm-relaxed-naad-s32-ml-500} for the ICs, namely the initial  CND model adopted in our runs, and for a later stage of one of our representative runs (see caption). The figure shows that adopting the IC to assign a scale height in the gap-opening criterion is sensible, as both the initial and the evolved vertical profile can be fitted by the same function. Typically, the FWHM of the vertical density profile is used for the determination of the disk height. However, this definition is highly sensitive to the peak value of the column density. The half-mass width of the disk is somewhat more resilient to these fluctuations and it will be preferred in some of the upcoming analysis.

\begin{figure*}
	\centering
	\includegraphics[height=0.4\textheight]{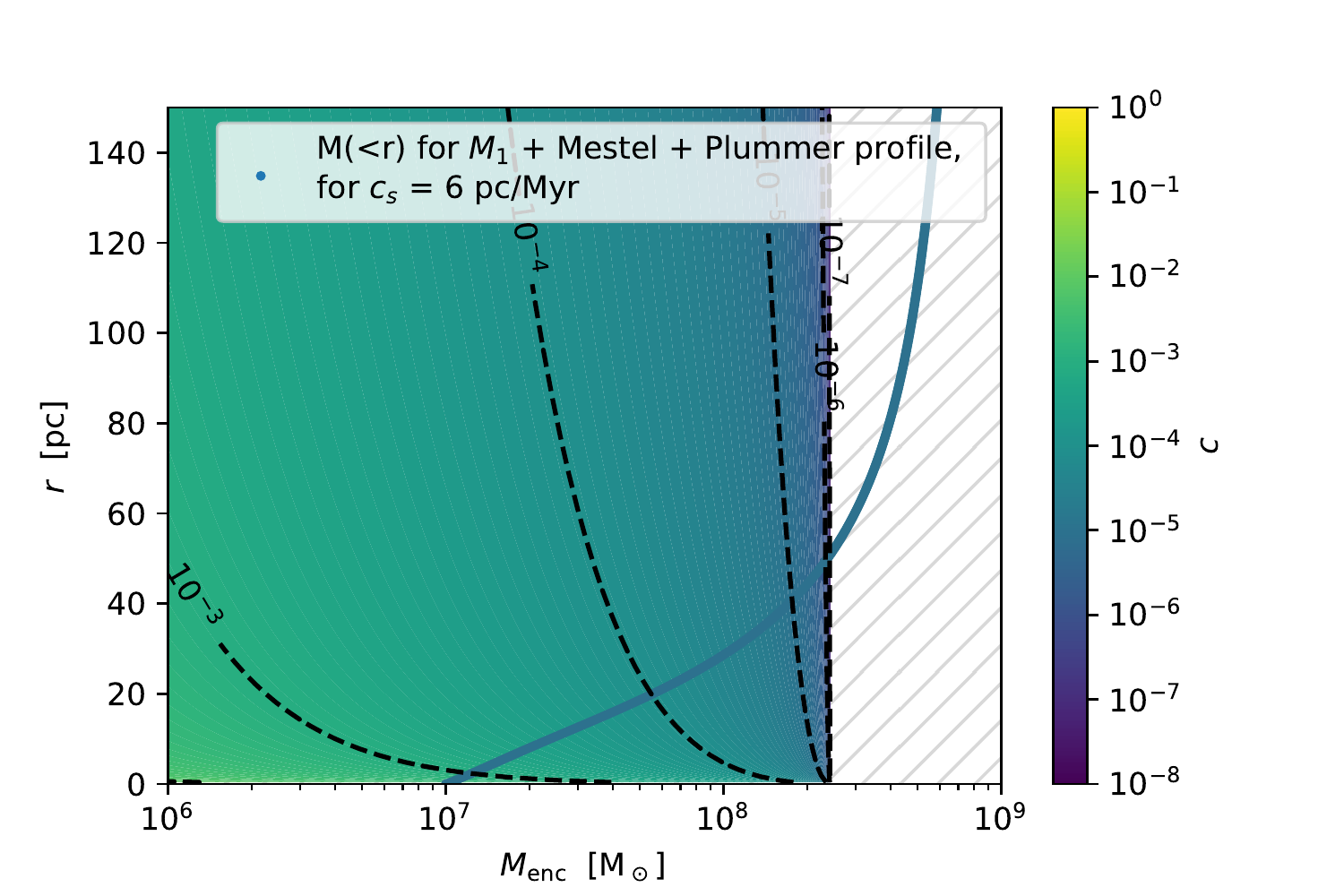}
	
 	\includegraphics[height=0.4\textheight]{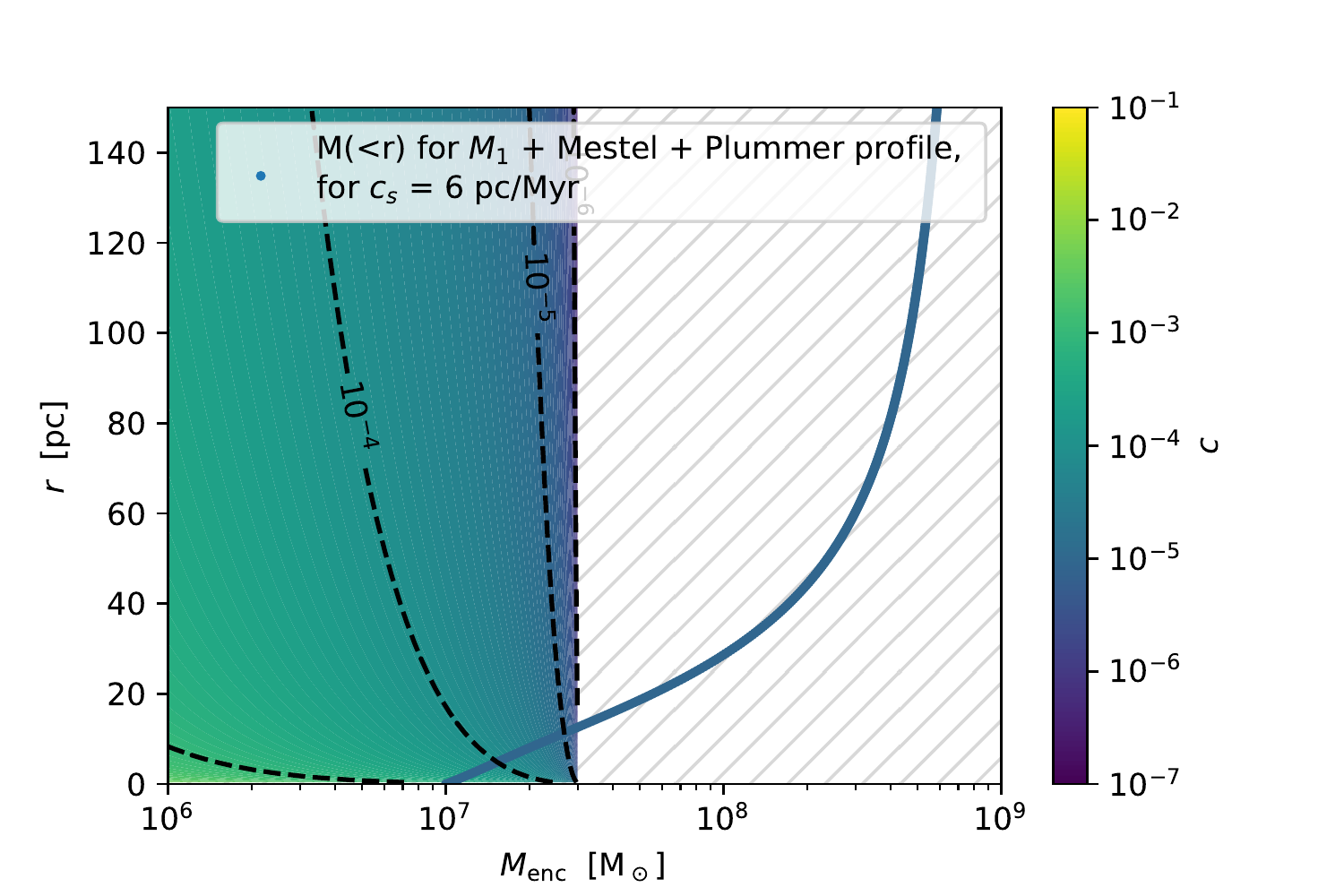}
	\caption{
	The $c$-surface defined according to inequality~\eqref{eq:c}, for a system with $M_2 = 5 \times 10^5~M_{\odot}$ and $H = 2 \sqrt{2 \ln{2}}\ 0.05 r$. In the top panel the surface is computed assuming	$g=1$ for the fudge factor, whereas in the bottom panel it is computed for $g=0.5$. This representation can be used to infer, for a given mass profile (pinpointed by the enclosed mass on the vertical axis), at which distance a gap can be opened in the chosen system's setup. The hashed region corresponds to the values where the right-hand side of the inequality assumes negative values, indicating the impossibility to open a gap according to the adopted model. The figure also shows the blue curve defined by the quasi-equilibrium equations used to generate the ICs of the simulations, thus showing the prediction valid if the relaxed system did not evolve strongly away from  them, which was verified to apply here. Different mass profiles will simply shift the curve vertically and horizontally, whereas changes in the gas temperature, adiabatic index, or viscosity will change the color of the profile; changes in $g$, $M_2$, and $H$ will affect the $c$-surface itself.
	}
	\label{fig:c-surface}
\end{figure*}

The {\it gap-opening area} is represented in Figure~\ref{fig:c-surface} for $H$ taken as the FWHM of the ICs' vertical mass profile. The figure also indicates the curve defined by the quasi-equilibrium equations used to generate the ICs of the simulations, in the vicinity of which data points were expected to fall if the relaxed system did not strongly deviate from the ICs. Finally, the actual data points collected at different times of the simulation are also plotted.

For a fixed radius (implying, in this case, also a fixed  scale height), Figure~\ref{fig:c-surface} shows that the $c$-parameter decreases with $M_{\rm{enc}}$, suggesting that more massive CNDs are overall more resilient to opening a gap. For a fixed enclosed mass, an increase in $r$ corresponds to a decrease in $c$, also pointing to a system that is more resilient to opening a gap. One might expect, as a result, that less dense CNDs, which have lower values of  enclosed mass at fixed size, or more compact CNDs of the same mass, are more prone to gap opening. Also, at smaller distances from the primary MBH, the gap/cavity-opening likelihood increases, consistent with the fact that a cavity is more likely observed for small, pc-scale separations between the two MBHs. However, the figure suggests that the dependence on the disk height at the perturber's position is even stronger. The temperature of the gas, and thus the chosen EoS, is thus an important parameter. Indeed, we find that a softer EoS, closer to isothermal, favours gap opening, as the pressure scale height is reduced owing to the lower resulting temperature (compare the third and second row of Figure~\ref{fig:temperature-maps}). As a result, the transition to a cavity-opening stage is more evident in the {\it MReps23g11R} run relative to the other medium-resolution runs (see Figure~\ref{fig:density-maps} and compare the third and second row; see also Figure~\ref{fig:sigma-profiles}). Note that, in the inner few pc, the temperature is higher in the low-resolution runs (Figure~\ref{fig:temperature-maps}), which probably reflects extra heating by artificial viscosity at lower resolution (see previous section). The decrease of temperature with increased mass resolution is thus among the reasons why the highest-resolution run of the set, the {\it HReps05g14} run, is that in which  the opening of a cavity is most effective, leading ultimately to the formation of a CBD. However, in this paper, we have gathered significant evidence supporting a rather stochastic evolution of the torques, and thus of the resulting orbital decay, which  prevents us from arguing that cavity opening will be generally more likely with increasing resolution.

Figure~\ref{fig:c-surface} also shows the same $c$-map for a smaller value of the fudge factor ($g = 0.5$). The higher $g$ maps (not shown) depart even further from our results, whereas for $g=0.5$  the prediction becomes more in line with the numerical results, as the {\it gap-opening area}, corresponding to scales of a few pc and below, is indeed the region in which the cavity arises in some of the runs. We can thus recover the simulation results by simply adjusting the value of $g$ (e.g. $g=0.5$). This approach should be regarded as a simple phenomenological one that allows to hide additional effects, such as concurrent migration (see next paragraph), which are not included in the standard gap-opening conditions. Such $g$ correction will be easy to implement in semi-analytical population synthesis models of the  evolving MBH binary population, which so far neglect the transitions between decay regimes analyzed in this paper \citep[e.g.][]{DeRosa_et_al_2020}.

So far it seems that the ability of the secondary MBH  to open a gap/cavity is substantially suppressed relative to predictions based on the conditions ~\eqref{eq:crida}--\eqref{eq:c}. We decided to investigate whether or not the main reason behind  this is the concurrent fast migration observed in the second phase of the decay. To address this, we should recall that, for massive perturbers in massive self-gravitating disks, the disk mass plays a role in the torque, hence on migration. In particular, in our case the secondary MBH's mass is about  $10^{-2}$ the mass of the CND, which is similar to the mass ratio in many of the self-gravitating disk simulations for migrating  Super-Jupiters and brown dwarfs considered in \citet{malik2015}. There it was found that the gap-opening timescale, computed  rigorously by turning off migration by hand, was in the range of 10--50 orbits, hence quite long compared to migration timescales for massive bodies in self-gravitating disks, which was found to be $< 10$ orbits (defined as the number of orbits completed before reaching the center of the disk or the central star). In our simulations, the gap-opening conditions are satisfied during the second, fastest decay phase (compare the distance $r$ in Figure~\ref{fig:separation} with Figure~\ref{fig:c-surface}), during which the secondary MBH drops to pc-scale separation in only a few orbits (less than 10--20~Myr), hence too fast to open any gap. A different situation applies to the final stages of the {\it MReps23g11R} and  {\it HReps05g14} runs, which also satisfy the gap-opening conditions based on Figure~\ref{fig:c-surface}. In the latter run, in particular, the net negative torque quickly decreases  (Figure~\ref{fig:relative-torque-time-averaged}) so that the secondary MBH nearly stalls at 0.5--2~pc separations, and can thus complete more than 10 orbits while remaining more or less at the same location. As a result, towards the end, a cavity, and thus a CBD, form. {\it Hence, cavity formation is the result, not the cause of a strong reduction of the net torque}.

In summary, the analysis presented in this section supports the view that it is the torque regime in which the secondary MBH finds itself (rapid or slow/stalling decay phase) that determines whether or not a CBD can form, and not viceversa. Since migration in the fast, Type-III-like phase should depend on the local gas surface density as it is driven by the co-orbital torque, it follows that, indirectly, the local gas surface density should enter a generalized  gap-opening criterion. The latter dependence can also be  hidden by choice of an appropriate value for the fudge factor $g$.

\section{Discussion}\label{sec:Discussion}

A first important result of this work is that, invariably, at pc separations the decay rate diminishes drastically, leading in some cases, such as in the {\it HReps05g14} run, to an almost complete stalling of the binary. Before that,  there are phases of very fast migration, during which the torque appears to be predominantly due to material  in the co-orbital region,  resembling the situation in Type~III migration, and in agreement with indications in previous work \citep[e.g.][]{mayer13}.

If the torque is quasi-periodic, changing sign on a timescale given by the orbital frequency of the gas within or just below the Hill sphere, the net torque over an orbit of the secondary around the primary MBH will likely cancel out. This is reminiscent of the behaviour of the horseshoe torque for migrating planets, although the analogy is only qualitative, as here the gas is self-gravitating and non-isothermal. Whether or not the torque will exactly cancel out will depend on the details of the local gas flow, in particular how asymmetric this can remain over a single orbital cycle. In a self-gravitating gas disk, even with fixed thermodynamics and resolution, the local gas flow's asymmetries and overdensities will vary if one varies slightly the ICs \citep[][]{durisen2007,fiacconi13,malik2015}, as even minor IC variations will induce slight variations in the orbital evolution of the secondary. This, in turn, will induce a slightly different perturbation on the surrounding gas, as shown by the sensitivity to changes in just initial azimuthal location of the perturber in \citet{malik2015}. Therefore, it is not surprising at all that, in our suite of runs, in which we also vary mass and spatial resolution, there is no convergent behaviour at small separations, where the torque is driven by the local gas flow properties. Note that this is very different from the large-scale regime in which the wake by dynamical friction dominates the drag, because in that case going to higher resolution allows to simply resolve the wake better, thus increasing the drag force, as shown in \citet{chapon13}. 

On the other end,  at galactic scales, a very recent analysis based on high-resolution cosmological simulations of galaxy formation suggests that other dynamical effects, such as global torques by bars, might dominate the orbital decay, introducing more complexity and stochasticity in how torques depend on resolution and on various physical processes in the interstellar medium (Bortolas et al., in preparation). On the other extreme of the orbital separation ladder, in CBD simulations probing separations of $10^{-3} - $0.1~pc, a high sensitivity of the net torque on the flow conditions has been found (see, e.g. \citealt{Munoz_et_al_2019}; and the review by \citealt{DeRosa_et_al_2020}), leading to inconclusive results on the orbital-decay timescale, and even on whether or not the net torque over many cycles is positive or negative. We can thus state that a general picture is emerging from all these different and complementary studies of the evolution of MBH binaries, namely that the binary hardening process in gaseous media is inherently stochastic.

Another important result of our work is that, at least in a subset of our simulations,  a cavity, and thus a CBD, is seen to form at separations of or slightly below one pc, as a result of the reduction of the torques driving orbital decay. This again agrees with the results of \citet{malik2015} for gaps opened by massive perturbers (Super-Jupiters and brown dwarfs): the migration rate controls gap opening because it requires the perturber to complete at least ten orbits at the distance where the gap-opening conditions are satisfied. This has important implications. While at first glance it validates the assumption made by simulations of tight MBH binaries embedded in CBDs, since the torques are very sensitive to the details of the flow, it means that the opening of a cavity, and therefore the appearance of a CBD is not simply controlled by standard gap/cavity opening conditions, so that it might arise at separations smaller than those indicated by such conditions (eventually it should happen as the secondary repels  enough of the surrounding gas with its gravitational perturbation). This means that the way the ICs of CBDs with embedded binary MBHs are normally setup (see \citealt{DeRosa_et_al_2020} for a review) might only apply to a subset of possible outcomes, because it neglects the ``memory effect'', namely the previous orbital history and the associated governing torques.  While computationally costly, it would be advisable to run a large number of high-resolution simulations of CNDs with binary MBHs starting at intermediate separations, of order of a few pc, and study the formation of a CBD and its configuration in a large sample of different ICs.

It is still possible that other mechanisms can arise in the gas and induce negative torques on the secondary MBH. For example, \cite{delvalle12} reported a trailing central ellipsoidal deformation developing at small separation in response to the perturbation of the binary. This should be captured in our highest-resolution simulations, but it is not seen. The difference might have to do with differences in the setups of the models, as we know that the orbital history of the MBH pair is highly sensitive to small differences in the gas flow at any stage of the evolution. Also, the mass ratio of the MBH binary will play a role. Here we used a fixed mass ratio $q = 1/20$, hence a very light secondary which probably induces a too small  perturbation on the surrounding disk material to generate a significant ellipsoidal perturbation in the local gas flow (note that in \citealt{escala05} the effect was first noted for very low mass ratio MBH binaries).

In the final stage of the highest-resolution run ({\it HReps05g14}), we do see that mini-disks are being assembled around both MBHs. These have been shown to play a role in extracting further orbital angular momentum at small separations, promoting decay \citep[e.g.][]{tang17}. Our simulations are not run for long enough, yet, to study the effect of the mini-disks. Shocked gas launched by them is also important for possible observations of electromagnetic counterparts. Regarding the disk material outside the orbit at small separations, in particular when the CBD cavity arises, resonances with the orbital period of the secondary may occur \citep{tang17}. The increasing periodicity of the ``torquing features'' described above  might reflect a role of resonances with the disk gas.

The mass ratio has been  kept fixed in this work because individual simulations are expensive enough to prevent a thorough exploration of the parameter space. However, one of the runs ({\it LReps50g12A}, see Table~\ref{tab:simulations}) includes accretion using the standard Bondi--Hoyle--Littleton sub-grid model and capping the accretion rate at Eddington (see also Paper~I). The mass ratio after $>$10~Myr changes by less than $10\%$, as both MBHs grow similarly in mass. However, as we know that, with increasing resolution, the amount of gas accumulating around the primary MBH due to numerical angular momentum loss diminishes, it is conceivable that the mass ratio would increase with increasing resolution. In addition, radiative cooling and feedback processes would also affect the evolution of the mass ratio. Simulations at larger scales, probing MBH separations from tens of pc to kpc, have typically found an increase of the mass ratio, by factors of even a few \citep[][]{callegari11,DeRosa_et_al_2020}, especially in the case of initially minor mergers \citep[][]{Capelo_et_al_2015}. Had we started with a larger mass ratio, the orbital decay of the secondary MBH would have been faster in the second phase -- the {\it migration} phase -- if we extrapolate the findings of \citet{malik2015}. On the other end, the third {\it slow} phase, which is the bottleneck in terms of timescales, might have started even earlier, as a more massive secondary MBH will be more effective at carving a gap or a CBD cavity (see Section~\ref{sec:methods}).

In addition to the resonance modes, the interaction between the MBHs and stars via three-body scattering could foster the orbital decay of the binary down to the scale of GW emission \citep[e.g.][]{sesana15, Bortolas2016, Bortolas2018tr, khan18}. Such process is not accounted for in our simulations; however, we can follow \citet{sesana15} to infer an order of magnitude for the hardening and coalescence time. First of all, binaries at the end of the presented simulations can be considered to be in the hardening phase, as they are either very close or below the hard binary separation $a_{\rm h}$ at which they eject stars with positive energies. Following, e.g. \citet{Merritt2013}, $a_{\rm h} = GM_2/(4\sigma^2)\sim 0.1$ pc for velocity dispersions $\sigma=30$--$90 {\rm \ km \ s^{-1}}$ (such range of velocity dispersions is compatible with the choice of density profiles about the primary MBH detailed in the next paragraphs). The timescale for the hardening and GW emission can be estimated as

\begin{equation} \label{eq:thard}
    t_{\rm gw}\approx \frac{\sigma_{\rm infl}}{GH_{\rm scat}\rho_{\rm infl}a_{\rm */gw}},
\end{equation}

\noindent where $H_{\rm scat} \approx 15$  is a parameter obtained via scattering experiments \citep{sesana15},  $\rho_{\rm infl}$ and $\sigma_{\rm infl}$ are, respectively, the stellar density and velocity dispersion at the binary's influence sphere, and $a_{\rm */gw}$ is a length-scale at which the binary shrinking rate transits from being dominated by three-body scatterings to being driven by GW emission; such scale can be estimated to be \citep{sesana15}

\begin{equation}
    a_{\rm */gw}^5 = \frac{64G^2M_1M_2(M_1+M_2)\sigma_{\rm infl}}{5c^5H_{\rm scat}\rho_{\rm infl}}f(e),
\end{equation}

\noindent where $c$ is the speed of light in vacuum and $f(e)=(1-e^2)^{-7/2}(1+73/24e^2+37/96e^4)$ is the eccentricity enhancement function for GW emission \citep{Peters1963}.

The inspiral timescale crucially depends on the assumptions for $\rho_{\rm infl}$, $\sigma_{\rm infl}$, and $e$. In the most pessimistic scenario, one could assume that only the stellar bulge would contribute to the hardening. The Plummer bulge considered in the present simulations yields central values for $\rho_{\rm infl}\approx 10^3 M_\odot {\rm pc}^{-3}$ and $\sigma_{\rm infl} \approx 80 {\rm \ km \ s^{-1}}$, so that the decay timescale is as long as 500--600~Myr if $e<0.5$. On the other hand, one would have to account at least for two additional important factors that could impact the decay timescale: (i) the enhancement of eccentricity during the hardening phase and (ii) the possible development of a dense stellar region about the primary MBH.

During the hardening phase, the eccentricity of a binary evolving in a spherical and isotropic or counter-rotating background has been found to grow with time prior to the GW induced decay \citep{Sesana2010, Sesana2011}. Such effect is particularly relevant for mass ratios in the range $10^{-1}$--$10^{-2}$ \citep{Bonetti2020}, and can bring binaries with initial $e\approx0.1$, which, if anything, is somewhat smaller than the eccentricity at the final stage of our runs, to eccentricities as large as 0.7--0.9 at the onset of the GW phase \citep{Sesana2010}. As a consequence, the decay timescale in the simulated Plummer bulge would drop to 100--300~Myr. The decay timescale for large $(e>0.7)$ eccentricities in the assumption of a different choice for the galactic bulge density profile, whose inner density behaves as $\rho(r) \propto r^{-\gamma_{b}}$, would remain in the range 100--300~Myr for $\gamma_b\lesssim 1.25$, and drop to less than $50$ Myr if $\gamma_b>1.7$\footnote{More in detail, in this situation $\sigma_{\rm infl}$ is given by  the $M_{\rm BH}$--$\sigma$ relation $(M_1/10^9M_\odot)=0.310(\sigma/200{\rm \ km\ s^{-1}})^{4.38}$  and $\rho_{\rm infl}$ is computed to be the density at the MBH influence radius, assuming a total stellar mass $M_*$ for the bulge given by the $M_{\rm BH}$--$M_{\rm bulge}$  relation $(M_1/10^9M_\odot)=0.49(M_{*}/10^{11} M_\odot)^{1.17}$ \citep{Kormendy2013}.}.

The binary hardening could also be aided by the development of a  dense nuclear cluster about the primary MBH (see, e.g. \citealt{Ogiya_et_al_2019}; see also discussions in \citealt{Tamfal_et_al_2018} and \citealt{Biava_et_al_2019}). In fact, it is expected that an increase in stellar density would be surely generated by star formation in the dense core of the CND, resulting from the ubiquitous gas inflows, both at large \citep[e.g.][]{wassenhove14, Capelo_Dotti_2017} and small scales (as seen for example in Figure~\ref{fig:density-maps}). We note that, in the simulations of CNDs with radiative cooling and star formation reported in Paper~I, the central stellar density of the CND was indeed reaching several thousand $M_\odot {\rm pc}^{-3}$ in the inner pc region. In addition, a nuclear stellar cluster can be generated or enhanced if stellar clusters  that form within the bulge efficiently decay by dynamical friction near the center of the system \citep{Antonini2012}, enhancing its central density and boosting the binary hardening rate \citep{Bortolas18, Arca-Sedda2019}. Let us assume stars in this dense cluster are distributed in a spherical and isotropic profile with power-law density   $\rho(r)\propto r^{-\gamma_b}$. The typical density and velocity dispersion within the primary sphere of influence, of radius $r_{\rm infl} = 10 (M_1/10^7 M_\odot )^{0.56} {\rm pc}$ \citep{Merritt2009},  can be estimated as $\rho_{\rm infl} = 2M_1/(4/3\pi r_{\rm infl}^3) \approx 5\times 10^3 M_\odot {\rm pc}^{-3}$ and $\sigma_{\rm infl}=(1+\gamma_b)^{-1}GM_1/r_{\rm infl}\approx 70$--$30 {\rm \ km \ s^{-1}}$ \citep{Alexander2005} for  $\gamma_b \in (0,2)$. It follows that the decay timescale would always be shorter than 130~Myr even assuming $e=0$.  Note that, in this situation, the eccentricity enhancement due to stellar scatterings can lead to timescales as short as $\sim$10~Myr, unless a significant fraction of interacting stars (e.g. the ones formed via the fragmentation of the CND) co-rotate with the binary: in this latter case, the binary is expected to circularize \citep{Sesana2011}.

In summary, it is likely that hardening down to the GW emission phase will be ultimately driven  by stellar dynamical processes, but aided by gas dissipation and star formation ensuing high enough  nuclear densities for such processes to be efficient, in line with the findings of \citet{khan2016} for the case of the merger remnants of massive galaxy hosts at high redshift.

\section{Conclusions}

We have studied the evolution of  MBHs in CNDs with a suite of adiabatic simulations that, overall, probes smaller spatial separations than in previous work. We have focused on unequal-mass binaries with masses in the LISA detection window for MBH mergers. Some of our simulations do indeed probe separations below $0.1$ pc, hence reaching the scale of CBD simulations of tight sub-pc scale MBH binaries. We have analyzed the character of the orbital decay process and the opening of gaps and cavities as a function of resolution, both mass and spatial, and of the EoS. Our findings can be summarized as follows:

\begin{itemize}
    \item The orbital decay of MBH binaries in smooth gaseous CNDs is fundamentally stochastic in character, as  small variations in setup, resolution, or thermodynamical description can lead to a significantly different behaviour of the decay process.
    
    \item Despite the observed stochasticity, three phases are generally identified in the orbital decay: an initially slow one governed by dynamical friction, in which the orbit circularizes; a fast phase driven by co-orbital disk-driven torques; and a third slow phase, below pc separations, which can lead to an almost complete stalling of the binary as torques from fluid elements at a range of distances combine to yield a negligible net negative torque.
     
    \item Gaps and cavities are always suppressed by fast migration due to co-orbital torques, except in the final slow phase, in which the secondary MBH can complete more than ten orbits under conditions favourable to opening a gap. Therefore, it is likely the onset of slow migration or stalling to allow a cavity to form, and not viceversa, in agreement  with results found for massive migrating planets in massive self-gravitating protoplanetary disks.
      
    \item At higher resolution, the CND is colder, as it is when the EoS is softer, which allows to open a cavity in the final slow decay phase. This results in the formation of a CBD and mini-disks around the two MBHs, qualitatively akin to what is seen in CBD simulations with embedded MBH binaries.
       
    \item Standard gap opening criteria can be easily recalibrated using a phenomenonological parameter $g$ accounting implicitly for the complex effects of migration, self-gravity, and thermodynamics.
       
    \item Despite the initial circularization, the MBH binary always maintains a non-zero eccentricity towards the end, including in cases where a CBD and a cavity form.
      
    \item While subsequent hardening might be driven by the complex torquing action between the mini-disks and the CBD, the conditions in the core of the CND, even inside the cavity, are of enough high stellar density to ensure efficient hardening down to the GW emission scale via three-body encounters between stars and the MBH binary.
     
\end{itemize}

We caution that our results were obtained within an idealized framework, wherein gas cooling, star formation, and feedback (from both stars and MBHs) were not modelled. Adding these ingredients would possibly affect some of the conclusions, as shown in previous work (e.g. Paper~I; \citealt{park17}).
In particular, the inclusion of radiative cooling, even in the presence of heating by feedback processes, can trigger fragmentation due to gravitational
instability, leading to a clumpy rather than a smooth medium, and consequently to other phenomena stifling orbital decay such as scattering and ejection
of the secondary MBH (see Peper~I). Note that the clumpy regime is not necessarily more realistic then the smooth CNDs simulated here because the dominance
of one or the other regime depends on the very uncertain balance between radiative cooling and heating by the several possible feedback effects,
from direct and indirect heating via stellar irradiation, to SN explosions and AGN feedback.

We have argued that the gap opening criterion can be adjusted via the fudge parameter $g$ to be predictive in scenarios that do not 
match the assumptions under which the criterion was derived. It is worth noting, however, that the criterion is a function of several 
other parameters. Most of them are correlated and may vary depending on stable/relaxed disk configurations. A caveat in our work, for instance, 
is that we limited our analysis to one disk height profile. We expect the height profile of a relaxed self-gravitating, 
pressure-supported disk to be dependent on its EoS, temperature, turbulence, energy dissipation mechanisms, and the 
gravitational potential it lies in (thus, also the force resolution in the case of simulations), among others. Here, perhaps 
the main value of our analysis is in the proposition of a way to visualize the dependency of the gap-opening criterion on key parameters, 
in order to gain insight on a range of different scenarios. For instance, we could imagine a system with a different enclosed mass  profile
from that considered here  and infer the temperature that would be required to yield a similar prediction of the allowed gap-opening region.

To conclude, we should recall that our 
 study is one of the first to explore the orbital decay of MBHs in CNDs from relatively large scales ($\sim$100~pc) down to the hardening scale ($\sim$10$^{-2}$~pc). Our results highlight that the orbital decay in purely gaseous frameworks becomes inefficient if small scales are well resolved, suggesting that stellar hardening or other physical mechanisms have to be invoked to guarantee the inspiral down to the GW-emission phase.


\bigskip

This work is supported by the STARFORM Sinergia Project funded by the Swiss National Science Foundation. EB, PRC, and LM acknowledge support from the Swiss National Science Foundation under the Grant 200020\_178949. Some of the analysis was performed using the open source \textsc{Pynbody} package \citep{pontzen13}.


\bibliographystyle{aasjournal}
\bibliography{references}

\appendix

\section{Additional plots}\label{sec:plots}

\begin{figure}
    \centering
    \includegraphics[width=.95\linewidth]{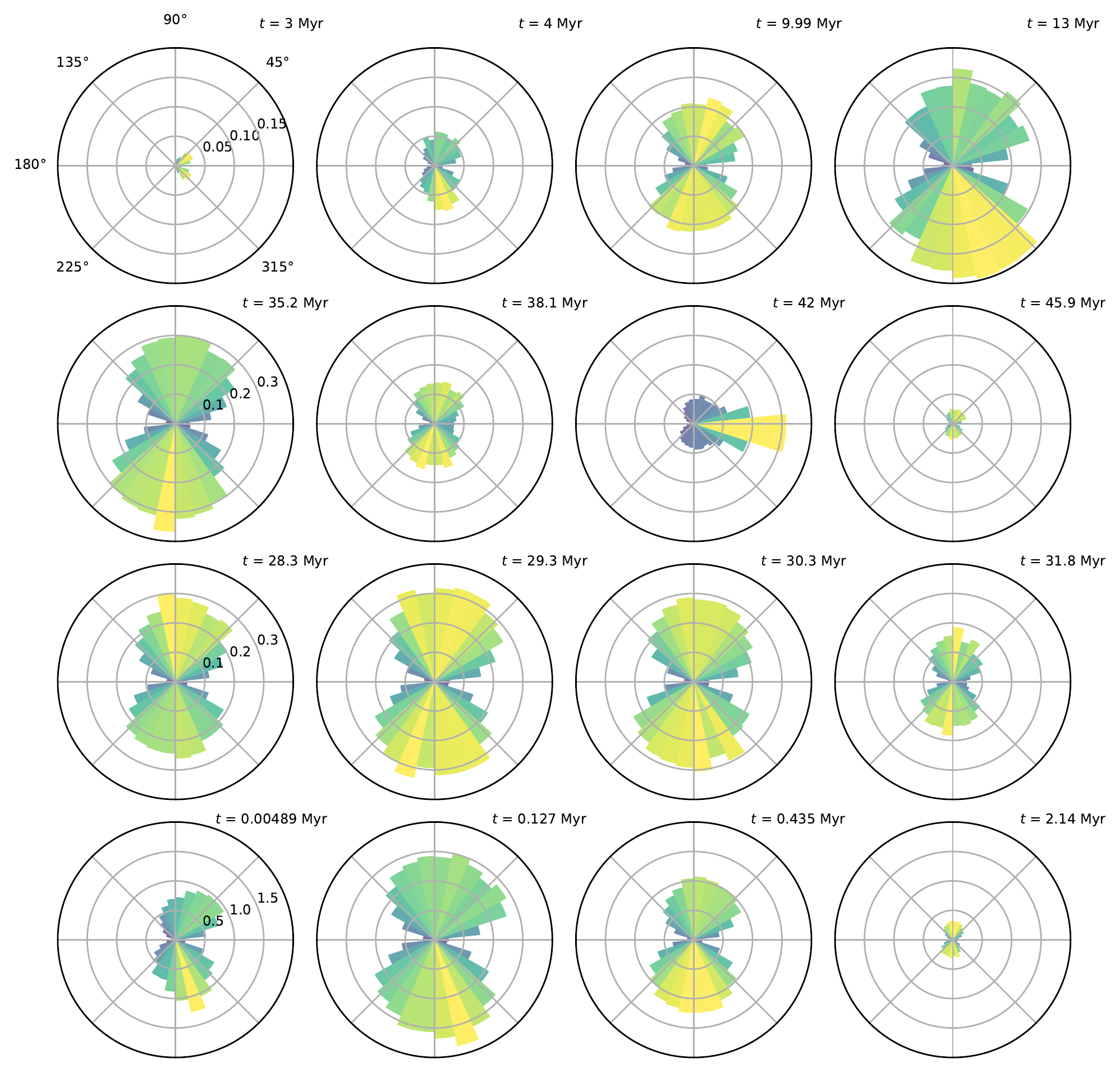}
    \caption{Polar histogram of the torque relative to MBH2's angular momentum for snapshots as in Figure~\ref{fig:torque-maps}, namely, in each row: \emph{LReps50g14}, \emph{MReps23g14R}, \emph{MReps23g11R}, and \emph{HReps05g14}. The angular space is binned by $12^{\circ}$ angle intervals with origin on the secondary MBH and measured from the direction of the primary MBH in the binary orbital plane. The ``radial'' axis represents the absolute value of $\tau/L$ in units of Myr$^{-1}$ (the colors are just a visual cue for this quantity). Its range -- from the origin to the black circle -- is divided in equal-sized intervals and the values corresponding to the grey circles are indicated in first panel of each row. The values corresponding to positive torques are displayed on the upper half of each panel, and negative ones are on the lower half. All ``resolved'' gas particles in an angular bin are considered, irrespective to their distance to the secondary MBH.}
    \label{fig:torque-directional-map-row}
\end{figure}

\begin{figure}
    \centering
    \includegraphics[width=.95\linewidth]{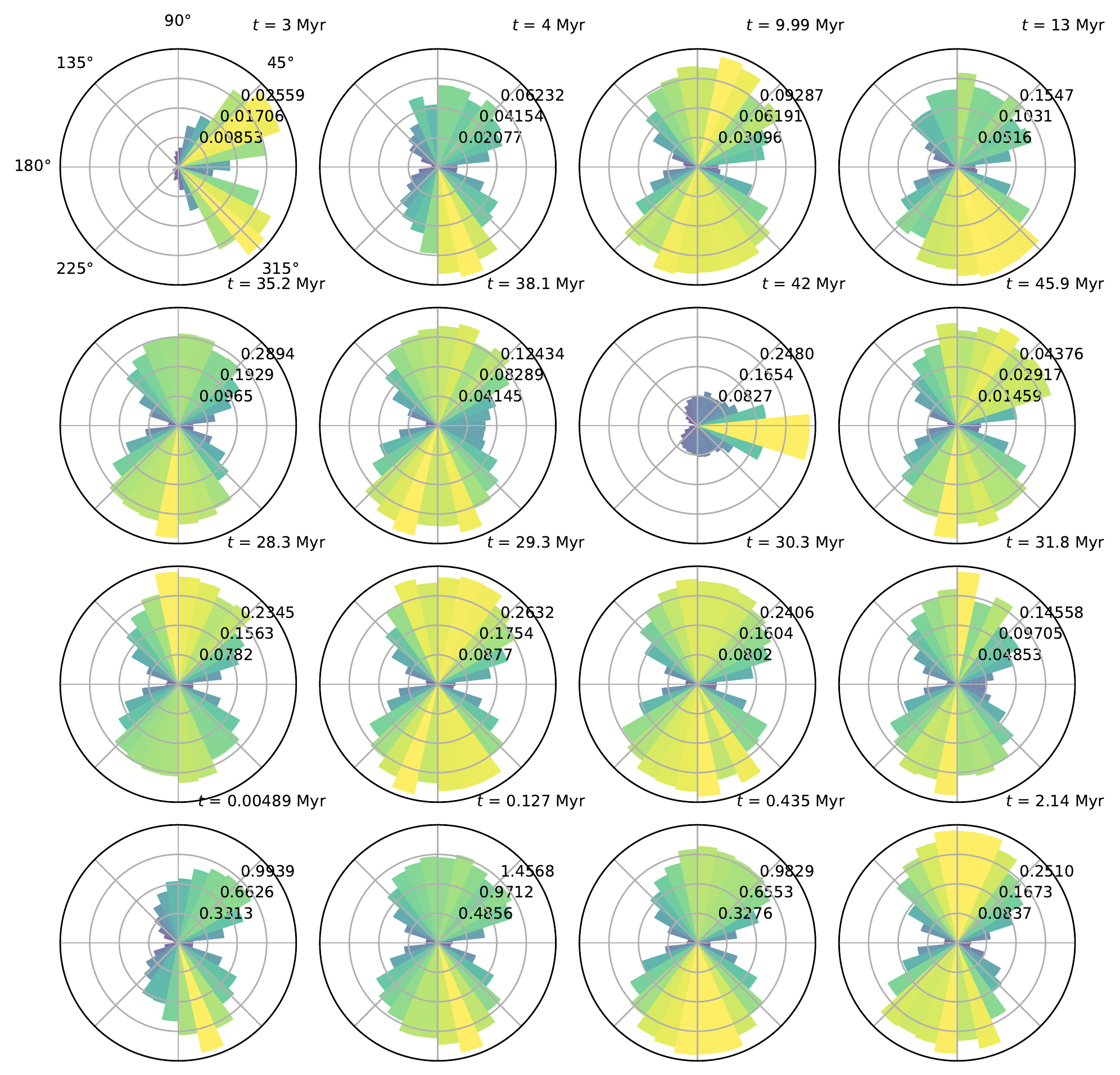}
    \caption{Polar histogram of the torque relative to MBH2's angular momentum for snapshots as in Figure~\ref{fig:torque-directional-map-row}. Here, the radial range (i.e. the numbers associated to the gray circles) is adapted to each plot, instead of being fixed for panels referring to the same simulation (as in Figure~\ref{fig:torque-directional-map-row}).}
    \label{fig:torque-directional-map-panel}
\end{figure}

\begin{figure}
    \centering
    \includegraphics[width=.95\linewidth]{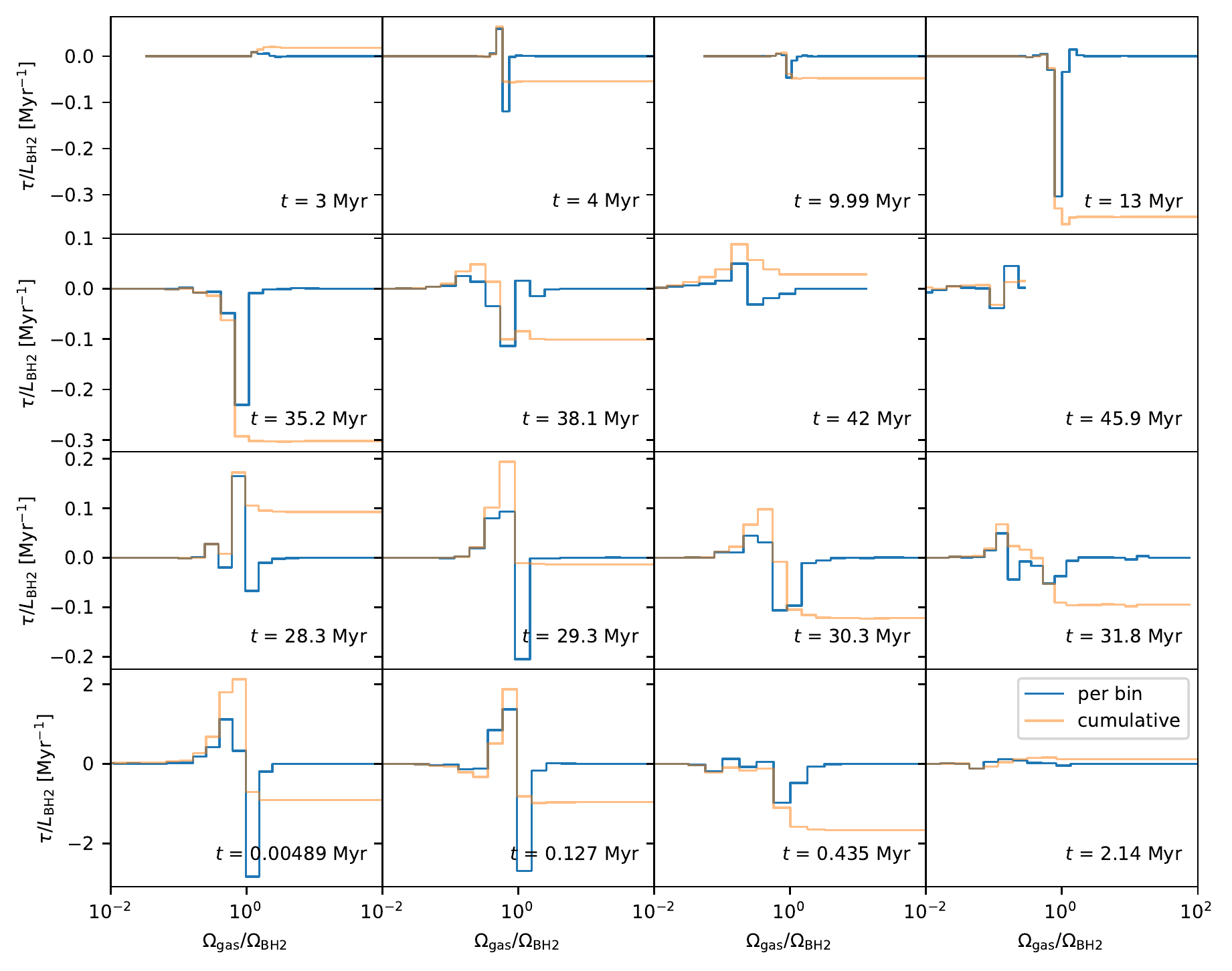}
    \caption{
    Relative torque profiles binned by the gas orbital frequency. The frequency range in each panel is determined by its minimum and maximum value in a snapshot, then it is binned in 51 intervals in logarithmic scale. The yellow lines show cumulative torque profiles, the blue lines refer to differential torques (\textit{i.e.}, the total torque in per orbital frequency bin). Each panel in a row shares the y-axis, which is indicated by the first panel of each row. The x-axis range in all panels is the same as indicated by the bottom-right panel. When the graphs are truncated on the left side, it indicates the outskirts of the disk; truncation to the right indicates lack of data due to reaching the softening scale (resolution limit).}
    \label{fig:torque-profiles-by-orbital-frequency}
\end{figure}

\end{document}